\def\cm{cm$^{-1}$}
\documentclass[rmp,twocolumn]{revtex4}
\usepackage{graphics}
\usepackage{epsfig}
\usepackage{amsmath,amssymb,graphicx,bm}
\usepackage{color}

\renewcommand{\vr}{{\mathbf{r}}}
\newcommand{\vk}{{\mathbf{k}}}

\newcommand{\vq}{{\mathbf{q}}}

\newcommand{\Tr}{\mathrm{Tr}}

\renewcommand{\Im}{\textrm{Im}}
\renewcommand{\Re}{\textrm{Re}}

\begin{document}
\author{Dimitri N. Basov}
\affiliation{Department of Physics, University of California San Diego, La Jolla, California 92093-0319, U.S.A.}
\author{Richard D. Averitt}
\affiliation{Department of Physics, Boston University, Boston, Massachusetts 02215, U.S.A.}
\author{Dirk van der Marel}
\affiliation{D{\'e}partment de Physique de la Mati{\`e}re Condens{\'e}e, Universit{\'e} de Gen{\`e}ve, CH-1211 Gen{\`e}ve 4, Switzerland}
\author{Martin Dressel}
\affiliation{1.~Physikalisches Institut, Universit{\"a}t Stuttgart, Pfaffenwaldring 57, 70550 Stuttgart, Germany}
\author{Kristjan Haule}
\affiliation{Department of Physics, Rutgers University, Piscataway, NJ 08854, USA}
\title{Electrodynamics of Correlated Electron Materials}

\begin{abstract} We review studies of the electromagnetic response of
various classes of correlated electron materials including
transition metal oxides, organic and molecular conductors,
intermetallic compounds with $d$- and $f$-electrons as well as
magnetic semiconductors. Optical inquiry into correlations in all
these diverse systems is enabled by experimental access to the
fundamental characteristics of an ensemble of electrons including
their self-energy and kinetic energy. Steady-state spectroscopy
carried out over a broad range of frequencies from microwaves to
UV light and fast optics time-resolved techniques provide
complimentary prospectives on correlations. Because the
theoretical understanding of strong correlations is still
evolving, the review is focused on the analysis of the universal
trends that are emerging out of a large body of experimental data
augmented where possible with insights from numerical studies.
\end{abstract}

\maketitle \tableofcontents

\section{Introduction}
\label{sec:Introduction}
In their report on the Conference on the Conduction of Electricity
in Solids held in Bristol in July 1937 Peierls and Mott wrote:
``Considerable surprise was expressed by several speakers that in
crystals such as NiO in which the $d$-band of the metal atoms were
incomplete, the potential barriers between the atoms should be high
enough to reduce the conductivity by such an enormous factor as
10$^{10}$'' \cite{mott1937a}. The ``surprise'' was quite
understandable. The quantum mechanical description of electrons in
solids - the band theory, developed in the late 1920-s
\cite{sommerfeld1928a,bethe1928a,bloch1929a} - offered a
straightforward account for distinctions between insulators and
metals. Furthermore, the band theory has elucidated why interactions
between 10$^{23}$ cm$^{-3}$ electrons in simple metals can be
readily neglected thus validating inferences of free electron
models. According to the band theory NiO (along with many other
transition metal oxides) are expected to be metals in conflict with
experimental findings. The term ``Mott insulator'' was later coined
to identify a class of solids violating the above fundamental
expectations of band theory. Peierls and Mott continued their
seminal 1937 report by stating that ``a rather drastic modification
of the present electron theory of metals would be necessary in order
to take these facts into account'' and proposed that such a
modification must include Coulomb interactions between the
electrons. Arguably, it was this brief paper that has launched
systematic studies of interactions and correlations of electrons in
solids. Ever since, the quest to fully understand correlated
electrons has remained in the vanguard of condensed matter physics.
More recent investigations showed that strong interactions are not
specific to transition metal oxides. A variety of $d-$ and
$f-$electron intermetallic compounds as well as a number of
$\pi$-electron organic conductors also reveal correlations. In this
review we attempt to analyze the rich physics of correlated
electrons probed by optical methods focusing on common attributes
revealed by diverse materials.

Central to the problem of strong correlations is an interplay
between the itineracy  of electrons in solids originating from
wavefunction hybridization and localizing effects often rooted in
electron-electron repulsion \cite{millis2004a}. Information on this
interplay is encoded in experimental observables registering the
electron motion in solids under the influence of the electric field.
For that reason experimental and theoretical studies of the
electromagnetic response are indispensable for the exploration of
correlations. In Mott insulators Coulomb repulsion dominates over
all other processes and blocks electron motion at low
temperatures/energies. This behavior is readily detected in optical
spectra revealing an energy gap in absorption. If a conducting state
is induced in a Mott insulator by changes of temperature and/or
doping, then optical experiments uncover stark departures from
conventional free electron behavior.

\begin{figure}
\centering
\includegraphics[width=3.0in]{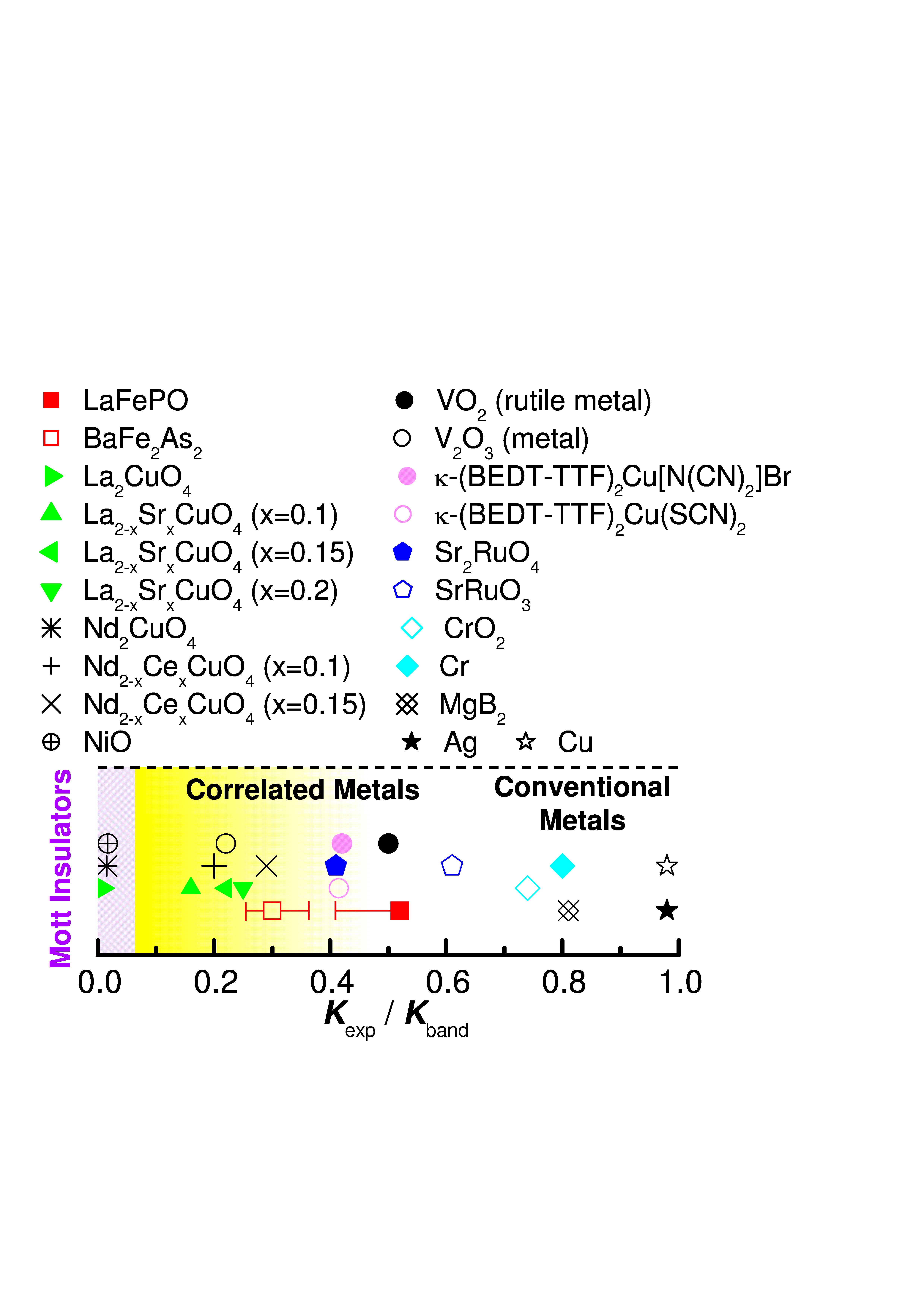}
\caption{(Color online)
The ratio of the experimental kinetic energy and the kinetic
energy from band theory $K_{\rm exp}/K_{\rm band}$
for various classes of correlated metals and also for conventional
metals. The data points are offset in the vertical direction for
clarity. Adapted from \textcite{qazilbash:075107}.}
\label{fig:KE-all}
\end{figure}

Of particular interest is the kinetic energy $K$ of mobile electrons
that can be experimentally determined from the sum rule analysis of
optical  data (Section \ref{subsec:Sum rules}) and theoretically
from band structure calculations.  As a rule, experimental results
for itinerant electronic systems are in  good agreement with the
band structure findings leading to $K_{\rm exp}/K_{\rm band}\simeq
1$  in simple metals (see Fig.1). However, in correlated systems
strong Coulomb interaction which has spin and orbital components
\cite{slater1929a} impedes the motion of electrons leading to the
breakdown of the simple single-particle picture of transport.  Thus
interactions compete  with itinerancy of electrons favoring their
localization, and specifically suppress the $K_{\rm exp}/K_{\rm
band}$ value below unity (see Fig.~\ref{fig:KE-all}). This latter
aspect of correlated systems appears to be quite generic and  in
fact can be used as a working definition of correlated electron
materials. Correlation effects are believed to be at the heart of
many yet unsolved enigmas of contemporary physics including
high-T$_c$ superconductivity (Section \ref{subsec:Cuprates}), the
metal-insulator transition (Section \ref{sec:Probing Metal Insulator
Transition in Frequency and Time Domain}, electronic phase
separation (Section \ref{subsec:Electronic phase separation}), and
quantum criticality (Section \ref{subsec:Power law behaviour of
optical constants and quantum criticality}).

Optical methods are emerging as a primary probe of correlations.
Apart from monitoring the kinetic energy, experimental studies of
the electromagnetic response over a broad energy range
(Section \ref{subsec:Spectroscopy in frequency domain}) allow one to examine
all essential energy scales in solids associated both  with
elementary excitations and collective modes (Section
\ref{sec:Excitations in Correlated Electron Matter}). Complementary
to this are insights inferred from time domain measurements allowing
one to directly investigate dynamical properties of correlated
matter (Section \ref{sec:Probing Metal Insulator Transition in
Frequency and Time Domain}). For these reasons, optical studies have
immensely advanced the physics of some of the most fascinating
many-body phenomena in correlated electron systems.

Importantly, spectroscopic results provide an experimental
foundation for tests of theoretical models. The complexity
of the the problem of correlated electrons poses difficulties
for the theoretical analysis of many of their properties.
Significant progress has been recently achieved by computational
techniques including the Dynamical Mean Field Theory (DMFT) offering
in many cases an accurate perspective on the observed behavior
(Section \ref{subsec:Dynamical Mean Field Theory}).  The ability of
the DMFT formalism to produce characteristics that can be directly
compared to spectroscopic observables is particularly relevant to
the main topic of this review.

\begin{figure}
\centering
\includegraphics[width=3.3in]{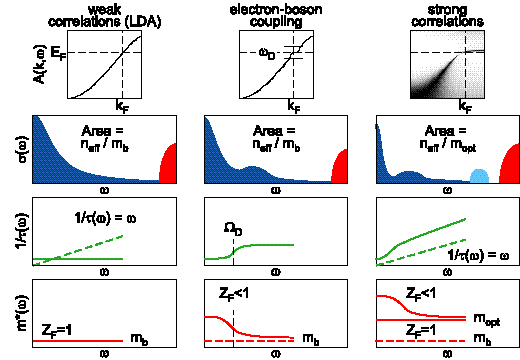}
\caption
{(Color online) Schematic diagram revealing complimentary approaches to
probing electronic correlations using IR and optical methods.
Top panels show the momentum-resolved spectral function in a non-interacting
metal (left), weakly interacting system (middle) and strongly correlated
system (right). Characteristic forms of the real
part of the conductivity $\sigma_1(\omega)$, the frequency-dependent scattering rate
$1/\tau(\omega)$ and effective mass $m^*(\omega)$ are displayed.
The Drude intraband contribution to the conductivity (blue) develops
a ``side-band''  in a system with strong electron-boson coupling.
The corresponding enhancement of $m*(\omega)$ at energies below a
characteristic bosonic mode $\Omega_0$ can be registered through the
extended Drude analysis (\ref{subsec:general optic theory}). The
magnitude of $m^*(\omega\rightarrow 0)$ is related to the
quasiparticle renormalization amplitude $Z$ introduced in
\ref{subsec:general optic theory}.  In a strongly correlated system (right panels)
the oscillator strength of the entire intraband contribution is
suppressed  with the spectral weight transfer to the energy scale of
the order of $U$ (light blue). The strength of this effect can be
quantified through the ration of $K_{\rm exp}/K_b$ as in
Fig.\ref{fig:KE-all} or equivalently through the ratio of optical
and band mass: $m_{b}/m_{\rm opt}$. Quite commonly this renormalization
effect and strong electron-boson interaction act in concert yielding
further enhancement of $m^*$ over the $m_{\rm opt}$ at
$\omega<\Omega_0$. }
\label{fig:correlations}
\end{figure}

In Fig.~\ref{fig:correlations} we schematically show possible
approaches towards an optical probe of interactions. It is
instructive to start this discussion with a reference to Fermi
liquids (left panels) where the role of interactions is reduced to
mild corrections of susceptibilities of the free electron gas
\cite{Mahan00}. The complex optical conductivity
$\tilde{\sigma}(\omega) =\sigma_1(\omega) + {\rm i}\sigma_2(\omega)$ of FL
quasiparticles residing in a partially filled parabolic band is
adequately described by the Drude model (see Subsection
\ref{subsec:Spectroscopy in frequency domain} for the definition of
the complex conductivity). The model prescribes the Lorentzian form
of the real part of the conductivity associated with the intraband
processes\cite{Drude00,DresselGruner02,Dressel06}:
\begin{equation}
\tilde{\sigma}(\omega)= \frac{N_{eff}e^2\tau_D}{m_b}\frac{1}{1-{\rm i}
\omega\tau_D} =  \frac{\sigma_{\rm dc}}{1- {\rm i} \omega\tau_D} \quad ,
\label{eq:Drude}
\end{equation}
where $e$ is the electronic charge, $N_{eff}$ is the relevant density, and
$m_b$ the band mass of the carriers which is generally different
from the free electron mass $m_e$, $1/\tau_D$ is the scattering rate
and $\sigma_{\rm dc}$ is the DC conductivity.  In dirty metals
impurities dominate and the scattering rate $1/\tau_D $ is
independent of frequency thus obscuring the quadratic form of
$1/\tau(\omega)$ that is expected for electron-electron scattering
of a Fermi liquid.\footnote{See, for example,
\textcite{AshcroftMermin76,Pines66,Abrikosov63}} Nevertheless, this
latter behavior of $1/\tau(\omega)$ has been confirmed at least for
two elemental metals (Cr and $\gamma$-Ce) through optical
experiments \cite{PhysRevB.65.054516,PhysRevLett.86.3407} using the
so-called extended Drude analysis (Section \ref{subsec:Spectroscopy
in frequency domain}). Another characteristic feature of Fermi
liquids in the context of infrared data is that the relaxation rate
of quasiparticles at finite energies is smaller than their energy:
$1/\tau(\omega)<\omega$ (at temperature $T\rightarrow 0$). The
contribution of interband transitions is sketched in red and is
usually adequately described through band structure calculations.
The band structure results also accurately predict the electronic
kinetic energy of a Fermi liquid that is proportional to the area
under the intraband Drude contribution to the conductivity spectra
(Section \ref{subsec:Sum rules}).

One of the best understood examples of interactions is the
Eliashberg theory of the electron-boson coupling
\cite{RevModPhys.62.1027}. Interactions with a bosonic mode at
$\Omega_{0}$ modify the dispersion of electronic states near the Fermi energy $E_{F}$
(top panel in the middle row of Fig.~\ref{fig:correlations}). The spectra of
$1/\tau(\omega)$ reveal a threshold near $\Omega_0$ reflecting an
enhancement of the probability of scattering processes at
$\omega>\Omega_{0}$. The spectral form of $\sigma_1(\omega)$ is
modified as well, revealing a the development of a ``side band'' in $\sigma_1(\omega)$ at $\omega>\Omega_{0}$. However, the total spectral weight including the coherent Drude like structure and side bands (dark blue area in Fig.\ref{fig:correlations}) is nearly unaltered compared to a non-interacting
system and these small changes are usually neglected. Thus, electron-boson interaction alone does not modify $K_{\rm exp}$ with respect to $K_{\rm band}$.  Importantly, characteristic features of the bosonic spectrum
can be extracted from the optical data
\cite{PhysRevB.10.2799}. Various analysis protocols employed for
this extraction are reviewed in Section \ref{subsec:Electron-boson interaction}. Coupling to other excitations, including magnetic resonances, also leads to the formation of side bands that in complex system may form a broad incoherent background in $\sigma_1(\omega)$.

The right panels in Fig.~\ref{fig:correlations} exemplify the
characteristic electronic dispersion and typical forms of the
optical functions for a correlated metal. Strong broadening of the
dispersion away from $E_F$ indicates that the concept of weakly
damped Landau quasiparticles may not be applicable to many
correlated systems over the entire energy range. An optical
counterpart of the broadened dispersion is the large values of
$1/\tau(\omega)$. Finally, the low-energy spectral weight is
significantly reduced compared to band structure expectations
leading to $K_{\rm exp}/K_{\rm band}$ that is substanitially less
than unity. Suppression of the coherent Drude conductivity implies
transfer of electronic spectral weight to energies of the order of
intrasite Coulomb energy $U$ and/or energy scale of interband
transitions. These effects are routinely found in doped Mott
insulators for example (Section \ref{sec:Transition Metal Oxides})
as well in other classes of correlated materials.\footnote{In
transition metal oxides the magnitude of the onsite Coulomb
repulsion can be both smaller or larger than the energy scale of
interband transitions~\cite{zaanen85}. In organic conductors the
hierarchy of energy scales is consistent with a sketch in
Fig.\ref{fig:correlations}.}

It is instructive to discuss dynamical properties of correlated
electron systems in terms of the effective mass which in general
is a tensor $m_k$. For a general dispersion $\epsilon_k$ the mass
is defined as  $m_k^{-1}=
\frac{1}{\hbar^2}\frac{\partial^2\epsilon_k}{\partial k^2}$, which
reduces to a constant for free electrons with a parabolic
dispersion. Deviations of $m_k$ from the free electron mass in
simple metals are adequately described by band structure
calculations yielding $m_{b}$. This quantity is frequency
independent (bottom left frame in Fig.~\ref{fig:correlations}) and
enters the Drude equation for the complex conductivity
Eq.~(\ref{eq:Drude}). Electron-boson interaction leads to the
enhancement of the effective mass compared to the band mass
$m_{b}$ at $\omega<\Omega_0$ as $m^*(\omega)=m_b[1+\lambda(\omega)]$,
quantifying the strength of the interaction $\lambda$ (middle panel in the
bottom row). The frequency dependence of $m^*(\omega)$ can be evaluated from the effective Drude analysis of the optical constants. Strong electron-electron interaction can
radically alter the entire dispersion so that $m_{\rm opt}$ is
significantly enhanced over $m_{b}$ (right bottom panel). An
equivalent statement is that $K_{\rm exp}$ is reduced compared to
$K_{\rm band}$ (see also Fig.\ref{fig:KE-all}). Additionally, electron-boson interactions may be
operational in concert with the correlations in modifying the
dispersion at $\omega<\Omega_0$. In this latter case one finds the behavior
schematically sketched in the bottom right panel of
Fig.~\ref{fig:correlations} with the thick red line.

Because multiple interactions play equally prominent roles in
correlated systems the resulting many-body state reveals a delicate
balance between localizing and delocalizing trends. This balance can
be easily disturbed by minute changes in the chemical composition,
temperature, applied pressure, electric and/or magnetic  field. Thus
correlated electron systems are prone to abrupt changes of
properties under applied stimuli and reveal a myriad of phase
transitions (Sections \ref{sec:Excitations in Correlated Electron
Matter}, \ref{sec:Transition Metal Oxides}). Quite commonly, it is
energetically favorable for correlated materials to form spatially
non-uniform electronic and/or magnetic states occurring  on diverse
length scales from atomic to mesoscopic. Real space inhomogeneities
are difficult to investigate using optical techniques because of the
fairly coarse spatial resolution imposed by the diffraction limit.
Nevertheless, methods of near field sub-diffractional optics are
appropriate for the task (Section \ref{subsec:Vanadium oxides}).

Our main objective in this review is to give a snapshot of recent
developments in the studies of electrodynamics of correlated
electron matter focusing primarily on works published over the last
decade. Introductory sections of this article are followed by the
discussion of  excitations and collective effects (Section
\ref{sec:Excitations in Correlated Electron Matter}) and metal-insulator transition physics
(Section \ref{sec:Probing Metal Insulator Transition in Frequency and Time Domain}) exemplifying
through optical properties  these essential aspects of correlated
electron phenomena. The second half of this review is arranged by specific
classes of correlated systems  for convenience of readers seeking a
brief representation of optical effects in a particular type of
correlated compounds.  Given the abundant literature on the subject,
this review is bound to be incomplete both in terms of topics
covered and references cited.  We conclude this account by outlining
unresolved issues.

\section{Experimental Probes and Theoretical Background}
\label{sec:Experimental Probes and Theoretical Background}

\subsection{Steady-state spectroscopy}
\label{subsec:Spectroscopy in frequency domain}
\begin{figure}
\centering
\includegraphics[width=3.3in]{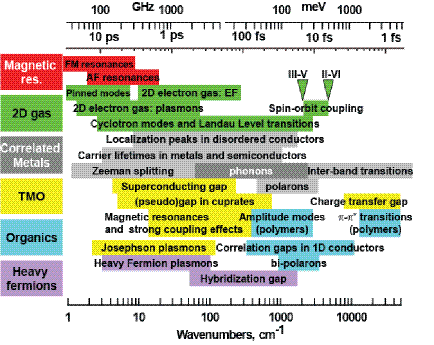}
\caption
{(Color online) Schematic representation of characteristic energy scales
in correlated electron systems. These different processes
give additive contributions to the dissipative parts of optical constants. TMO: transition metal oxides. }
\label{fig:introduction}
\end{figure}

Optical spectroscopy carried out in the frequency domain from 1 meV
to 10 eV has played a key role in establishing the present physical
picture of semiconductors and Fermi liquid metals
\cite{DresselGruner02,burch2008a} and has immensely contributed to
uncovering exotic properties of correlated materials
\cite{RevModPhys.70.1039,Degiorgi99,basov:721,millis2004a}.
Spectroscopic measurements in the frequency domain allow one to
evaluate the optical constants of materials that are introduced in
the context of ``materials parameters'' in Maxwell's equations. The
optical conductivity is the linear response function relating the
current $j$ to the applied electric field $E$:
$j(\omega)=\sigma(\omega)E(\omega)$. Another commonly employed
notation is that of the complex dielectric function
$\tilde{\epsilon}(\omega)=\epsilon_1(\omega)+{\rm i}\epsilon_2(\omega)$. The real
and imaginary parts of these two sets of optical constants are
related by $\sigma_{1}(\omega)=(\omega/4\pi)\epsilon_{2}(\omega)$
and $\sigma_{2}(\omega)=-(\omega/4\pi)[\epsilon_{1}(\omega)-1]$.\footnote{\label{footnote3} In general higher-energy contributions from interband transitions $\epsilon_b(\omega)$ ("bound charge" polarizability) are present apart from the quasi-free electrons that are summarized in $\epsilon_{\infty}$ replacing the factor 1 in this expression of $\sigma_2(\omega)$. The static "bound charge" polarizability is defined as the zero-frequency limit of  $\epsilon_b(\omega)$, {\em i.e.} $\epsilon_{\infty}=\epsilon_b(0)$.}
Absorption mechanisms associated with various excitations and
collective modes in solids (Fig.~\ref{fig:introduction}) give rise
to additive contributions to spectra of $\sigma_1(\omega)$ and thus
can be directly revealed through optical experiments. In anisotropic
materials the complex optical constants acquire a tensor form. For
instance, time reversal symmetry breaking by an applied magnetic
field introduces non-diagonal components to these tensors implying
interesting polarization effects \cite{zvezdin1997}. In the vast
majority of optics literature it is assumed that the magnetic
permeability of a material $\mu$=1 with the exception of magnetic
resonances usually occurring in microwave and very far infrared
frequencies.\footnote{This common assertion has recently been
challenged by the notion of ``infrared and optical magnetism''
\cite{ISI:000220000100034,ISI:000242622700020,ISI:000246293200016}
realized primarily in lithographically prepared metamaterial
structures but also in bulk colossal magneto-resistance manganites
\cite{PhysRevLett.95.247009,pimenov:197401}. For inhomogeneous media,
however, spatial dispersion becomes relevant that
in general mixes electric and magnetic components \cite{Agranovich84}.}

The complex optical constants can be inferred from one or several
complementary procedures \cite{DresselGruner02}: (i) a combination
of reflectance $R(\omega)$ and transmission $T(\omega)$ spectra
obtained for transparent materials can be used to extract the
dielectric function through analytic expressions, (ii)
Kramers-Kronig  (KK) analysis of $R(\omega)$  for opaque systems or
of $T(\omega)$  for transparent systems, (iii) ellipsometric
coefficients $\psi$ and $\Delta$ for either transparent or opaque
materials can be used to determine the dielectric function through
analytic expressions,\footnote{This is straightforward only in the case of isotropic bulk materials; in the case of anisotropic materials or films some models have to be assumed.}   (iv) various interferometric approaches, in
particular Mach-Zehnder interferometry,  and (v) THz time-domain
spectroscopy directly yield optical constants. These experimental
techniques have been extensively applied to correlated matter.
Extension of ``optical'' data to the microwave region is often
desirable especially for superconductors and heavy electron
materials which show interesting properties below 1 meV (Section
\ref{subsec:Intermetallic Compounds}).

\subsection{Pump probe spectroscopy}
\label{subsec:Spectroscopy in time domain}
Ultrafast optical spectroscopy provides the possibility to
temporally resolve phenomena in correlated electron matter at the
fundamental timescales of atomic and electronic motion.
Subpicosecond temporal resolution combined with spectral selectivity
enables detailed studies of electronic, spin, and lattice dynamics,
and crucially, the coupling between these degrees of freedom. In
this sense, ultrafast optical spectroscopy complements
time-integrated optical spectroscopy and offers unique possibilities
to investigate correlated electron materials. This includes, as
examples, phenomena such as electron-phonon coupling, charge-density
wave dynamics, condensate recovery, and quasiparticle formation.

In time resolved optical experiments, a pump pulse photoexcites a
sample initiating a dynamical response that is monitored with a time
delayed probe pulse. Experiments on correlated electron materials
fall into two categories as determined by the photoexcitation
fluence \cite{hilton2006a}. In the low-fluence regime
($\lesssim$100~$\mu$ J/cm$^{2}$) it is desirable to perturb the
sample as gently as possible to minimize the temperature increase.
Examples of low fluence experiments discussed below include
condensate dynamics in conventional and high-temperature
superconductors (Sections \ref{subsec:BCS Superconductors} and
\ref{subsec:Cuprates}, respectively), spin-lattice relaxation in
manganties (Section \ref{subsec:Manganites}), and electron phonon
coupling in Heavy-Fermions (Section \ref{subsec:VO$_{2}$ Time-Domain
Dynamics}). At the other extreme are high-fluence non-perturbative
experiments where goals include photoinducing phase transitions or
creating non-thermally accessible metastable states having a
well-defined order
parameter.\footnote{\textcite{nasu2004a,Yonemitsu08,
averitt2002a,kaindl_averitt2007a,
kuwatagonokami_koshihara2006a,hilton2006a}.} This is an emerging
area of research that is quite unique to ultrafast optical
spectroscopy. The coupling and interplay of correlated electron
materials are of considerable interest in these high-fluence
experiments as discussed in more detail in Sections
\ref{subsec:Photoinduced phase transitions} on photoinduced phase
transitions and \ref{subsec:VO$_{2}$ Time-Domain Dynamics} on the
vanadates.

Low and high-fluence time-resolved experiments have been made
possible by phenomenal advances in ultrashort optical pulse
technology during the past fifteen years which have enabled the
generation and detection of subpicosecond pulses from the
far-infrared through the visible and into the x-ray region of the
electromagnetic spectrum \cite{UPXIV2005}. Formally, ultrafast
optical spectroscopy is a nonlinear optical technique. In the
low-fluence regime, pump-probe experiments can be described in terms
of the third order nonlinear susceptibility. However, more insight
is often obtained by considering ultrafast optical spectroscopy as a
modulation spectroscopy where the self-referencing probe beam
measures the induced change in reflectivity $\Delta R/R$ or
transmission $\Delta T/T$ \cite{CardonaMS69,sun1993a}. This provides
an important connection with time-integrated optical spectroscopy
where the experimentally measured reflectivity and the extracted
dielectric response are the starting point to interpret and analyze
the results of measurements. Further, this is applicable to
high-fluence experiments from the perspective of temporally
resolving  spectral weight transfer (Section \ref{subsec:Sum
rules}). In femtosecond experiments, the dynamics can be interpreted
using the equation
\begin{equation}
\frac{\Delta R }{R }(t)=\frac{\partial \ln (R )}{%
\partial \epsilon _{1}}\Delta \epsilon _{1}(t)+\frac{\partial \ln (R)%
}{\partial \epsilon _{2}}\Delta \epsilon _{2}(t)
\end{equation}%
where $R$ is the reflectivity, and $\Delta \epsilon _{1\text{,
}}\Delta \epsilon_{2}$ are the induced changes in the real and
imaginary parts of the dielectric function, respectively
\cite{sun1993a}. Insights into the electronic properties obtained
from time-integrated measurements of $\epsilon_{1}+{\rm i}\epsilon
_{2}$ (or the complex conductivity $\sigma _{1}+{\rm i}\sigma_{2}$)
serve as a useful starting point to understand the quasiparticle
dynamics measured using time-resolved techniques. Further, the
development of time-gated detection techniques have enabled direct
measurement of the electric field which, in turn, permits the
determination of the temporal evolution of $\sigma _{1}+{\rm
i}\sigma _{2}$ over the useful spectral bandwidth of the probe
pulse.\footnote{See \textcite{averitt2002a,kaindl_averitt2007a} and
references therein for details.}

\begin{figure} [ptb]
\begin{center}
\includegraphics[width=2.8in,keepaspectratio=true]%
{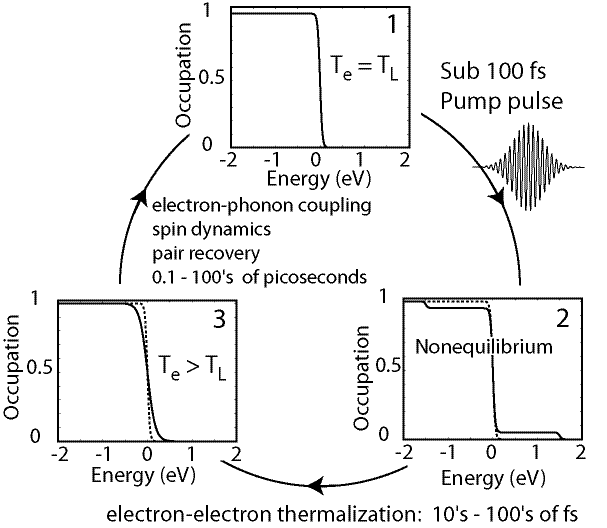}%
\caption{Schematic description of dynamics in condensed matter
probed with femtosecond spectroscopy. Prior to photoexcitation (1)
the electrons, lattice, and spins are in thermal equilibrium.
Photoexcitation creates (2) a nonthermal electron distribution. The
initial relaxation proceeds primarily through electron-electron
thermalization. Following thermalization, the electrons have excess
energy which is transferred to other degrees of freedom on
characteristic timescales ranging from approximately 1 ps for
electron-phonon relaxation to tens of picoseconds for processes such
as pair recovery across a gap. After \textcite{averitt2002a}.}
\label{averitt1}
\end{center}
\end{figure}

The foundation for ultrafast experiments on correlated electron
materials (at any fluence) is based on efforts during the past 25
years in understanding quasiparticle dynamics in semiconductors and
metals.\footnote{\textcite{shah1999a,chemla2001a,allen87a,Groeneveld95,sun1993a,axt2004a,beaurepaire1996a}.}
In ultrafast optical experiments, an incident
pump pulse perturbs (or prepares) a sample on a sub-100 fs time
scale. This induced change is probed with a second ultrashort pulse
that, depending on the wavelength and experimental setup, measures
pump induced changes in the reflectivity, transmission, or
conductivity. In the majority of experimental studies in condensed
matter to date, the pump pulse creates a nonthermal electron
distribution [Fig.~\ref{averitt1} (1 $\rightarrow$ 2)] fast enough
that, to first order, there is no coupling to other degrees of
freedom. During the first  100 fs, the nonthermal (and potentially
coherent) distribution relaxes primarily by electron-electron
scattering [Fig.~\ref{averitt1} (2 $\rightarrow$ 3)]
\cite{allen87a,fann1992a,Groeneveld95,sun1993a}. Subsequently, the
excited Fermi-Dirac distribution thermalizes through coupling to the
other degrees of freedom (3 $\rightarrow$ 1).

There are of course, important aspects that Fig.~\ref{averitt1} does
not capture. Of particular importance are coherence effects where
the impulsive nature of the initial photoexcitation leads to a
phase-coherent collective response \cite{shah1999a}. This can
include coherent phonons or magnons
\cite{ThomsenPRB1986,DekorsyPRB1996,koopmansPRL2005,talbayev2005a}.
However, even in the coherent limit, the results can often be
interpreted as a dynamic modulation of the optical conductivity
tensor though the connection with Raman scattering is important for
certain experiments \cite{MerlinCPR1997,MisochkoCPR2001}.

\begin{figure} [ptb]
\begin{center}
\includegraphics[width=2.8in,keepaspectratio=true]%
{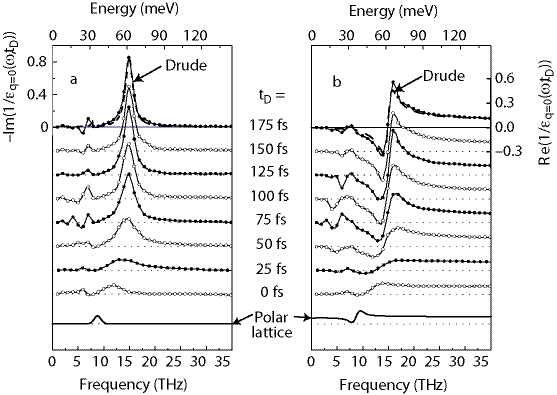}%
\caption{Quasiparticle formation in GaAs at room temperature following
excitation with 10 fs 1.55 eV pulses. The dynamic
loss function is plotted as a function of frequency
at various delays following photoexcitation. The
response evolves to a coherent Drude response on a
timescale of ~175 fs as dressed quasiparticles are
formed from an initially uncorrelated state at zero delay. From \textcite{huber2001a}.}%
\label{averitt2}
\end{center}
\end{figure}

An example that embodies what is possible with ultrafast optical
spectroscopy we consider recent results on the formation of
quasiparticles following above band-gap photoexcitation in undoped
GaAs \cite{huber2001a}.\footnote{These results provide a striking
example of the onset of correlation following photoexcitation. In
this experiment, pulses with 1.55 eV photon energy and approximately
10 fs duration excited an electron-hole plasma at a density of
$10^{18}$ cm$^{-3}$. Monitoring the dynamics requires probe pulses
with sufficient temporal resolution and with a spectral bandwidth
extending beyond 160 meV. This was achieved using a scheme based on
difference-frequency generation in GaSe combined with ultrabroadband
free-space electro-optic sampling \cite{huber2001a}.} The
experimental results are displayed in Fig.~\ref{averitt2}, where
spectra of the dynamic loss function
$1/\tilde{\epsilon}(\omega,\tau_{D})$ are plotted at various delays
$\tau_{D}$ between the optical pump and THz probe pulses. The
imaginary part of $1/\tilde{\epsilon}(\omega,\tau_{D})$ is plotted
in panel (a) and the real part in panel (b). This is a particularly
useful form to display the data as it highlights what this
experiment is actually measuring, namely, the evolution of particle
interactions from a bare Coulomb potential $V_{q}$ to a screened
interaction potential $W_{q}(\omega,\tau_{D})$, where q is the
momentum exchange between two particles during a collision.

In essence, $V_{q}$ becomes renormalized by the longitudinal
dielectric function leading to a retarded response associated with
the polarization cloud about the carriers. This is a many-body
resonance at the plasma frequency where the loss function peaks at
the plasma frequency with a width corresponding to the scattering
rate. Thus, the results of Fig.~\ref{averitt2} show the evolution
from an uncorrelated plasma to a many-body state with a well-defined
collective plasmon excitation. This is evident in
Fig.~\ref{averitt2}(a) where, prior to photoexcitation, there is a
well-defined peak at 36 meV corresponding to polar optical phonons.
Following photoexcitation, a broad resonance appears at higher
energies that evolves on a 100 fs timescale into a narrow plasma
resonance centered at 60 meV. The response is described by the Drude
model only at late delay times. These results are consistent with
quantum kinetic theories describing nonequilibrium Coulomb
scattering \cite{huber2001a}.

In wide-band materials, it is possible to consider the dynamics
largely in terms of the bandstructure where photoexcitation leads to
changes in band occupation followed by subsequent relaxation
processes. The example in Figure \ref{averitt2} is along these lines
highlighting the dynamical evolution of low-energy spectral weight
following photoexcitation of carriers across the badgap. In many
ways, this can be considered as a model example of measurements in
correlated electron materials in that it is the dynamical evolution
of spectral weight (even if only over a narrow spectral range) that
is monitored. The situation can be considerably more complicated in
correlated materials starting with the fact that the electronic
structure varies with occupancy. Thus, an excitation pulse can
initiate a sequence of dynamical events quite different in
comparison to the relaxation of a non-thermal electron distribution
in a rigid band. For example, a change in orbital occupancy upon
photoexcitation can near-instantaneously relax the need for a
coherent lattice distortion (e.g. cooperative Jahn-Teller effect in
the manganites) \cite{tokura2000a,Polli2007a} thereby launching a
coherent lattice excitation that will in turn couple to other
degrees a freedom. More generally, a delicate balance between
various degrees of freedom occurs. Consequently, many such materials
teeter on the edge of stability and exhibit gigantic
susceptibilities to small external perturbations
\cite{dagotto2003a,dagotto-science05}. Short optical pulses can play
an important role as the external perturbation yielding a powerful
tool to investigate dynamical interactions which determine the
macroscopic response. Many examples will be encountered in the
following sections.

The results presented in Fig.~\ref{averitt2} represent the
state-of-the art of what is currently feasible both in terms of
experiment and theory of ultrafast optical spectroscopy as applied
to condensed matter. The challenge is to utilize such experimental
tools to investigate more complicated materials. This includes, as
discussed in more detail below, the cuprates, manganites, heavy
fermions, organics, and others. Interesting experimental  insights
have been obtained, but there is a need for theoretical studies
focused on interpreting the results of time-resolved measurements.
While theoretical studies on dynamics in wide-band semiconductors
and heterostructures is relatively mature \cite{axt2004a}, to date,
there have been relatively few theoretical studies on dynamics in
correlated electron materials.\footnote{For example,
\textcite{unterhinninghofen2008a,howell2004a,takahashi2002a,Ahn04,Carbotte2004a}.}
As described in this review, DMFT is a promising approach to analyze
time-domain optical experiments and recently, a DMFT study along
these lines has been published \cite{eckstein2008a}.

\subsection{Theoretical Background}
\label{subsec:general optic theory}
\begin{figure}[!bt]
\centering{
\includegraphics[width=0.75\linewidth]{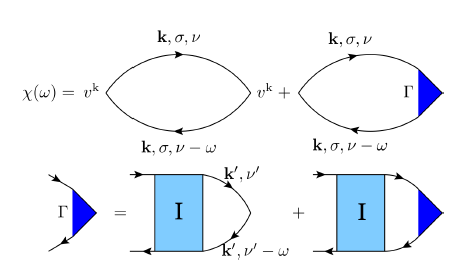}
}
\caption{(Color online) Diagrammatic representation of the
current-current
correlation function and the Bethe-Salpeter equation for the vertex
correction to the optical conductivity.  }
\label{optdiag}
\end{figure}

In an optical experiment a current is induced  in the solid by
the electric (proportional to $\partial\bf{A}/\partial t$ where
$\bf{A}$ is the vector potential) and magnetic ($\bf{B}$)
components of the electromagnetic field\cite{Cohen-Tannoudji}. The
coupling in leading order of $\bf{A}$ and $\bf{B}$ is
\begin{equation}
\label{eq:magnons1}
H^i=\frac{e}{2mc}\left[\bf{A}(\bf{r})\cdot\bf{p}+ \bf{p}\cdot\bf{A}(\bf{r})\right] - \frac{e\hbar g}{2mc} \bf{B}(\bf{r})\cdot \bf{S}
\end{equation}
The $\bf{A}\cdot \bf{p}$ term of the interaction couples the
angular momentum of the photon ($\hbar$) to the orbital degree of
freedom of the electron, leaving the spin unaffected. The
$\bf{B}\cdot \bf{S}$ term couples the photon angular momentum to
the spin of the electron. In the  absence of spin-orbit coupling
these two couplings lead to the electric- and magnetic dipole
selection rules respectively. Spin-orbit coupling relaxes these
rules, which provides a channel for optically induced spin-flip
processes through the $\bf{A}\cdot \bf{p}$ term. Since this
coupling contributes typically $1/\alpha^2$ times the oscillator
strength from the $\bf{B}\cdot \bf{S}$ term, the latter coupling
is usually neglected; here $\alpha=e^2/\hbar c=1/137$ is the
fine-structure constant. The optical conductivity
is then computed by the linear response
theory\cite{Mahan00}
\begin{eqnarray}
\tilde{\sigma}^{\mu\nu}_{\vq}(\omega) = \frac{{\rm i} e^2
n}{m\omega}\delta_{\mu\nu}+\frac{1}{{\rm i}\omega}\chi_{\mu\nu}(\vq,\omega+{\rm i}{\rm i}\delta)
\label{sigmar}
\end{eqnarray}
where
\begin{eqnarray}
\chi_{\mu\nu}(\vq,{\rm i}\omega_n)=\int_0^{1/T} e^{{\rm i}\omega_n\tau}\langle
j^p_{\mu,\vq}(\tau) j^p_{\nu,-\vq}(0) \rangle {\rm d}\tau
\label{curr-curr}
\end{eqnarray}
is the current-current correlation function, and $j^p$ is the
paramagnetic current density
$\textbf{j}^p(x)=-\frac{i e}{2m}\sum_\sigma
[\psi_\sigma^\dagger(x)\nabla\psi_\sigma(x)-(\nabla
\psi_\sigma^\dagger(x))\psi_\sigma(x)].$
$T$ is temperature.
Calculation of the current-current correlation function
Eq.~(\ref{curr-curr})  requires the full solution of the many-body
problem. Usually Eq.~(\ref{curr-curr}) is then expressed in terms
of the one-particle Green's function $G_{\vk}(\omega)$, the two
particle vertex function $\Gamma(\vk\nu,\vq\omega)$ and electron
velocities $v^{\vk\mu}$ by
\begin{widetext}
\begin{equation}
\chi_{\mu\nu}(\vq,i\omega_n)=
-e^2T \sum_{\vk\nu_m}
\textrm{Tr}[G_{\vk-\vq/2}({\rm i}\nu_m-{\rm i}\omega_n)
v^{\vk\nu}G_{\vk+\vq/2}(i\nu_m)v^{\vk \mu}
+G_{\vk-\vq/2}({\rm i}\nu_m-i\omega_n)
\Gamma(\vk\nu_m,\vq\omega_n) G_{\vk+\vq/2}({\rm i}\nu_m)v^{\vk \mu}],
\label{opt_chi}
\end{equation}
\end{widetext}
as diagrammatically depicted in Fig.~\ref{optdiag}.  All three
quantities are matrices in the band index, i.e., $v^{\vk\mu}_{ij}$,
$G_{\vk,ij}$, and $\Gamma(\vk,\vq)_{ij}$. The velocities are
$v^{\vk\mu}_{i j}=-\frac{i}{m}\langle\psi_{\vk
i}|\nabla_{\mu}|\psi_{\vk j} \rangle$, where $\psi_{\vk i}(\vr)$ are
a set of one particle basis functions.

Within a single band approximation, the Green's function
$G_{\vk}(\omega)$, the spectral function of electronic excitations
$A_{\vk}(\omega)$ and electronic self-energy $\Sigma_{\vk}(\omega)$
are related by:
\begin{eqnarray}
A_{\vk}(\omega ) =-\frac{1}{\pi}\Im G_{\vk}(\omega)=
-\frac{1}{\pi} \Im\frac{1}{\omega
-\epsilon_{\vk}-\Sigma_{\vk}(\omega)}, \label{eq:spectr-funct}
\end{eqnarray}
The self-energy $\Sigma_{\vk}(\omega)$ in
Eq.(\ref{eq:spectr-funct}) contains information on all possible
interactions of an electron with all other electrons in the system
and the lattice. In the absence of interactions the spectral
function is merely a $\delta$-peak at $\omega=\epsilon_\vk$ whereas
$\Re{\Sigma_{\vk}(\omega})$ = $\Im{\Sigma_{\vk}(\omega)}$ =0.
Interactions have a two-fold effect on the spectral function.
First, the energy is shifted with respect to non-interacting case
by the amount proportional $\Re{\Sigma_{\vk}(\omega})$. Second,
the spectral function acquires a Lorentzian form with the width
given by $\Im{\Sigma_{\vk}(\omega})$. The corresponding states are
referred to as dressed states or quasiparticle states.  The
spectral function as well as complex self-energy are both
experimentally accessible through photoemission
measurements \cite{damascelli2003a}.

Finally, the two particle vertex function $\Gamma(\vk,\vq)$ (dark
blue triangle in Fig.~\ref{optdiag}) can be computed from the
fully irreducible two particle vertex function $I(\vk\vk';\vq)$
(light blue square in Fig.~\ref{optdiag}) through the
Bethe-Salpeter equation depicted in the second line of
Fig.~\ref{optdiag}. A consequence of this vertex is that an
electron-hole pair can form a bound neutral particle, {\em i.e.}
an exciton.  In wide band insulators such as
rocksalts \cite{rocksalt_excitons},
semiconductors \cite{semiconductor_excitons}, or organic
materials \cite{organic_excitons}, the exciton binding energies
form a Rydberg series below the excition gap of unbound
electron-hole pairs. In transition metal compounds the
interaction is often strong enough to bind an electron-hole pair
on a single atomic site (Section \ref{subsec:charge_tansfer and
excitons}).

\subsection{Sum rules}
\label{subsec:Sum rules}
The response functions including optical constants of materials obey
numerous sum rules \cite{kubo1957a}. The most frequently used sum
rule is the f-sum rule for the real part of the optical conductivity
$\sigma_1(\omega)$:

\begin{equation}
\int_{0}^{\infty }\sigma_{1}(\omega ){\rm d}\omega ={\frac{{\pi ne^{2}}}{2{m}_{e}%
}} .
\label{eq:SR-global}
\end{equation}
This expression relates the integral of the dissipative part of the
optical conductivity to the density of particles participating in
optical absorption and their bare mass. The optical conductivity of
a solid is dominated by the electronic response and therefore an
integration of the data using Eq.~(\ref{eq:SR-global}) can be compared
to the total number of electrons including both core and valence
electrons.

A special role is played by the following sum rule for the optical conductivity of a single-band system
governed by a Hamiltonian $H$:
\begin{equation}
\int_{0}^{\infty }\sigma_{1}(\omega ){\rm d}\omega = - {\frac{{\pi
e^{2}}}{{2\hbar^2}}} \langle K\rangle . \label{eq:SR-KE}
\end{equation}
Here
$K=\delta^2 H/\delta A^2$ and the brackets $\langle\ldots\rangle$ denote  the
thermal average. In a tight binding model $K$ is the kinetic energy\cite{PhysRevB.16.2437,PhysRevB.35.8391}:
\begin{equation}
K= -4/N \sum_{\vk} \frac{d^2\epsilon_\vk}{dk_x^2}n_\vk,
\label{eq:KE-definition}
\end{equation}
where $n_{\bf k}$ is the electron momentum distribution function.
Since $K$ must accommodate the entire free carrier response ({\em i.e.}
Drude peak and all side-bands due to interactions), one has to extend
the integration to an energy above the free carrier response while
still below the interband transitions. Kinetic energy
Eq.~\ref{eq:KE-definition} quantifies the oscillator strength of
intraband transitions that can be equivalently characterized with the
plasma frequency $\omega_p^2=\frac{ne^2}{m_b}$ in weakly interacting
systems or $\frac{ne^2}{m_{\rm opt}}$ in a strongly interacting
material where correlations renormalize the entire dispersion so that
$m_{\rm opt}>m_{b}$ and $K_{\rm exp}/K_{\rm band}<1$.

\textcite{devreese1977} obtained the following ``partial sum rules'' for electrons occupying
a band with a $k$-independent mass, $m_b$, coupled to phonons
causing band mass $m_b$ to increase to the renormalized value m$^*$ at energies below the phonon frequencies:
\begin{eqnarray}
\int_{0}^{\infty}\sigma_{\rm Drude}(\omega){\rm d}\omega =\frac{\pi
n_{\rm eff}e^2}{2m^*}
\label{eq:SR-partial1}
\\
\int_{0}^{\infty}\sigma_{\rm intra}(\omega){\rm d}\omega =\frac{\pi
n_{\rm eff}e^2}{2m_b} \label{eq:SR-partial}
\end{eqnarray}
where $\sigma_{\rm Drude}(\omega)$ is the narrow Drude peak alone whereas
$\sigma_{\rm intra}(\omega)$ is the complete intraband contribution involving both the Drude peak and side-bands
resulting from electron-phonon coupling
(middle panels of Fig.\ref{fig:correlations}). A caveat: as pointed out above in many correlated electron materials the entire dispersion is modified by correlations leading to a suppression of the total intraband spectral weight \cite{qazilbash:075107}. This implies that $m_b$ in Eq.\ref{eq:SR-partial} has to be replaced with higher optical mass $m_{opt}$ and electron-boson mass renormalization is also executed with respect to $m_{opt}$, not $m_b$.  Following
\textcite{PhysRevB.16.2437} it is customary to define the
effective spectral weight $N_{\rm eff}(\omega)$:
\begin{equation}
N_{\rm eff}(\omega)=\frac{2 m_e}{\pi e^2}\int_{0}^{\omega }\sigma _{1}(\omega ' ){\rm
d}\omega ' \label{eq:n-eff}
\end{equation}
which has the meaning of the effective number of electrons
contributing to electromagnetic absorption at frequencies below
$\omega$.

Of special significance for superconductors is the
Ferrell-Glover-Tinkham (FGT) sum rule \cite{tinkham1996a}:
\begin{equation}
\frac{\rho _{s}}{8} =  \int_{0+}^{\infty }{\rm d}\omega \lbrack
\sigma _{1}(\omega ,T>T_{c})-\sigma _{1}(\omega ,T<T_{c})]
\label{eq:FGT}.
\end{equation}
This equation relates the spectral weight ``missing'' from the
real part of the conductivity upon transition to the
superconducting state to the superfluid density $\rho_s$ which is
proportional to the density of superconducting pairs $n_s$ and
inversely related to their effective mass $m^*$ as: $\rho
_{s}=4\pi n_{s}e^{2}/m^{\ast }$. Often, for practical reasons, the
integration is limited to the free carrier response. Validity of
the FGT sum-rule in this restricted sense requires that the
electronic kinetic energy is unchanged below $T_c$ (see Section
\ref{subsec:Cuprates} which discusses sum rule anomalies in
high-$T_c$ cuprates). The superfluid density is of fundamental
importance for the electrodynamics of superconductors. The sum
rule [Eq.~(\ref{eq:FGT})]  allows one to evaluate all three
diagonal components of the superfluid density in anisotropic
superconductors such as cuprates.
\footnote{\textcite{PhysRevB.50.3511,PhysRevLett.74.598,liu1999a,PhysRevLett.86.4144,PhysRevB.65.134511,ISI:000222946100041}
and iron pnictides\cite{li:107004}.}

Experimental access to the quasiparticle kinetic energy is one of
the important virtues of optical probes of correlations. An analysis
of the one dimensional Hubbard Hamiltonian is particularly
instructive in this regard \cite{PhysRevB.33.7247}. Exact results
obtained for a half-filled band reveal that the electronic kinetic
energy is monotonically reduced with the increase of the on-site
repulsion $U$ and tends to zero as $K\propto 1/U$. This result, along with
the analysis of the spectral weight within the two-dimensional
Hubbard model \cite{Millis:1990}, reinforces the notion that $K_{\rm
exp}$ reported in Fig.~\ref{fig:KE-all} can be used as a
quantitative measure of correlation effects.

Equation \ref{eq:SR-KE} is derived for a hypothetical single-band
system where the kinetic energy may depend on temperature $T$,
magnetic field $B$ or other external stimuli. Strong variations of
the electronic spectral weight  commonly found in correlated
electron systems upon changes of temperature $T$ or magnetic field
$B$  may signal interesting kinetic energy effects. Consider, for
example, a $\sigma_1(\omega)$ data set collected for a conducting system over the spectral
range that is at least of the order of the width of the electronic band $W$ where the Fermi energy resides.
The kinetic energy interpretation of Eq.~(\ref{eq:SR-KE}) applied to
such a data set is highly plausible. Quite commonly, one finds that
the sum rule results in this case are temperature dependent
\cite{Molegraaf2002,ortolani:067002}. The only source of such a
temperature dependence in a non-interacting system pertains to
thermal smearing of the Fermi-Dirac distribution function leading to
fairly weak effects scaling as $T^2/W$
\cite{Benfatto-05,benfatto:155115}. In correlated electron systems
this temperature dependence can become much more pronounced. This
latter issue has been explicitly addressed within the framework of
several scenarios of interacting electrons.\footnote{\cite{toschi-prl05,benfatto:155115,norman:220509, Karakozov200680,kuchinskii2008a,marsiglio:064507,PhysRevB.70.100504}.}
We wish to pause here to strike a note of caution and stress that
apart from intrinsic origins the variation of the electronic
spectral weight may be caused by ambiguities with the choice of
cut-off for integration of experimental
spectra~\cite{benfatto:533,norman:220509}. Indeed, in many realistic
materials including transition metal oxides intra- and inter-band
contributions to the conductivity spectra commonly overlap unlike
idealized schematics in
Fig.~\ref{fig:correlations}.\footnote{Examples of extensive experimental
literature on sum rule anomalies in correlated systems can be found
in the following references:
\cite{Basov-sci1999,PhysRevB.61.5930,PhysRevB.63.134514,Molegraaf2002,
Kuzmenko2003,Homes2004a,Santander2004a,boris-science-ISI:000221105300037,
laforge:097008,LaForge2009a}.}

\subsection{Extended Drude formalism and infrared response of a Fermi liquid}
\label{subsec:extended Drude}

In a conducting system physical processes responsible for renormalization
of electronic lifetimes and effective masses also lead to
deviations of the frequency dependent conductivity from conventional
Drude theory. These deviations can be captured through the
extended Drude formalism\cite{gotze1972,PhysRevB.15.2952}:
\begin{eqnarray}
4\pi\tilde{\sigma}(\omega,T)=\frac{{\rm i}\omega_p^2}{\omega + M(\omega)}
=\frac{\omega_p^2}{1/\tau(\omega)-{\rm i}\omega[1+\lambda(\omega)]}
\label{eq:memory}
\end{eqnarray}
The complex memory function $M(\omega)$ has causal analytic
properties and bears strong similarities with the electron self-energy for $\bf k$-points averaged over the Fermi surface. This
analysis is particularly useful for the exploration of
electron-boson coupling (Section \ref{subsec:Electron-boson
interaction}) and of power law behavior in quantum critical
systems (Section \ref{subsec:Power law behaviour of optical
constants and quantum criticality}). The subtle differences
between $M(\omega)$ and the self-energy are discussed in a number
of publications\cite{allen71,shulga91,dolgov1995a}.


In the absence of vertex corrections, the following approximate
relation between $M(\omega)$ of an isotropic Fermi liquid and the
single particle self-energies was derived~\cite{allen71}
\begin{equation}
\frac{M(\omega)}{\omega}=
\left\{
 {\int \frac{f(\omega')-f(\omega'+\omega)}{\omega+\Sigma^*(\omega')-\Sigma(\omega'+\omega)}{\rm d}\omega'}\right\}^{-1}
- 1,
\label{eq:electron-boson1}
\end{equation}
where $\Sigma(\xi)$ is the self-energy of electrons with binding
energy $\xi$, and $f(\xi)$ is the Fermi-Dirac distribution.
The coupling of electrons to phonons or other bosonic fluctuations is
described by the boson density of states multiplied with the
square of the coupling constant, $\alpha^2F(\omega)$ for phonons,
$I^2\chi(\omega)$  for spin fluctuations, and $\tilde{\Pi}(\omega)$
in general. The self-energy is within this approximation
\begin{equation}
\Sigma(\omega,T)=\int \tilde{\Pi}(\omega') K(\omega,\omega',T)
{\rm d}\omega'
\label{eq:electron-boson2}
\end{equation}
where the Kernel $K(\omega,\omega',T)$ is a material independent
function given by the Fermi and Bose distributions \cite{allen71}.
In this set of expressions a double integral relates
$\tilde{\Pi}(\omega)$  to $M(\omega)$ and the optical
conductivity, which is reduced to a single integral by the
-reasonably accurate- Allen approximation \cite{allen71,shulga91}
\begin{equation}
\begin{array}{l}
1/\tau(\omega) = \frac{\pi}{\omega}\int_0^{\infty}
\tilde{\Pi}(\omega') K'(\omega,\omega',T){\rm d}\omega'
\end{array}
\label{eq:electron-boson3}
\end{equation}
where $K'(\omega,\omega',T)$ is a material independent
Kernel, different from $K(\omega,\omega',T)$.  \textcite{marsiglio} derived in the limit of weak
coupling and zero temperature
\begin{equation}
\frac{1}{2\pi}\frac{{\rm d}^2}{{\rm d}\omega^2}\frac{\omega}{\tau(\omega)} = \tilde{\Pi}(\omega) \quad,
\label{eq:marsiglio}
\end{equation}
which for the optical spectra of K$_3$C$_{60}$ \cite{degiorgi1994}
resulted in the qualitatively correct electron-phonon spectral
function.

If the low energy band structure can be approximated by a single
``effective'' band and the scattering $\gamma$ is small, one may
approximate the electron self-energy by a Fermi liquid expansion
$\Sigma=\Sigma(0)+(1-1/Z_F)\omega-i\gamma$, with $\gamma\ll T$. Here
$Z_F$ is the quasiparticle renormalization amplitude.  The low energy
conductivity of such a Fermi liquid is given by
\begin{eqnarray}
\tilde{\sigma}(\omega) = \frac{(\omega_p^0)^2/(4\pi)}{2\gamma - {\rm i}(\omega/Z_F)} +
\sigma_{\rm reg}(\omega)
\label{slowe}
\end{eqnarray}
where $(\omega_p^0)^2=4\pi e^2 \sum_\vk
(v^{\vk})^2\delta(\epsilon_\vk-\mu_0)$ is the non-interacting plasma
frequency, $\mu_0=\mu-\Sigma(0)$ is the non-interacting chemical
potential, and $\sigma_{\rm reg}$ is the regular part of the conductivity.

It is evident from Eq.~(\ref{slowe}) that the Drude weight is reduced
by the quasiparticle renormalization amplitude $Z_F$, {\it i.e.},
$\omega_p^2=(\omega_p^0)^2 Z_F$.
Within the band structure method the Drude weight can be
characterized by the effective density $n_{\rm eff}$ and the band mass
$m_b$ by $(\omega_p^0)^2=n_{\rm eff}e^2/m_b$. The renormalized Drude
weight, defined in Eq.~(\ref{eq:SR-partial1}), can be similarly
expressed by $\omega_p^2=n_{\rm eff}e^2/m^*$. Hence the renormalized
quasiparticle mass is $m^*=m_b/Z_F$. As expected, the quasiparticle
dispersion $\epsilon_\vk Z_F$, measured by ARPES, is also renormalized
by the same amount.

The spectral form of the optical conductivity is usually more
complicated then the Drude term alone and in addition contains both
the incoherent spectral weight as well as many sidebands due to
coupling to various excitations including magnetic and bosonic modes.
These additional contributions are contained in $\sigma_{\rm
reg}(\omega)$. The plasma frequency is hence modified due to renormalization
of quasiparticles and presence of other excitations by
\begin{equation}
\omega_p^2=8\int_0^{\Lambda}\sigma_1(\omega)d\omega = (\omega_p^0)^2
Z_F + 8 w_{reg}
\end{equation}
where $w_{\rm reg}$ is the integral of the regular part of $\sigma_1$ up to a cutoff
$\Lambda$. The cutoff should exclude the interband transitions, but
should be large enough to include the intraband transitions of some
low energy effective Hamiltonian.
The total spectral weight $\omega_p^2$, which is closely related to
the kinetic energy of a corresponding low energy Hamiltonian, defines
the optical effective mass $m_{opt}$ via
$m_{opt}=n_{\rm eff}e^2/\omega_p^2$, as depicted by a dark blue area in
Fig.~\ref{fig:correlations}. Hence the optical mass renormalization
over the band mass is $m_{opt}/m_b=1/(Z_F+8 w_{\rm reg}/(\omega_p^0)^2)$,
which is smaller then the enhancement of the low energy quasiparticle
mass $m^*/m_b=1/Z_F$, measured by ARPES. The optical mass enhancement
is also sketcked in Fig.~\ref{fig:correlations} as the high energy
limit of the effective mass $m^*(\omega)$. The low energy quasiparticle
effective mass is further enhanced by an amount $1+8
w_{reg}/((\omega_p^0)^2 Z_F)$. This additional enhancement can be
obtained using the extended Drude analysis. Comparing
Eq.~(\ref{eq:memory}) with Eq.~(\ref{slowe}) in the zero frequency limit,
we see that $1+\lambda(\omega=0)=\omega_p^2/(Z_F(\omega_p^0)^2)=1+8
w_{\rm reg}/[(\omega_p^0)^2 Z_F]$. Hence the quasiparticle effective mass
is
\begin{equation}
m^*(\omega=0) = m_{\rm opt}[1+\lambda(\omega=0)] = m_b/Z_F,
\end{equation}
which is equal
to the renormalization of the quasiparticle dispersion, as meassured
by ARPES. Hence the optical effective mass $m_{opt}$ of a correlated
metal can be obtained from optical conductivity data by comparing the
total spectral weight below some cutoff $\Lambda$ with the band
structure method. To obtain the quasiparticle effective mass $m^*$,
one needs to further renormalize the mass by the factor $[1+\lambda]$,
which can be obtained by the extended Drude model analysis.

Finally, for a very anisotropic Fermi liquid with strong variation of
quasiparticle weight $Z_F(\vq)$ across the Fermi surface, the formula
for the effective mass needs to be corrected. As shown
by~\textcite{Drew:2008}, the quasiparticle effective mass measured by
optics is roughly proportional to $1/\langle Z_F(\vq)\rangle$, where
$\langle \rangle$ stands for the average over the Fermi surface. The
effective mass measured by other probes can be different. In
particular, the Hall effect experiments measure the effective mass
proportional to $\langle Z_F(\vq)\rangle/\langle Z^2_F(\vq)\rangle$,
and the quantum oscillations experiments measure the effective mass
proportional to $\langle 1/Z_F(\vq)\rangle$
\cite{Drew:2008}.

\subsection{Dynamical mean field theory}
\label{subsec:Dynamical Mean Field Theory}

The theoretical modeling of correlated materials proved to be a very
difficult challenge for condensed matter theorists due to the absence
of a small parameter for a perturbative treatment of correlations,
such as the small ratio between the correlation energy and the kinetic
energy, or a small electron radius $r_s$ in the dense limit of
the electron gas.

For realistic modeling of weakly correlated solids, the local density
approximation (LDA) turn out to be remarkably successful in predicting
the electronic band structure, as well as the optical constants.
However LDA can not describe very narrow bands, found in many heavy
fermion materials, nor Hubbard bands. Not surprisingly, it fails to
predicts the insulating ground state in several Mott insulators and
charge transfer insulators.
The combination of LDA with static Hubbard $U$ correction, so called
LDA+U \cite{Anisimov:1991}, was able to predict the proper insulating
ground state in numerous correlated insulators. Being a static
approximation, LDA+U works well for many correlated insulators with
long range magnetic or orbital order. But the exaggerated tendency to
spin and orbital order, the inability to describe the correlated
metallic state, or capture the dynamic spectral weight transfer in
correlated metals hindered the applicability of the method.
A perturbative band structure method was developed over a course of
several decades, named GW method \cite{Hedin:1965}, and proved to be
very useful for moderately correlated materials. In particular, its
quasiparticle self-consistent version \cite{Schilfgaarde:2006}
successfully predicted band gaps of several semiconductors. However,
its perturbative treatment of correlations does not allow one to
describe Mott insulators in paramagnetic state, nor strongly
correlated metals.

The theoretical tools were considerably advanced in the last two
decades mostly due to the development of the practical and powerful
many body method, the dynamical mean field theory
(DMFT) \cite{Georges:1996}. This technique is based on the one
particle Green's function and is unique in its ability to treat
quasiparticle excitations and atomic-like excitations on the same
footing. The dynamic transfer of spectral weight between the two is
the driving force for the metal insulator transition in Hubbard-like
models as well as in transition metal oxides.

Historically, it was not the photoemission, but optical conductivity
measurements, in combination with theory \cite{PhysRevLett.75.105},
that first unraveled the process of the temperature dependent spectral
weight transfer.
In these early days it was
difficult to probe the bulk photoemission due to the issues with the
surface states that precluded the detection of the the quasiparticle
peak and its temperature dependence.
On the other hand, the optical conductivity measurements on
V$_2$O$_3$ \cite{PhysRevLett.75.105} unambiguously proved that a
small decrease in temperature results in a redistribution of the
optical spectral weight from high energy (of the order of few eV)
into the Drude peak and mid-infrared peak. It was nearly a decade
later before photoemission \cite{Mo:2003} detected the subtle
effects of the spectral weight transfer between the quasiparticle
peak and Hubbard band.

The accuracy of DMFT is based on the accuracy of the local
approximation \cite{Georges:1996} for the electron self-energy. It
becomes exact in the limit of infinite lattice coordination (large
dimension), and is very accurate in describing the properties of
numerous three dimensional materials \cite{Kotliar:2006}.

Just as the Weiss mean field theory~\cite{Weiss} for an Ising model
reduces the lattice problem to a problem of a spin in an effective
magnetic field, the DMFT approximation reduces the lattice problem to
a problem of a single atom embedded in a self-consistent electronic
medium. The medium is a reservoir of non-interacting electrons that
can be emitted or absorbed by the atom.  The local description of a
correlated solid in terms of an atom embedded in a medium of
non-interacting electrons corresponds to the celebrated Anderson
impurity model, but now with an additional self-consistency condition
on the impurity hybridization $\Delta(\omega)$
\cite{Georges:1996}. The central quantity of DMFT, the one particle
Green's function, is thus identified as an impurity Green's function
of a self-consistent Anderson impurity problem. Diagrammatically,
the DMFT approximation can be viewed as an approximation which sums
up \textit{all local} Feynman diagrams. Hence, the mapping to the
Anderson impurity problem can be viewed as a trick to sum all local
diagrams.

A second theoretical advance came when DMFT was combined with band
structure methods \cite{Anisimov:1997}, such as the Local Density
Approximation (LDA), in an approximation dubbed LDA+DMFT
\cite{Kotliar:2006}. This method does not require one to build the low
energy model to capture the essential degrees of freedom of a specific
material, a step, which is often hard to achieve. In LDA+DMFT the
extended $sp$ and sometimes $d$ orbitals are treated at the LDA level,
while for the most correlated orbital, either $f$ or $d$, one adds to
the LDA Kohn-Sham potential all local Feynman diagrams, the diagrams
which start at the specific atom and end at the same atom
\cite{Kotliar:2006}.

The LDA+DMFT approach allows one to compute both the one particle
Green's function and the current vertex entering Eq.~(\ref{opt_chi})
for the optical response. These quantities are normally expressed in
the Kohn-Sham basis in which the one-particle part of the
Hamiltonian is diagonal.  The DMFT one-particle Green's function
$G^{ij}_{\vk}$ (propagator in Fig.~\ref{optdiag}) in the Kohn-Sham
basis is
\begin{eqnarray}
G_{\vk}^{ij}&=&\langle\psi_{\vk,i}|
[({\rm i}\omega+\mu+\nabla^2-V_{KS})\delta(\vr-\vr')- \nonumber\\
&&\Sigma_\omega(\vr,\vr')]^{-1}|\psi_{\vk,j}\rangle,
\end{eqnarray}
where $V_{KS}$ is the Kohn-Sham potential, and
$\Sigma_\omega(\vr,\vr')$ is the DMFT self-energy. The procedure of
embedding the DMFT impurity self-energy to the Kohn-Sham basis was
extensively discussed in \textcite{HauleEmbed:2009}.
Finally, the two particle vertex function $\Gamma(\vk,\vq)$ (dark
blue triangle in Fig.~\ref{optdiag}) can be computed from the fully
irreducible two particle vertex function $I(\vk\vk';\vq)$ (light
blue square in Fig.~\ref{optdiag}) through the Bethe-Salpeter
equation depicted in the second line of Fig.~\ref{optdiag}.  Within
the DMFT approximation, the two particle irreducible vertex
$I(\vk\vk';\vq)$ is local, i.e., does not depend on $\vk$, $\vk'$ or
$\vq$, and hence can be computed from the solution of the DMFT
impurity problem \cite{Georges:1996}. It was first noticed by
\textcite{Khurana:1990}, that the vertex corrections to the optical
conductivity within DMFT approximation vanish in the one-band
Hubbard-like model. This is because the electron velocity $v_\vk$ is
an odd function of momentum $\vk$, $I(\vk,\vk')$ and does not depend
on $\vk$ and $\vk'$, and hence the vertex corrections to conductivity
vanish. In general, for multiband situations encountered in LDA+DMFT,
the vertex corrections do not necessarily vanish even though the two
particle irreducible vertex $I$ is purely local in this
approximation. This is because, in general, velocities are not odd
functions of momentum, which is easy to verify in the strict atomic
limit. Nevertheless, the vertex corrections are small in many
materials because they vanish at low energy, where a single band
representation is possible, and are also likely sub-leading at
intermediate and high energy, where the itinerant interband
transitions dominate. To date, a careful study of the vertex
correction effects within LDA+DMFT is lacking. In the context of the
Hubbard model, \textcite{Lin:2009} demonstrated that vertex
corrections substantially contribute to the optical conductivity at the
scale of the Coulomb repulsion $\omega\sim U$, whereas negligible
contributions were found to the Drude and the mid-infrared peaks.

In the absence of vertex corrections, the optical constants,
Eq.~(\ref{sigmar}) on the real axis takes a simple form
\begin{eqnarray}
\mbox{Re}\left\{{\sigma}^{\mu\nu}(\omega)\right\}=\pi e^2\sum_\vk \int
{\rm d}\varepsilon\frac{f(\varepsilon-\omega)-f(\varepsilon)}{\omega}\times
\nonumber\\
\Tr\left\{\rho_{\vk}(\varepsilon)v^{\vk\mu}\rho_\vk(\varepsilon-\omega)v^{\vk\nu}\right\}.
\label{sigmaB}
\end{eqnarray}
where
${\rho_\vk}(\varepsilon)=(G_\vk^\dagger(\varepsilon)-G_\vk(\varepsilon))/(2\pi
{\rm i})$ and the trace needs to be taken over all the
bands~\cite{Haule:2005}. Eq.~(\ref{sigmaB}) has been used in the
majority of the LDA+DMFT calculations.

\section{Excitations and collective effects}
\label{sec:Excitations in Correlated Electron Matter}

\subsection{Free charge carriers}
\label{subsec:Free charge carriers}
\label{sec:Drude}
The electrical conduction of a material is
governed by how freely charge carriers can move throughout it. In
his seminal model, \textcite{Drude00} considered the charge carriers
to propagate independently. The span between two scattering events
has an exponentially decaying probability characterized by the time
$\tau$ and the mean free path $\ell$. This scattering or relaxation
time fully describes the dynamical response of the entire system to
external an electric field, summarized in the complex
frequency-dependent conductivity [Eq.~(\ref{eq:Drude})]. The Drude
model does not take into account interactions with the underlying
lattice, with electrons, or other quasi-particles. In his
Fermi-liquid theory \textcite{Landau57} includes electronic
correlations, yielding an effective mass $m^*$  and also an
effective scattering time \cite{Pines66}.

\begin{figure}
\centering
\includegraphics[width=6cm]{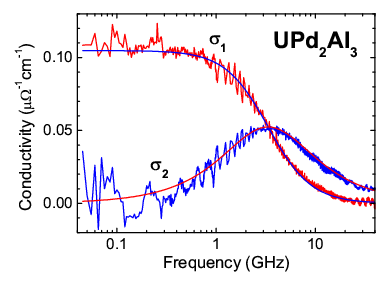}
\caption{\label{fig:Drude}(Color online)
Optical conductivity spectrum (real and
imaginary parts) of UPd$_2$Al$_3$ at temperature $T=2.75$~K. The fit by
Eq.~(\ref{eq:Drude}), with
$\sigma_{\rm dc}= 1.05\times 10^{5}~(\Omega{\rm cm})^{-1}$ and $\tau =
4.8\times 10^{-11}$~s, documents the excellent agreement of experimental
data and the Drude prediction. The characteristic relaxation rate
$1/(2\pi\tau)$ is marked by the decrease in $\sigma_1$ and the maximum
in $\sigma_2$ around 3~GHz \cite{Scheffler05}.}
\end{figure}

In heavy fermion materials the hybridization of nearly localized
$f$-shell electrons with quasi-free conduction electrons leads to
an effective mass orders of magnitude larger than the bare
electron mass \cite{Fisk88,Grewe91}. Accordingly, the spectral
weight [proportional to $n/m^*$ according to the sum rule
Eq.~(\ref{eq:SR-global})] and the scattering rate
$1/\tau^*=(m/m^*)(1/\tau)$ are significantly reduced
\cite{Varma85a,Varma85b,Millis87a,Millis87b}. Hence, the charge
carriers are extremely slow due to electron-electron interactions
which shifts the relaxation rate into the microwave regime. As
demonstrated in Fig.~\ref{fig:Drude}, \textcite{Scheffler05}
probed the real and imaginary parts of the Drude response in
UPd$_2$Al$_3$ and UNi$_2$Al$_3$ \cite{Scheffler06,Scheffler10}
over three orders of magnitude in frequency and verified that the
actual shape is perfectly described by Eq.~(\ref{eq:Drude}),
because impurity scattering still dominates over electron-electron
scattering in spite of the strong renormalization.

More specific to the GHz range, Fermi-liquid theory predicts a
renormalized frequency-dependent scattering rate
\cite{AshcroftMermin76,Pines66,Abrikosov63}:
\begin{equation}
1/\tau^*(\omega,T)=A(k_BT)^2 + B(\hbar\omega)^2
\label{eq:Fermiscattering}
\end{equation}
with the prefactors increasing as the square of the effective mass
\cite{Kadowaki86}, and $A/B$ depending on the material properties
\cite{Rosch05,Rosch06}. An experimental confirmation of relation (\ref{eq:Fermiscattering}) is still missing.

\subsection{Charge transfer and excitons}
\label{subsec:charge_tansfer and excitons}
Optical transparency of insulating compounds is a consequence of the
energy gap in the spectrum for electron-hole pair excitations,
which, if final-state interactions between the electron and the hole
can be neglected, corresponds to the gap between the valence and the
conduction band. Different physical origins of the gap are known,
and the corresponding insulators can be classified accordingly. For
the purpose of the review we make a distinction between two main
classes: (i) A gap caused by the periodic potential of the lattice.
Standard semiconductors and insulating compounds fall in this class.
(ii) A gap opened by on-site Coulomb repulsion (Hubbard $U$) on the
transition metal ion with an odd number of electrons per site. A
further distinction in the latter group is made according to the
value of $U$ compared to the charge transfer energy $\Delta$ needed
for the excitation process $d^n\rightarrow d^{n+1} \underline{L}$
where $\underline{L}$ denotes a hole in the anion valence band
\cite{zaanen85}. When $U<\Delta$, processes of the type  $d_i^n
d_j^n \rightarrow d_i^{n+1} d_j^{n-1}$ are the dominant charge
fluctuation corresponding to the optical gap at an energy $U$. On
the other hand, when $U>\Delta$, $d^n\rightarrow d^{n+1}
\underline{L}$ corresponds to the optical gap at energy $\Delta$ and
fluctuations $d_i^n d_j^n \rightarrow d_i^{n+1} d_j^{n-1}$ at an
energy $U$ fall inside the interband transitions. The case
$U<\Delta$ corresponds to the limit of a Mott-Hubbard insulator, and
is found on the left hand side of the $3d$ series, {\em i.e.}
vanadates and titanates, as well as organic compounds. The situation
$U>\Delta$, indicated as `charge transfer insulator' is common on
the right hand side of the $3d$ series; the cuprates and nickelates
fall in this class. Coupling between different bands mixes the
character of the bands on either side of the gap, which softens the
transition from Mott-Hubbard insulator to charge transfer insulator
as a function of $U/\Delta$. This is of particular relevance for
substances with $U$ and $\Delta$ of the same size, {\em e.g.} in Cr,
Mn and Fe oxides \cite{RevModPhys.70.1039,zaanen85}.

The Coulomb interaction can bind an electron and a hole to form an
exciton, the energy of which is {\em below} the excitation threshold
of unbound electron-hole pairs. This is illustrated by the example
of cuprous oxide (Cu$_2$O). This material is important in the quest
for Bose-Einstein condensation of excitons \cite{snoke1990}, a goal
which until now has remained elusive \cite{denev2002}. Cu$_2$O is a
conventional band-insulator with a zone center gap of 2.17 eV. The
valence and conduction bands have the same (positive) parity at the
zone-center, rendering direct transitions across the gap optically
forbidden. The optical spectrum is therefore dominated by the $2p$,
$3p$, $4p$, and $5p$ exciton lines situated 2 to 22 meV below the
gap. The excitonic $1s$ ground state is split by the electron-hole
exchange interaction into an optically forbidden singlet, and a
triplet situated respectively 151 meV and 139 meV below the gap. The
triplet corresponds to a weakly dipole allowed transition at 2.034
eV, whereas the singlet (2.022 eV) can be optically detected in a
finite magnetic field \cite{fishman2009}. Detection schemes
employing THz radiation generated by $3p$-$2s$ transitions
\cite{huber2006} or THz absorption by $1s$-$2p$ excitations
\cite{PhysRevLett.101.246401,fishman2006} of excitons created by laser excitation,
allow to monitor the internal conversion of the excitons to the $1s$ ground
state as a function of time.

In organic molecular crystals electron-hole pairs can be bound on a
single molecule. Due to the larger band mass as compared to typical
semiconductors, the exciton binding energy is relatively large: In a
two-photon absorption experiment \cite{janner95} the ground state
exciton of C$_{60}$ was observed at an energy 0.5 eV below the
threshold of the electron-hole continuum at 2.3 eV.

\begin{figure}
\centering
\includegraphics[width=7cm]{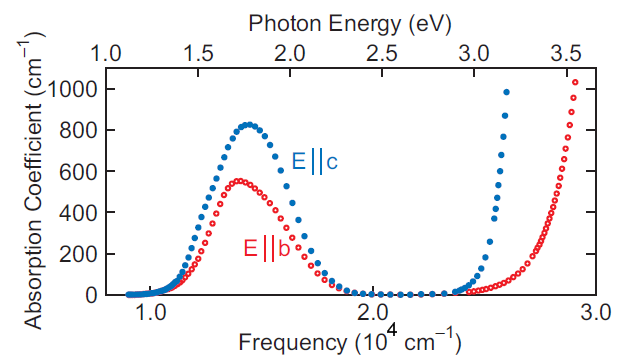}
\caption{(Color online) Absorption spectrum of CuGeO$_3$ measured at 300 K for two
different polarizations of the light. The bandgap energy is 3
eV. The peak at 1.75 eV is a phonon assisted copper $d-d$
exciton \cite{bassi96}.}
\label{fig:cugeo3}
\end{figure}
When a gap is opened by the on-site Coulomb repulsion, a special
situation arises due to the fact that the energy of a charge-neutral
local configuration change can be smaller than the correlation gap.
The result is again an excitonic bound state below the electron-hole
continuum. For example, in the spin-Peierls system CuGeO$_3$ the upper Hubbard band is
separated from the occupied oxygen $2p$ states by a 3 eV correlation
gap. \textcite{bassi96} observed a Cu $d-d$ exciton at 1.75 eV
(Fig.~\ref{fig:cugeo3}), far below the onset of the electron-hole
continuum at 3 eV. This weak absorption is responsible for the
transparent blue appearance of this compound. In the one-dimensional
compound Sr$_2$CuO$_3$ \cite{kim08} sharp peaks observed at 10 K
were attributed to weakly bound excitons. The $3d^8$ ground state in
NiO is three-fold degenerate, and the remaining 42 $3d^8$ states are
spread over about 10 eV, grouped in 7 multiplets. About half of
these are below the 4 eV correlation gap \cite{sawatzky84}. These
excitons have been observed in optical absorption
\cite{newman59,tsuboi94}. In KCuF$_3$ crystal field excitons were
observed at 0.7, 1.05, 1.21, and 1.31 eV corresponding to a local
d-d excitation from the $d_{x^2-y^2}$ groundstate to
$d_{z}$,$d_{xy}$,$d_{xz}$ and $d_{yz}$ excited states
\cite{deisenhofer08}.

For La$_2$CuO$_4$ the electron-hole threshold is at 1.9 eV;
\textcite{ellis08} observed a crystal field exciton at 1.8 eV, as
well as a peak at 2.2 eV which they attribute to a quasi-bound
electron-hole pair occupying neighboring copper and oxygen atoms.
YTiO$_3$ (SmTiO$_3$) has a 0.6 eV Mott-Hubbard gap;
\textcite{gossling2008} reported excitons corresponding to processus
of the type $d^1d^1\rightarrow d^0d^2$ on two neighboring Y-atoms,
at 1.95 (1.8) eV, as well as other $d^0d^2$ configurations at higher
energies, having strongly temperature dependent spectral weight in
the vicinity of the magnetic ordering transitions
\cite{kovaleva2007}. \textcite{khaliulin2004} showed that, as a
consequence of the temperature dependent orbital correlations, both
superexchange and kinetic energy have strong temperature and
polarization dependences, leading to the observed temperature
dependence of the spectral weight.

\subsection{Polarons}
\label{subsec:polarons}
Electron-phonon coupling quite generally renormalizes the mass,
velocity, and scattering processes of an electron. The
quasiparticles formed when phonons dress the bare electrons are
referred to as polarons. However, different conditions in the solid
require different theoretical approaches to the electron-phonon
interaction. If the electron density is high, the
Migdal-approximation holds and standard Holstein-Migdal-Eliashberg
theory is applied \cite{Mahan00}. Historically, the concept of a
polaron started from the opposite limit, {\em i.e.} a low density
electron system interacting strongly with lattice vibrations. In
this case the starting point is that of individual polarons, out of
which a collective state of matter emerges when the density of
polarons is increased. In many ways a polaron is different from an
undressed electron. The polaron mass is higher and the Fermi
velocity lower compared to those of the original electron, and a
phonon-mediated polaron-polaron interaction arises in addition to
the Coulomb interaction.

The original description by Landau and Pekar considers that an
electron polarizes the surrounding lattice, which in turn leads to
an attractive potential for the electron \cite{feynman,Mahan00}. The
situation where the electron-phonon interaction is local, is
decribed by the Holstein model \cite{holstein1,holstein2}. This
potential is capable of trapping the electron, and a bound state is
formed with binding energy $E_p$. In the literature a distinction is
usually made between large and small polarons. Both in the Holstein
and the Fr\"ohlich model the polaron diameter varies continuously
from large to small as a function of the electron-phonon coupling
parameter, but typically the Holstein (Fr\"ohlich) model is used to
describe small (large) polarons \cite{alexandrov-mott}. The
Fr\"ohlich model uses optical phonon parameters such as the
longitudinal phonon frequency $\hbar\omega_{LO}$, which can be
measured spectroscopically \cite{calvani01}. In transition metal
oxides the dominant coupling is to an oxygen optical mode
$\omega_{LO} \sim 0.1$ eV. The binding energy and mass enhancement
factor in the weak and strong coupling limits are summarized in
Table \ref{table:polaron-binding-energy}
\begin{table}
\begin{tabular}{|l|l|l|}
\hline \hline
& weak coupling& strong coupling\\
\hline
&&\\
$\alpha^2$&$\frac{Ry}{\tilde{\varepsilon}_{\infty}^2\hbar\omega_{LO}}\frac{m_{b}}{m_{e}}$&
$\frac{Ry}{\tilde{\varepsilon}_{\infty}^2\hbar\omega_{LO}}\frac{m_{b}}{m_{e}}$\\
&&\\
$E_p$&$\alpha\hbar\omega_{LO}$ &  $0.1085\alpha^2\hbar\omega_{LO}$\\
&&\\
$\frac{m^{*}}{m_{b}}$&$1+\frac{1}{6}\frac{E_p}{\hbar\omega_{LO}}$&$1.8
\left(\frac{E_p}{\hbar\omega_{LO}}\right)^2$\\
\hline \hline
\end{tabular}
\caption{Expressions for the Fr\"ohlich coupling constant $\alpha$,
polaron binding energy $E_p$, and mass enhancement $m^{*}/m_{b}$ in
the weak and strong coupling limits \cite{alexandrov99b}.}
\label{table:polaron-binding-energy}
\end{table}
where
$\tilde{\varepsilon}_{\infty}^{-1}=\varepsilon_{\infty}^{-1}-\varepsilon(0)^{-1}$.
In transition metal oxides the band mass is typically $m_b\sim
2m_e$, and $\epsilon_{\infty}\sim 4$. The corresponding strong
coupling values provide the upper limit for the binding energy ($E_p
\sim 0.17$ eV) and the mass enhancement ($m^{*}\sim 5 m_b$).

In general, if the electrons interact with a single Einstein mode,
the spectrum consists of a zero frequency mode and a series of sharp
side-bands that describe the incoherent movement of a polaron
assisted by $n=1,2,3 \ldots$ phonons \cite{devreese98}. In real
solids these sharp side-bands are smeared out due to the fact that
phonons form bands, and usually only the envelope function is
expected \cite{alexandrov99}. In a pump-probe experiment it is
possible to move the electron suddenly away from the centre of
the surrounding lattice distortion. This sets up coherent lattice
vibrations, which have recently been observed in GaAs using a
midinfrared probe pulse \cite{gaal08}. Predictions of the energy of
the midinfrared peak using the Fr\"ohlich model are as high as
$4.2E_p$ in the strong coupling limit \cite{myasnikova08}, and
$2E_p$ in the Holstein model \cite{fratini06}. Consequently, in the
case of the transition metal oxides, the Fr\"ohlich coupling
predicts a midinfrared peak at 0.7 eV at most.
\begin{table}
\begin{tabular}{|c|l|l|c|}
\hline \hline
\mbox{Compound}& \mbox{$E_{MIR}$}(eV)&\mbox{Ref.}&\mbox{T increase}\\
\hline
\mbox{La$_{1.5}$Sr$_{0.5}$NiO$_{4}$}&\mbox{0.75 }&a&\mbox{weak redshift}\\
\mbox{Fe$_{3}$O$_{4}$}&\mbox{0.6 }&b&\mbox{--}\\
\mbox{La$_{1-x}$Sr$_{x}$NiO$_{4}$}&\mbox{0.5 }&c&\mbox{--}\\
\mbox{Pr$_{1/2}$Sr$_{1/2}$MnO$_{3}$}&\mbox{0.5 }&d&\mbox{blueshift}\\
\mbox{Ba$_{1-x}$K$_{x}$BiO$_{3}$ }&\mbox{0.4 \& 1.2 }&e&\mbox{no shift}\\
\mbox{La$_{7/8}$Sr$_{1/8}$MnO$_{3}$}&\mbox{0.4 }&f&\mbox{intensity-loss}\\
\mbox{$\alpha\prime$Na$_{0.33}$V$_2$O$_5$}&\mbox{0.38 }&g&\mbox{no shift}\\
\mbox{La$_{1-y}$Ca$_y$TiO$_{3.4\pm\delta}$}&\mbox{0.31-0.38 }&h&\mbox{--}\\
\mbox{LaTiO$_{3.41}$}&\mbox{0.31 }&i&\mbox{blueshift}\\
\mbox{V$_3$O$_5$}&\mbox{0.38 }&j&\mbox{blueshift}\\
\mbox{Bi$_{1-x}$Ca$_{x}$MnO$_{3}$}&\mbox{0.25 }&k&\mbox{intensity-loss}\\
\mbox{SrTiO$_3$}&\mbox{0.25 }&l&\mbox{blueshift}\\
\mbox{Eu$_{1-x}$Ca$_{x}$Ba$_2$}Cu$_3$O$_6$&\mbox{0.15 }&m&\mbox{--}\\
\mbox{Nd$_{2}$CuO$_{4-\delta}$ } &\mbox{0.1 }&n&\mbox{--}\\
\hline \hline
\end{tabular}
\caption{Midinfrared peak positions for various
compounds. a \cite{jung01}, b \cite{park98}, c \cite{bi93}, d
\cite{jung00}, e\cite{puchkov95,ahmad05}, f \cite{jung99}, g
\cite{presura03}, h \cite{thirunavukkuarasu06}, i
\cite{kuntscher03}, j \cite{baldassarre:245108}, k \cite{liu98}, l
\cite{mechelen08}, m \cite{mishchenko08}, n \cite{lupi99}.}
\label{table:MIR}
\end{table}

If we now consider Table \ref{table:MIR}, we observe that in most
cases the peak maximum is below 0.75 eV. An exception is formed by
the high T$_c$ superconductor Ba$_{1-x}$K$_{x}$BiO$_{3}$ where, in
addition to a weaker peak between 0.33 and 0.45 eV, a strong peak
has been observed at 1.2 eV. The latter peak was originally
interpreted as a small-polaron mid-infrared peak \cite{puchkov95}
and more recently as a purely electronic transition \cite{ahmad05}.
The formalism has been extended to arbitrary density of Fr\"ohlich
polarons by \textcite{tempere01}. By fitting a moderate electron
phonon coupling ($m^*/m_b=1.3$), they obtained an excellent
agreement with the optical data for Nd$_{2}$CuO$_{3.996}$
\cite{lupi99}. In contrast, the one-polaron model does not capture
the optical line shape near the maximum of these data, despite the
the very low doping level.

\begin{figure}
\centering
\includegraphics[width=\columnwidth]{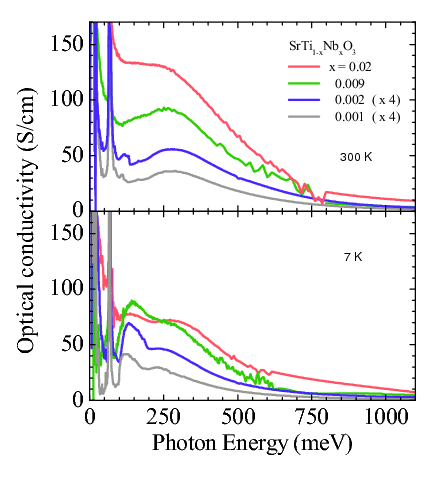}
\caption{(Color online) Optical conductivity of SrTi$_{1-x}$Nb$_{x}$O$_3$ for $x=$
0.01, 0.002, 0.009 and 0.02 at 300 K (top) and 7 K
(bottom) \cite{mechelen08}. The broad, temperature
dependent, mid infrared band between 100 and 750 meV
corresponds to (multi-) phonon sidebands of the Drude peak.
The narrow Drude peak contains approximately the same amount
of spectral weight as the sidebands, implying that $m^*/m_b\sim 2$. }
\label{fig:polarons}
\end{figure}

Electrons doped into the unoccupied Ti $3d$-band of SrTiO$_3$ are
believed to form polarons due to the Fr\"ohlich-type electron-phonon
coupling \cite{eagles95}. Indeed, a midinfrared band characteristic
of a polaron is observed at 0.25 eV \cite{calvani93,mechelen08},
which redshifts and splits when the temperature decreases (Fig.
\ref{fig:polarons}). The free carrier mass derived from the Drude
spectral weight is $m^{*}/m_b\approx 2$ implying moderate
electron-phonon coupling and large Fr\"ohlich polarons in this
material.
%
%
%
%
%

A clear trend in Table~\ref{table:MIR} is the large values of $E_p$
in the transition metal oxides containing Ni, Mn, or Fe, {\it i.e.}
materials where a transition metal has an open shell with more than
one electron or hole. Recent LDA calculations of the electron-phonon coupling strength of YBa$_2$Cu$_3$O$_7$ \cite{heid2009} gave $\lambda^{tr}$=0.26,0.27, and 0.23 along the $a$, $b$ and $c$-axis respectively. Addressing the problem of a single hole doped in the antiferromagnetic insulator, \textcite{cappelluti2007} and \textcite{mishchenko08} argued that the electron-phonon and exchange coupling conspire to self-trap a polaron. Adopting $\lambda_{e-ph}=0.39$ they predicted a double
structure in the midinfrared similar to the experimental data, {\em i.e.} a phonon sideband at 0.1~eV and a sideband at 0.5 eV of mixed phonon-magnon character. In a similar way, the high energy of the midinfrared
peak of the transition metal oxides in the top of
Table~\ref{table:MIR} may be a consequence of the combination of
electron-phonon coupling and magnetic correlation.


\subsection{Optical excitation of magnons}
\label{subsec:Optical excitation of magnons}

In correlated electron systems the spin degrees of freedom are
revealed by the collective modes emanating from the inter-
electronic correlations. Depending on the state of matter, these
modes can take the form of paramagnons for a regular metal
\cite{monthoux07}, spinons in the Luttinger liquid
\cite{Giamarchi04b}, triplons in spin-dimers \cite{giamarchi08},
triplet excitons in insulators (see subsection
\ref{subsec:charge_tansfer and excitons}), or magnons in a
ferromagnetic or antiferromagnetic state.

Ferromagnetic resonance (FMR) or antiferromagnetic resonance (AFR)
occurs by virtue of coupling of the electromagnetic field to
zone-center magnons. If inversion symmetry is not broken, the only
coupling to electromagnetic field arises from the $\bf{B}\cdot
\bf{S}$ term in Eq.~(\ref{eq:magnons1}). The selection rules are
then those of a magnetic dipole transition. Hence the resonance
features are present in the magnetic permeability $\mu(\omega)$,
while being absent from the optical conductivity $\sigma(\omega)$.
Asymmetry of the crystalline electric field upon the spins causes
the AFR frequency to be finite even at $k=0$, the interaction
occurring via the spin-orbit coupling. AFR and FMR allow to measure
magnetocrystalline anisotropy and spin-wave damping in the
hydrodynamic limit \cite{heinrich93}. \textcite{langner09} have
recently applied this technique to SrRuO$_3$, and demonstrated that
the AFR frequency and its damping coefficient are significantly
larger than observed in transition-metal ferromagnets. Technological
advances using synchrotron sources permit to measure the absorption
spectra as a function of magnetic field for $\omega>4$~cm$^{-1}$ and
fields up to 14 Tesla. The high sensitivity of this technique has
led to the discovery of a novel, strongly field and temperature
dependent mode in LaMnO$_3$ \cite{mihaly04}. Sensitive detection of
FMR by the time-resolved magnetooptic Kerr effect measures the
time-evolution of the magnetization following an optical pump pulse
\cite{hiebert97}.

Optical single-magnon excitations arise not exclusively from the
$\bf{B}\cdot \bf{S}$ coupling: spin-orbit interaction allows
photons to couple to magnons through the $\bf{A}\cdot \bf{p}$ term
of Eq.~(\ref{eq:magnons1}). Activation of this type of optical
process requires the breaking of inversion symmetry, which is
present in multiferroic materials due to their ferro-electric
polarization (section \ref{subsec:multiferroics}). The optical
excitation of a single magnon can be explained if the coupling to
the electric field is an effective operator of Dzyaloshinski-Moriya
symmetry \cite{cepas01}. In the ordered spin state one of the two
magnons in the Hamiltonian is replaced by the static modulation of
spin density. In cases where magnons are electric dipole active,
this has important consequences: Optical phonons and single magnon
waves of the same symmetry will mix. Moreover, two-magnon and
single magnons can be excited by the electric field component of
electromagnetic radiation \cite{katsura07}.

An excitation at 44.5 cm$^{-1}$ was observed by
\textcite{vanloosdrecht96} in the infrared transmission spectrum of
the spin-Peierls phase of CuGeO$_3$. The observed Zeeman splitting
identified it as a magnetic excitation \cite{uhrig97}. However, the
selection rules are those of an electric dipole
\cite{damascelli97a}. Extensive magnetic field studies of the
infrared spectra of $\alpha^{\prime}$NaV$_2$O$_5$,
SrCu$_2$(BO$_3$)$_2$ and Sr$_{14}$Cu$_{24}$O$_{41}$ indicated mixing
of phonon and magnon excitations in these compounds
\cite{room04a,room04b,huevonen07}. For these examples a dynamical
Dzyaloshinski-Moriya coupling has been proposed by
\textcite{cepas04}.

The first optical spectra of double magnon excitations were reported
by \textcite{silvera66} in FeF$_2$, and interpreted by
\textcite{tanabe65} as the coupling of the electric field vector to
the effective transition dipole moment
%
%
associated with a pair of magnons. The coupling is nonzero only in
the absence of a center of symmetry between the two neighboring
spins, as is indeed the case in FeF$_2$ rutile crystals. If the
crystal lattice itself is centro-symmetric, electronic charge
(dis)-order can still provide the inversion symmetry breaking field:
In the quarter-filled ladder compound $\alpha^{\prime}$NaV$_2$O$_5$
the ``charged'' magnon effect \cite{popova97,damascelli98}, lent
support to a symmetry breaking charge ordering transition at 34 K.
Later investigations favored a zigzag type of charge order without
the required inversion symmetry breaking. An alternative mechanism
proposed by Mostovoy requires a dynamically fluctuating symmetry
breaking field rather than a static one. In this process the photon
excites simultaneously one low-energy exciton and two spinons
\cite{mostovoy02}.

The inversion symmetry is not broken by
pairs of Ni$^{2+}$ ions in NiO. Yet, \textcite{newman59}
reported a magnetic absorption at 0.24 eV. \textcite{mizuno64}
attributed this to the simultaneous excitation of two magnons
and an optical phonon, a process which {\em is}
allowed by the electric dipole selection rules.
\begin{figure}
\centering
\includegraphics[width=6.5cm]{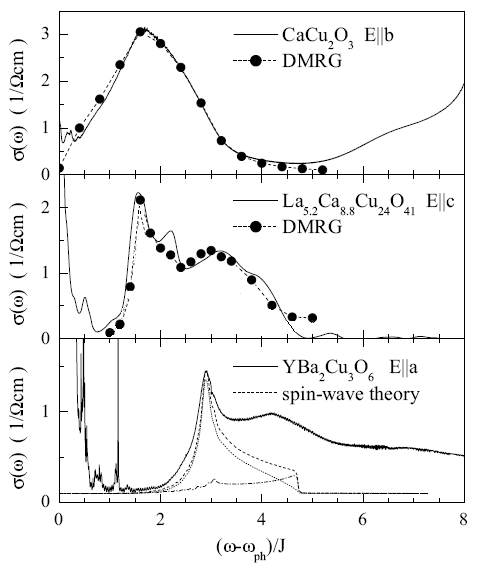}
\caption{\label{fig:magnons} Evolution of the optical conductivity
from weakly coupled chains via two-leg ladders to 2D layers at $T
= 4$~K. (Top) $\sigma_1(\omega)$ of CaCu$_2$O$_3$ for $E \parallel
b$ (solid line), DMRG result (circles) for
$J_{\parallel}/J_{\perp} = 5$ and $J_{\perp}$ = 1300 cm$^{-1}$.
(Middle) $\sigma(\omega)$ of La$_{5.2}$Ca$_{8.8}$Cu$_{24}$O$_{41}$
for E $\parallel  c$ (solid), DMRG calculation (closed symbols),
for $J_{\parallel}/J_{\perp} = 1.3$, $J_{cyc}/J_{\perp}$ = 0.2 and
$J_{\parallel} = 1000$~cm$^{-1}$ \cite{nunner02}. (Bottom)
$\sigma_1(\omega)$ of the 2D bilayer YBa$_2$Cu$_3$O$_6$ for $E
\parallel  a$ (solid). In a bilayer, the two-magnon
contribution from spin-wave theory (dashed) contains an in-
plane part (dotted) and an interplane part (dash-dotted).
Here, the in-plane exchange is J = 780 cm$^{-1}$ and the inter-
plane exchange amounts to $J_{12}/J = 0.1$ \cite{gruninger00}. The two-
magnon peak corresponds to 2.88$J$ for $J_{12}/J = 0.1$, and
to 2.73$J$ for $J_{12} = 0$ \cite{gruninger_advances}.}
\end{figure}
Strong renewed interest in the magnetic fluctuations in transition
metal oxides was revived following the discovery of high $T_c$
superconductivity in cuprates. The observation of a peak at 0.4 eV
in the anti-ferromagnetic Mott insulater La$_2$CuO$_4$ by Perkins
{\em et al.}, was initially interpreted as an intra-atomic $d-d$
exciton on the copper site \cite{perkins93}. However, the lower
bound of the $d-d$ excitons was expected at about twice that energy
based on microscopic calculations \cite{eskes90,mcmahan90}, as was
confirmed by resononant inelastic x-ray scattering experiments
\cite{kuiper98}. \textcite{lorenzana95a,lorenzana95b}  therefore
postulated that the 0.4 eV peak is due to a phonon assisted
two-magnon process similar to NiO \cite{mizuno64}, and developed a
theory for the optical conductivity spectra.
%
%
This interpretation was confirmed by the excellent agreement between
the experimentally observed optical spectra and the
two-magnon+phonon model for $S=1/2$ moments in  two dimensions
\cite{gruninger00,struzhkin00}. The line shape of the
phonon-assisted two-magnon optical absorption of the 1D spin chain
CaCu$_2$O$_3$ \cite{suzuura96} is very well described by the
two-spinon continuum \cite{lorenzana97}. In the ladder system
La$_{5.2}$Ca$_{8.8}$Cu$_{24}$O$_{41}$\cite{vuletic06} the spectrum
of the on-rung triplet bound state was found in perfect agreement
with the theory of two-triplon excitations, and it allowed the
precise determination of the cyclic exchange constant
\cite{windt01,nunner02}. The importance of quantum corrections to
the linear spin-wave theory are illustrated by the comparison in
Fig.~\ref{fig:magnons} of the two-magnon plus phonon optical
absorption spectra of chains, ladders and 2D planes with dynamical mean field renormalization group (DMRG)
calculations and linear spin wave theory \cite{gruninger_advances}.
The multi-magnon excitations in the lower panel
(YBa$_2$Cu$_3$O$_6$), having energies exceeding $5J$, are clearly
not captured by linear spin wave theory, an aspect which DMRG theory
describes rather well as is demonstrated by the upper two panels.

\subsection{Power law behavior of optical constants and quantum criticality}
\label{subsec:Power law behaviour of optical constants and quantum criticality}
In certain materials a quantum phase transition can occur at zero
temperature \cite{sondhi97}. A quantum critical state of matter has
been anticipated in the proximity of these transitions
\cite{sachdev99,varma02}. This possibility has recently attracted
much attention  because the response of such a state of matter is
expected to follow universal patterns defined by the quantum
mechanical nature of the fluctuations \cite{belitz05}. Candidates
are for example found in heavy-fermion systems
\cite{coleman05,lohneysen07} and high $T_c$ superconductors
\cite{varma89}. Quantum fluctuations play a dominating role in
one-dimensional systems causing {\em inter alia} the breakdown of
the Fermi-liquid into a Tomonaga-Luttinger (TL) liquid
\cite{Giamarchi04b}. Powerlaw behavior of the response functions is
a natural consequence. Since $\ln(\sigma)=\ln(|\sigma|)+i\arctan(\sigma_2/\sigma_1)$, the phase Arg$(\sigma(\omega))$ and $\ln(|\sigma(\omega)|)$ are related by a Kramers-Kronig transformation. Due to the fact that $\int_0^{\infty}\sigma_1(\omega)d\omega$ is subject to the f-sum rule, we need $\alpha-1< 0$ for the integration to converge for $\omega\rightarrow 0$. Since in addition $\alpha-1\ge 0$ is needed to have a convergent result for $\omega\rightarrow \infty$, the integral diverges for any value of $\alpha$. These divergencies can be avoided by limiting the powerlaw behavior to the range $\omega_{L}<< \omega << \omega_{H}$ as in the expression\cite{PhysRevB.60.R765}
\begin{equation}
\label{eq:branchcut}
\tilde{\sigma}(\omega) = \frac{\omega_p^2}{4\pi}\frac{1}{ (\omega_{L}-{\rm i}\omega)^{\alpha}(\omega_{H}-{\rm i}\omega)^{1-\alpha}}
\end{equation}

The optical conductivity follows
the relation $\sigma_1(\omega)\propto\omega^{4n^2K_{\rho}-5}$, where
the TL parameter $K_{\rho}$ characterizes the electron-electron
interaction ($K_{\rho}>1$ if the interaction is attractive), and $n$
is the order of commensurability ($n=1$ at half filling and $n=2$ at
quarter filling) \cite{giamarchi06}. This has been confirmed by experiments on the organic compound
(TMTSF)$_2$$X$, where powerlaw behavior of the optical conductivity
has been observed with $K_{\rho} = 0.23$, indicating a repulsive
electron-electron interaction \cite{schwartz98}. Recent pressure
dependend studies of (TMTSF)$_2$AsF$_6$ indicate a pressure
dependence where $K_{\rho}$ increases from 0.13 (ambient pressure)
to 0.19 (5 GPa), indicating a weakening of the electronic
interaction \cite{pashkin06}. A similar trend was reported by
\textcite{lavagnini09} for the CDW system LaTe$_2$, where the
exponent $\eta$ in $\sigma_1(\omega)\propto \omega^{-\eta}$ evolves
from 1.6 to 1.3 when the pressure increases from 0.7 GPa to 6 GPa.
\textcite{lee05} measured the optical conductivity for the chain
contribution in YBa$_2$Cu$_3$O$_y$, and observed a universal
exponent $\eta=1.6$ in the doping range from $6.3<y<6.75$.

No exact solutions are known up to date for interacting particles in
two or three dimensions. However, the preponderance of quantum
fluctuations diminishes as the number of dimensions is increased,
and consequently the breakdown of the Fermi-liquid is not expected
to be universal in dimensions higher than 1. An exception occurs
when the system is tuned to a quantum phase transition. In this case
a quantum critical state is approached, and powerlaw behavior of
the optical conductivity
\begin{equation}
\label{eq:powerlaw1}
\tilde{\sigma}(\omega) = C (1/\tau-{\rm i}\omega)^{\alpha}
\end{equation}
is a natural consequence for the response of charged bosons
\cite{Fisher1990}.\footnote{When $\alpha>-1$ the spectral weight integral $\int_0^{\Omega} \sigma_1(\omega){\rm d}\omega$ diverges for $\Omega\rightarrow\infty$. The power law behavior behavior is therefore necessarily be limited to frequencies below some finite ultraviolet cutoff.} Whether for fermions similar behavior should be
expected is subject of intensive theoretical research
\cite{cubrovic09}. The limit of zero dissipation is described by
$\alpha=-1$. Experimentally $\alpha=-0.4$ \cite{Lee:2002,Kamal:2006}
was observed for the paramagnetic metal \cite{Cao:1997} CaRuO$_3$
while $\alpha=-0.5$ \cite{Kostic:1998,Dodge00a} for the 3D
ferromagnet SrRuO$_3$ with Curie temperature of $165\,$K having a large
magnetization of $1.6\mu_B/$Ru
\cite{Randall:1959,Callaghan:1966,Longo:1968} and the 3D
helimagnetic metal MnSi \cite{Mena03}. Multiorbital correlations
were shown to lead to an orbital non-FL metal with the observed
frequency dependence \cite{laad08}.
\begin{figure}[!bt]
\centering{
\includegraphics[width=0.9\linewidth]{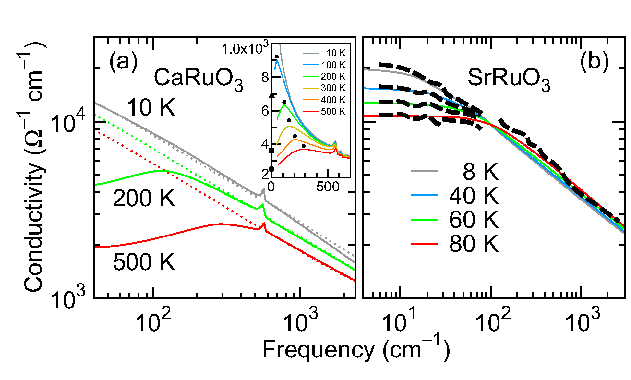}
}
\caption{(Color online) (a) Temperature dependent $\sigma_1(\omega)$ of CaRuO$_3$
shows the powerlaw scaling at three representative temperatures.
The symbols in the inset mark the energy scale where powerlaw
$\omega_1^{-1/2}$ cease to hold.
From \onlinecite{Lee:2002}.
(b) Logarithmic plot of $\sigma_1(\omega)$ for SrRuO$_3$. The curves
from the top correspond to temperatures $T=8\,$K, 40, 60, and
80$\,$K, respectively. Dotted lines are fits to
Eq.~\ref{eq:powerlaw1}.
From \onlinecite{Dodge:2000}. }
\label{RuO3}
\end{figure}
At higher temperature, the powerlaw dependence of SrRuO$_3$ is
cutoff at a scale proportional to temperature, marked by dots in
Fig.~\ref{RuO3}a. In contrast, in SrRuO$_3$ the deviation from
$\sigma\sim\omega^{-\alpha}$ occurs at $\omega\tau\gtrsim 1$ (see
Eq.~\ref{eq:powerlaw1}). At temperature higher than $95\,$K, authors
found a deviation from the formula (\ref{eq:powerlaw1}) due to
appearance of a downturn at low frequency. This gapping (not shown
in Fig.~\ref{RuO3}b) might be connected with the similar low
frequency downturn apparent also in CaRuO$_3$ (Fig.~\ref{RuO3}a),
which \onlinecite{Lee:2002} interpreted as a generic feature of the
paramagnetic state of ruthenates.

\begin{figure}
\centering
\includegraphics[width=6cm]{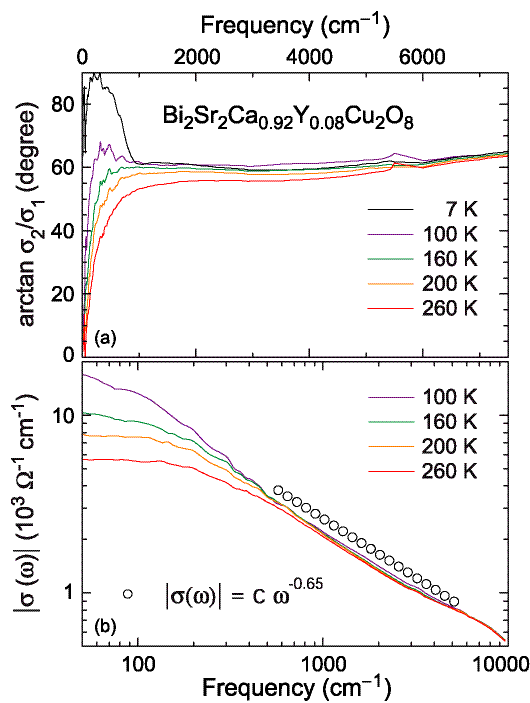}
\caption{\label{fig:powerlaw} (Color online) Universal power law of the optical
conductivity and the phase angle spectra of optimally doped
Bi$_2$Sr$_2$Ca$_{0.92}$Y$_{0.08}$Cu$_2$O$_8$. In $a$, the phase
function of the optical conductivity, Arg$(\sigma(\omega))$ is
presented. The error margins of $\epsilon_{\infty} = 4.5\pm 0.5$ propagate to an uncertainty of $\pm 2$ degrees of the phase of $\sigma(\omega)$ for $\omega=5000$ cm$^{-1}$. In $b$, the absolute value of the optical conductivity is
plotted on a double logarithmic scale. The open symbols correspond
to the power law $|\sigma(\omega) |=C\omega^{-0.65}$
\cite{vandermarel03}. }
\end{figure}

In cuprate high-T$_c$ superconductors one obtains, near optimal doping, the coefficient $\alpha=-2/3$ \cite{schlesinger90,Elazrak1994a,vandermarel03,hwang-JPCM-2007}. According to Eq.~(\ref{eq:powerlaw1}) the phase should be constant and equal to $-\pi\alpha/2$ \cite{baraduc96,anderson97}. A crucial check therefore consists of a measurement of the phase angle of the optical conductivity, $\mbox{Arg}\{\sigma(\omega)\}$. A constant phase angle of 60 degrees up to at least 5000 cm$^{-1}$ is observed in optimally doped Bi$_2$Sr$_2$CaCu$_2$O$_8$ \cite{vandermarel03,hwang-JPCM-2007}, shown in Fig.~\ref{fig:powerlaw}.\footnote{\label{footnotepowerlaws} The dielectric constant at finite frequencies is the superposition of the free carrier contribution, $4\pi i\sigma(\omega)/\omega$, which is the focus of this discussion, and "bound charge" polarizability (see footnote \ref{footnote3} of section \ref{subsec:Spectroscopy in frequency domain}), the onset of which is above 1.5 eV for the cuprates (see \ref{subsec:charge_tansfer and excitons}. Using ellipsometry between 0.8 and 4 eV, and reflectance data between 0.01 and 0.8 eV \textcite{vandermarel03} obtained $\epsilon_{\infty} = 4.5\pm 0.5$ for optimally doped Bi$_2$Sr$_2$Ca$_{0.92}$Y$_{0.08}$Cu$_2$O$_8$. Using reflectance spectra in a broad frequency range \textcite{hwang-JPCM-2007} obtained $\epsilon_{\infty}$ between 4.3 and 5.6 for Bi$_2$Sr$_2$CaCu$_2$O$_8$ samples with different dopings.}

\textcite{sachdev99} showed that for $k_BT>\hbar\omega$ the system
exhibits a classical relaxation dynamics. $\sigma_1(\omega,T)$ then
becomes a universal function
\begin{equation}
\label{eq:powerlaw2}
\sigma_1(\omega,T)=T^{-\mu} g(\omega/T).
\end{equation}
In the insulator-superconductor transition in two space dimensions,
the optical conductivity is characterized by a single exponent
$\mu=0$ \cite{Fisher1990}. Universal scaling was observed for
SrRuO$_3$ with $\mu=1/2$ and $g(\omega/T)=(\omega/T)^{1/2}\tanh(1.6\omega/T)$
\cite{Lee:2002}. This absence of any other characteristic scale but
temperature is usually associated with quantum criticality. On the
other hand, \textcite{Kamal:2006} argued that the $\omega/T$
scaling might be accidental in the temperature and frequency range
measured, and might not be enough to prove the closeness of the
quantum critical point. Another decade in the far infrared might be
needed to fully establish the $\omega/T$ scaling. For the optimally
doped cuprates scaling is found with $\mu=1$ and $g(\omega/T)=C/(1-iA\omega/T)$
\cite{vandermarel06}. Since $\mu=(2-d)/z$ where $d$ and $z$ are
dimension and critical exponent respectively, a positive value for
$\mu$ implies that $z$ is negative, which is unusual if not
impossible \cite{phillips05}. Combining the two aspects of quantum
critical behavior, the frequency powerlaw behavior
[Eq.~(\ref{eq:powerlaw1})] and $\omega/T$ scaling
[Eq.~(\ref{eq:powerlaw2})], required, in the case of the cuprates,
the introduction of a non-universal energy (approximately 50 meV)
where the cross-over takes place. This raises the question as to the
role of such energy scale in a quantum critical scenario. It has
been argued \cite{caprara07} that this non-universality occurs due
to broken Galilean invariance under the influence of the crystal
lattice, but that standard quantum criticality still provides the
correct framework. Using the conventional framework of
electron-boson coupling (see section \ref{subsec:Electron-boson
interaction}) \textcite{norman06} related the powerlaw behavior of
$\sigma(\omega)$ and the concave appearance of $1/\tau(\omega)$ to
an electron-boson coupling function with an upper cutoff scale of
the boson spectrum of about 300~meV, implying that spin fluctuations are involved in the electron-boson coupling function.

Power law optical conductivity is not limited to sublinear behavior as in the materials discussed above. Superlinear behavior or "universal dielectric response"\cite{jonscher1977} associated to disorder has been observed in a large variety of materials \cite{RevModPhys.72.873,PhysRevLett.91.207601}.


\subsection{Electron-boson interaction}
\label{subsec:Electron-boson interaction}
\begin{figure}[b!!]
\centering
\includegraphics[width=\columnwidth]
{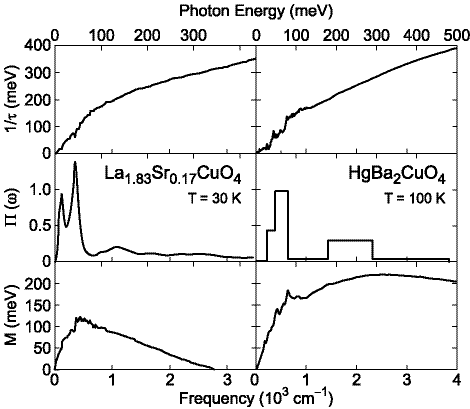} \caption{\label{fig:glue} Comparison at
$T\approx T_c$ of the imaginary part ($1/\tau(\omega)$) and the
real part ($M(\omega)$) of the optical self energy, and the
electron-boson coupling functions of La$_{1.83}$Sr$_{0.17}$CuO$_4$
(T$_c$= 31K) \cite{hwang:137005} and HgBa$_{2}$CuO$_ {4}$ (T$_c$=
97K) \cite{heumen:184512}. }
\end{figure}
The theoretical approaches to the high-T$_c$ pairing mechanism in
the cuprates are divided in two main groups \cite{maier08} (see
Section \ref{subsec:Cuprates}): According to the first school
electrons form pairs due to a retarded attractive interaction
mediated by virtual bosonic excitations in the solid
\cite{millis-PRB-1990}. These bosons can be lattice vibrations
\cite{shulga91}, fluctuations of spin-polarization
\cite{scalapino-PRB-1986,haslinger2000}, electric polarization or
charge density \cite{varma89}. The second school concentrates on a
pairing-mechanism entirely due to the non-retarded Coulomb
interaction \cite{anderson-sci-2007,phillips-annphys-2006}. This section deals with the first group of ideas.

\textcite{munzar1999} obtained good agreement between experimental
optical spectra for YBa$_2$Cu$_3$O$_{6.95}$ \cite{Puchkov1996a} and
spectra calculated from a spin fluctuation model. Analysis of the optical conductivity of
YBa$_2$Cu$_3$O$_{6.95}$ using Eq.\ref{eq:electron-boson3}
demonstrated a conspicuous peak in the coupling function at 60~meV
\cite{carbotte99,dordevic05}. In addition, angle resolved
photoemmision \cite{bogdanov-PRL-2000,lanzara-NAT-200,non-PRL-2006}
and tunneling spectroscopy
\cite{lee-nat-2006,levy-condmat-2007,zasad-PRL-2001} spectra show
clear indications of a peak in the electron-boson coupling function
at approximately the same energy. \textcite{hwang04} observed a peak
in Re$\{M(\omega)\}$ (defined in Eq. \ref{eq:electron-boson1}) of Bi$_2$Sr$_2$CuO$_6$, which makes its
appearance at the superconducting T$_c$ in overdoped copper oxides
and slightly above T$_c$ in the underdoped copper oxides. The
question whether the peak is due to a spin-resonance, a phonon, or
both, is still open. Its intensity weakens with doping before
disappearing completely at a critical doping level of 0.23 holes per
copper atom where T$_c$ is still 55 K. In addition, they found a
broad background in Re$\{M(\omega)\}$ at all doping levels, and
postulated that this provides a good candidate signature of the
``glue'' that binds the electrons \cite{hwang04}.

\textcite{norman06} explained the spectral shape of $M(\omega)$ (see
Section \ref{subsec:general optic theory}) with a model of electrons coupled to a broad
spectrum of spin-fluctuations extending to about 300~meV, {\em i.e.}
the scale of magnons in the insulating parent compounds
\cite{hayden96}. In the context of a discussion of the Hubbard
model, it has been pointed out \cite{maier08}, that the
``anomalous'' self-energy associated with the pairing has small but
finite contributions extending to an energy as high as $U$, an
aspect which is not captured by the approach of
Eq.~(\ref{eq:electron-boson1}). Alternative approaches based on
Eq.~(\ref{eq:electron-boson1}) assume that the bosonic spectral
function is provided by (near quantum critical) orbital current
fluctuations \cite{varma05}, or excitons \cite{little04}.

Extensive efforts have been made to infer $\tilde{\Pi}(\omega)$, introduced in section \ref{subsec:extended Drude}, from
the experimental optical spectra. Usually, as the first step
$1/\tau(\omega)$ is calculated from the original data. The most
direct approach then uses Eq.~(\ref{eq:marsiglio}) to calculate
$\tilde{\Pi}(\omega)$ from $1/\tau(\omega)$. However, since this
expression is only valid at zero temperature and for weak coupling,
\textcite{dordevic05} calculated $\tilde{\Pi}(\omega)$ from
$1/\tau(\omega)$ from the inverse transformation of the integral
equation Eq.~(\ref{eq:electron-boson1}) using the method of singular
value decomposition. To fit the Allen
approximation, Eq.~(\ref{eq:electron-boson3}), to the experimental
$1/\tau(\omega)$ data \textcite{maxent} implemented a maximum
entropy method. \textcite{carbotte_dwave} extended this analysis to
the $d$-wave superconducting state at finite temperature by
approximating the optical conductivity with a superposition of
$s$-wave gaps. However, $1/\tau (\omega)$ is not a purely
experimental quantity; to determine it a correction must be made for the
interband transitions (see Section \ref{subsec:Power law behaviour of optical constants and quantum criticality}) and a value of $\omega_p$ must
be assumed. Therefore \textcite{heumen2009jpconf} implemented a
method which analyzes the original reflectance and ellipsometry
spectra. A flexible parametrization of $\tilde{\Pi}(\omega)$ was
used to calculate, using Eq.~(\ref{eq:electron-boson1}), respectively
$M(\omega)$, $\sigma(\omega)$, reflectance and ellipsometry spectra.
The parameters describing $\tilde{\Pi}(\omega)$, $\epsilon_b(\omega)$, and $\omega_p$ were varied using a numerical
least-squares fitting routine with analytical derivatives, until
convergence was reached to the experimental reflectance and
ellipsometry data.

The universal description of optical and angle resolved
photoemission spectra (ARPES) was demonstrated by extracting this
way $\tilde{\Pi}(\omega)$ from optical and photoemission spectra of
a series of Bi$_2$Sr$_2$CuO$_6$ crystals with different carrier
concentrations \cite {heumen2009njp}. Close to $T_c$ the different
methods give by and large the same $\tilde{\Pi} (\omega)$. However,
for reasons which still need to be clarified, and as illustrated by
the case of HgBa$_{2} $CuO$_{4}$, stronger temperature dependence of
$\tilde{\Pi}(\omega)$ is obtained with the maximum entropy
method\cite{yangPRL09} than with least-squares
fitting\cite{heumen:184512}. Similar efforts were undertaken on the analysis of Raman spectra of cuprates \cite{Grilli09,Devereaux07}.

In Fig. \ref{fig:glue}, results are shown for two single layer
compounds, La$_{1.83}$Sr$_{0.17}$CuO$_4$ (T$_c$=
31K)\cite{hwang:137005} and  HgBa$_{2}$CuO$_ {4}$ (T$_c$=
97K)\cite{heumen:184512}. These spectra illustrate the main features
of the bosonic spectrum observed in all cuprates, namely a high
energy spectrum (extending to about 300 meV for the optimally and
overdoped samples), and a peak between 50-70 meV in all cuprates.
These aspects are also present in the spin fluctuation spectrum of
\textcite{casek05} in their analysis of the optical spectra. For
La$_{1.83}$Sr$_{0.17} $CuO$_4$, an additional peak was obtained at
18 meV, present only below 50 ~K \cite{hwang:137005}. The intensity
of the 50-70 meV peak decreases strongly as a function of doping,
and on the overdoped side of the phase diagram only the high energy
part $\tilde{\Pi}(\omega)$ can account for the observed high $T_c
$\cite{heumen:184512}.  These observations support the idea that the
pairing mechanism is, at least in part, of electronic nature, {\it
i.e.} involves spin, charge, or orbital fluctuations.

\subsection{Superconducting energy gap}
\label{subsec:BCS Superconductors}
Far-infrared spectroscopy has played an important role in the
characterization of superconductors ever since the BCS
phonon-mediated pairing mechanism was proposed and experimentally
verified \cite{bardeen1957a}. \textcite{mattis1958a} described the
electrodynamics of superconductors and the opening of a
spectroscopic gap 2$\Delta_{o}$ for quasiparticle excitations . In
weak-coupling BCS theory \cite{bardeen1957a}, the gap value is given
by 2$\Delta_{o}$ = 3.5$k_{B}T_{c}$. The electrodynamics were
experimentally measured by \textcite{glover1957a} using far-infrared
techniques \cite{tinkham1996a}. Far-infrared spectroscopy continues
to be an important tool to investigate superconductors especially
considering recent discoveries of new superconducting compounds
\cite{basov:721}.

\begin{figure} [ptb]
\begin{center}
\includegraphics[width=2.8in,keepaspectratio=true]%
{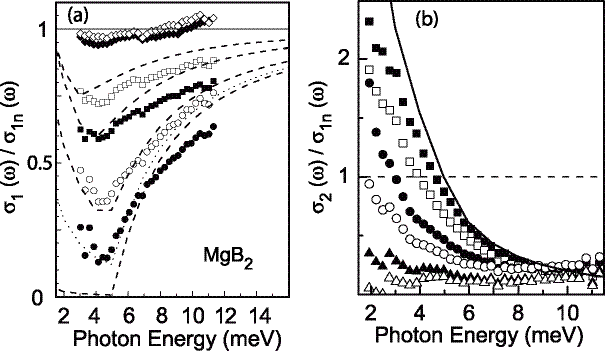}%
\caption{Far-infrared conductivity of MgB$_{2}$. (a)
$\sigma_{1}/\sigma_{n}$ and  (b) $\sigma_{2}/\sigma_{n}$ as a
function of frequency for various temperatures: 6~K (dots),
17.5~K (open circles), 24~K (solid squares), 27~K (open
squares), 30~K (solid diamonds), and 50~K (open diamonds).
The opening of a gap
is clearly observed in (a) while (b) shows
the characteristic 1/$\omega$ inductive response of the condensate. From \textcite{kaindl2002a}.}%
\label{averitt3}
\end{center}
\end{figure}

MgB$_{2}$ provides one such example. Discovered in 2001, this
material becomes superconducting below $T_{c}$= 39 K which was, at
the time of discovery, an unexpected result for a simple
intermetallic compound \cite{nagamatsu2001a}.  Shortly after its
discovery, experiments using time-domain spectroscopy from 2-11 meV
were performed on a 100-nm thick MgB$_{2}$ film as shown in
Fig.~\ref{averitt3} \cite{kaindl2002a}; complementary measurements
in the frequency domain yield equivalent results
\cite{pronin01PRL,pimenov2002PRB}.

A useful way to analyze the electrodynamic properties of
superconductors is to plot the ratio of the conductivity to its
normal state value $\sigma_{n}(\omega)$. This is shown in
Fig.~\ref{averitt3}(a) and (b) for the real and imaginary parts,
respectively. A strong depletion in the real part
$\sigma_{1}/\sigma_{n}(\omega)$ is evident in Fig.~\ref{averitt3}(a)
as the temperature is decreased below $T_{c}$. This is the signature
of the superconducting gap. The imaginary part
$\sigma_{2}/\sigma_{n}(\omega)$ in Fig.~\ref{averitt3}(b) shows the
buildup of a component in the superconducting state that strongly
increases with decreasing frequency and the characteristic
$1/\omega$  inductive condensate response. The lines in the figure
are calculations using Mattis-Bardeen theory for BCS superconductors
with an isotropic $s$-wave gap \cite{mattis1958a}, valid in ``dirty
limit'', which occurs when the superconducting gap 2$\Delta_{0}$ is
much smaller than the normal state scattering time. The overall
agreement is possible only when a gap value of 2$\Delta_{0}$ = 5 meV
is used in the calculations. This value is nearly a factor of two
smaller than the ratio expected from the known $T_{c}$ even in
weak-coupling BCS theory, which predicts 2$\Delta_{0} = 3.5
k_{B}T_{c}\approx 9$~meV. This small gap is a fundamental property
of MgB$_{2}$ explained by the existence of two superconducting gaps,
of which the smaller one dominates the optical
conductivity.\footnote{\cite{choi2001a,kortus2001a,liu2001b,gorshunov2001a,kuzmenko2002a,pimenov2002PRB}}
First-principle bandstructure calculations confirm this novel
physics and indicate that the dominant hole carriers in boron $p$
orbitals are split into two distinct sets of bands with quasi-2D and
3D character. The coupling between these bands leads to a novel
superconducting state with two gaps but a single $T_{c}$.

While far-infrared spectroscopy is clearly an important tool to
investigate the electrodynamics of the condensate response of new
superconductors (see, e.g., \ref{subsec:Iron-pnictides} on the
ferro-pnictides), there are also interesting studies on elemental
superconductors in recent years. These experiments include detailed
measurements of the microwave response of Al
\cite{steinberg2008a} for comparison with BCS
theory and probing the nonequilibrium condensate response of DyBa$_2$Cu$_3$O$_7$
\cite{feenstra1997a} and Pb
\cite{carr2000a} using time-resolved far-infrared techniques.

As mentioned above, far-infrared spectroscopy of the superconducting
gap served as an important test of BCS in the early days. However,
other predictions such as the coherence peak remained much more
difficult to experimentally verify. In Pb and Nb the expected
maximum in the temperature dependent conductivity was not observed
until the 1990s. The full frequency temperature and frequency
dependence of the coherence peak was recently measured in Al ($T_c =
1.9$~K) using microwave spectroscopy from 45~MHz - 40~GHz
\cite{steinberg2008a}.  The experiments were performed on a series
of films with mean free paths ranging from 1.8  to 5.0~nm. The
results are in agreement with the sum rule Eq.~(\ref{eq:FGT}) and
show a clear reduction in the coherence peak with increasing mean
free path as the clean limit is approached.

Another area which has its roots in 1960s efforts to experimentally
verify the BCS response of conventional superconductors are
nonequilibrium studies which originated in tunneling studies
\cite{rothwarf1963a,miller1967a}. Early nonequilibrium studies of
BCS superconductors showed that the time for quasiparticle
recombination $\tau _{R}$ to Cooper pairs is sensitive to the
magnitude of the superconducting gap
\cite{gray1981a,kaplan1976a,schuller1976a}. Quasiparticle
recombination (i.e.\ the formation of a Cooper pair) is a
fundamental process in a superconductor arising from the pairing of
two quasiparticles which are thermally or otherwise excited out of
the condensate. Thus, the goal of these initial experiments was to
determine the bare quasiparticle recombination where calculations
taking into account the electron-phonon coupling suggest a time of
$\sim$100 ps in BCS superconductors.

However, pair-breaking by excess phonons complicates matters as
described by the phenomenological rate equations of
\textcite{Rothwarf67}. This model consists of two coupled rate
equations describing the temporal evolution of the density of excess
quasiparticles and phonons injected into a superconductor. The
Rothwarf-Taylor equations are written as
\begin{subequations}
\begin{equation}
\label{roth-tay1} \frac{{\rm d}n}{{\rm d}t} = \beta N - R n^{2} - 2R nn_{T}
\end{equation}
\begin{equation}
\label{roth-tay2} \frac{{\rm d}N}{{\rm d}t} = \frac{1}{2}[R n^{2} - \beta N] -
\frac{N - N_{o}}{\tau_{p}}
\end{equation}
\label{eq:RTall}
\end{subequations}
Here, $n$ is the excess quasiparticle density, $n_{T}$ is the
thermal quasiparticle density, $N$ is the excess density of phonons
with energies greater than 2$\Delta$, $R$ is the bare quasiparticle
recombination coefficient, $\beta$ is the pair-breaking coefficient
for ($>$2$\Delta$) phonons, and $\tau_{p}$ describes the relaxation
time of the phonons either by anharmonic decay to phonons with
energies $<$ 2$\Delta$  or through phonon escape from the sample
(e.g.\ into the substrate).  As Eqs.~(\ref{eq:RTall}) reveal, in the
limit of small $n_{T}$, the recombination of quasiparticles
requires two quasiparticles to form one Cooper pair, hence the $n^{2}$ term. However, the direct
determination of $\tau_{R}$ is difficult since it is masked by
excess 2$\Delta$ phonons which break additional Cooper pairs (e.g.\
the $\beta N$ term). Thus, in many measurements, it is actually the
phonon decay $\tau_{p}$ that is measured. This has has been termed
the phonon bottleneck.\footnote{These phenomenological equations
have been utilized to understand the dynamics of other gapped
materials as described in
\textcite{kabanov2005a,chia2007a,chia2006a,demsar2006a}.}

\begin{figure} [ptb]
\begin{center}
\includegraphics[width=2.3in,keepaspectratio=true]%
{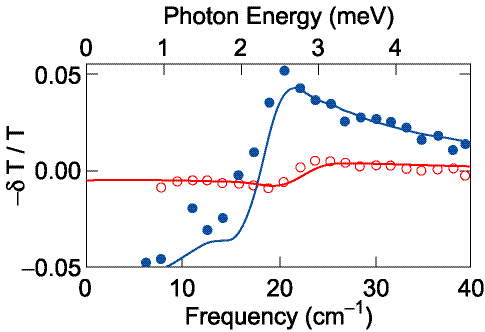}%
\caption{(Color online) Photoinduced change in the superconducting gap in Pb at $T=3.7$~K.
Photoexcitation reduces the condensate density which in turn leads
to a decrease in the magnitude of the gap and a modified
far-infrared transmission. The solid blue circles are data obtained
following photoexcitation with 1.8 nJ pulses while the open red
circles are for photoexcitation with 0.4 nJ pulses. The
corresponding solid lines are fits to the
change in transmission using BCS theory. From \textcite{carr2000a}.}%
\label{averitt4}
\end{center}
\end{figure}

A recent example on Pb using a femtosecond pump pulses synchronized
to the far-infrared beam line at NSLS at Brookhaven producing pulses
300 ps in duration \cite{carr2000a}. This permitted a direct probe
of the recovery of the superconducting gap following photoexcitation
with 800 nm pulses which reduce the condensate density by
approximately one percent which is comparable to the excess thermal
quasiparticle density. An exponential recovery with of $\sim$ 250ps
is related to the recombination of quasiparticles within the
bottleneck regime. This assignment was verified through
spectroscopic measurements of the induced change in the far-infrared
transmission ($\delta$\emph{T}/\emph{T}) as shown in
Fig.~\ref{averitt4}. The fits are using BCS theory with the spectral
change related to the decrease in the gap due to the reduction in
condensate density.

Similar experiments were performed on MgB$_{2}$ \cite{demsar2003a}.
In these studies, the increased temporal resolution ($\sim$1ps)
afforded by electro-optic techniques enabled measurements of the
quasiparticle recombination and the initial pair-breaking process.
The pair breaking process extended to 10 ps ($\sim$100 times the
duration of the pair breaking pulse) and is in contrast with Pb
where the pair-breaking dynamics are complete in approximately 1 ps
\cite{federici1992a}. Through analysis with Eqs.~(\ref{eq:RTall})
this delay in the condensate reduction was attributed to a
preferential phonon emission by the photoexcited quasiparticles
which subsequently break additional Cooper pairs. It was also
possible to extract $\beta$, $R$, and $\tau_{p}$ \cite{demsar2003a}.

These recent time-integrated and time-resolved experiments on BCS
superconductors provide new insights into these materials and point
the way towards experiments on more exotic superconductors such as
the cuprates and pnictides (sections \ref{subsec:Cuprates} and
\ref{subsec:Iron-pnictides}, respectively).

\subsection{Pseudogap and density waves}
\label{subsec:Pseudogaps/correlation gap}

\begin{figure}[!bt]
\centering{
\includegraphics[width=0.9\linewidth]{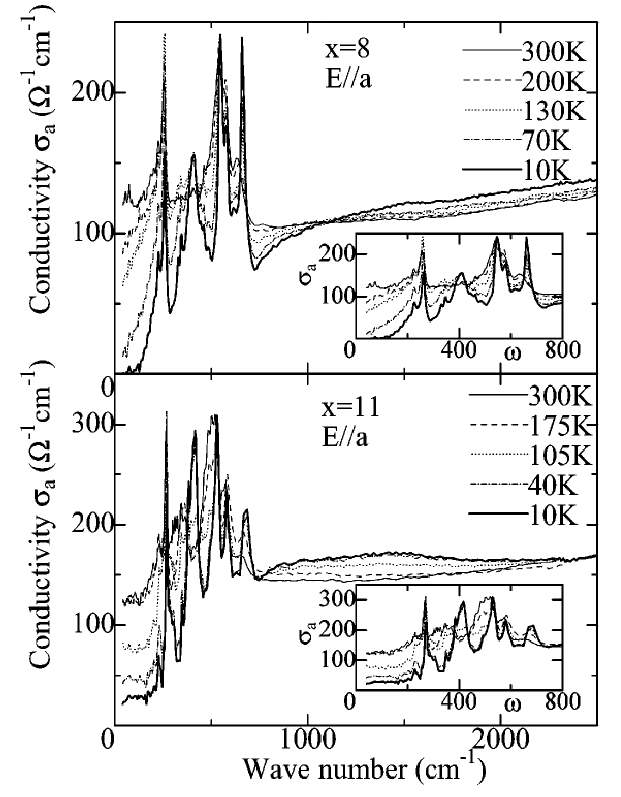}
}
\caption{
%
The in-plane $a$-axis
optical conductivity of the Sr$_{14-x}$Ca$_x$Cu$_{24}$O$_{41}$
compound for doping $x=8$ and
$x=11$.
The insets shows the enlargement of the low-frequency conductivity.
From \onlinecite{Osafune:1999}.}
\label{optCu}
\end{figure}

The term ``pseudogap'' is heavily used in the physics of
correlated electron materials in a variety of different contexts.
Most generally, pseudogap describes a partial or incomplete gap in
the density of states. Pseudogaps are common in doped Mott
insulators in the vicinity of the insulator-to-metal transition
(see Sec.~\ref{subsec:Emergence of conducting state in correlated
insulators}). The best studied class of materials displaying a
pseudogap are cuprate superconductors, discussed in
Sec.~\ref{subsec:Cuprates}. But cuprates are not the only class of
materials which show a pseudogap in the charge response. In this
regard, various one-dimensional analogs of cuprates have been
recently synthesized. A prominent example is
Sr$_{14-x}$Ca$_x$Cu$_{24}$O$_{41}$.

This material has a layered structure and consists of two distinct
one dimensional objects: 1D two-leg ladders and 1D chains.  The
conductivity is primarily determined by the two-leg ladders
\cite{Osafune:1997}.
Namely, the holes on the chains are dimerized with periodicity
$5c_{\rm chain}$, as shown by neutron scattering experiments
\cite{Matsuda:1997}. Hence the conductivity on the two-leg ladder
subsystem is up to 1000 times larger than the conductivity along the
chains.

The parent compound exhibits Mott-like insulating behavior,
exhibiting spin and charge gaps.
Due to chemical pressure the substitution of Sr$^{2+}$ by isovalent Ca$^{2+}$ introduces hole carriers,
which form a charge density wave (CDW) on the ladder subsystem
[for a review see \onlinecite{Vuletic:2006}]. At higher doping levels,
the CDW is gradually suppressed and the CDW gap eventually vanishes at
the critical doping.  \onlinecite{Vuletic:2003} estimated the critical
doping to be $x\approx 9$. For even larger doping $x \gtrsim 12$ under
pressure, the system exhibits superconductivity with
T$_{c}^{\rm max}\approx 12$~K \cite{Uehara:1996}.

Fig.~\ref{optCu} shows the $a$-axis optical conductivity for two
different hole doping levels $x=8$ and $x=11$ and few temperatures,
measured by \onlinecite{Osafune:1999}. Similar spectra for the same
and lower hole dopings were obtained by
\textcite{Vuletic:2003,Vuletic:2005}. Note however, that the gap
size at the same nominal doping is somewhat smaller in
\onlinecite{Vuletic:2005} than measured by
\onlinecite{Osafune:1999}. The system clearly develops a gap in the
charge response at low temperatures. The low frequency $a$-axis
optical conductivity decreases with decreasing temperature, and the
depressed spectral weight is transferred to much higher frequencies.
The energy scale of this suppressed conductivity due to the CDW is
very large, and comparable to pseudogap scale in cuprates. The
optical conductivity in $c$-direction (not shown) also displays a
gap \cite{Vuletic:2005}, although the gap size is somewhat smaller
in $c$ direction compared to $a$, which is consistent with the
conductivity being largest along the legs of the
ladders.\footnote{The optical conductivity in Fig.~\ref{optCu}
resembles the $c$-axis conductivity of cuprates, first measured by
\textcite{PhysRevLett.71.1645,Homes95}.}

The existence of a CDW for $x=0$ was well established from impedance
measurements, a giant dielectric constant, and non-linear
current-voltage curve \cite{Gorschunov:2002,Blumberg:2002}. A direct
detection of crystallization of holes with a periodicity of $5c_{\rm
ladder}$ was recently achieved by resonant x-ray scattering
\cite{Abbamonte:2004}. Note that the material is intrinsically
self-doped, and the $x=0$ sample is already doped with holes at
approximately 0.07~hole/Cu. The collective nature of the CDW is
reflected in large effective masses of the CDW condensate. From
$c$-axis conductivity, \onlinecite{Osafune:1999} estimated effective
mass to be of the order of $100-200 m_e$. \onlinecite{Vuletic:2003}
found somewhat smaller but still large mass of the order of $20-50
m_e$ from CDW phason mode. The CDW develops in the $a-c$ plane, while it
does not develop  true long range order in the $b$-direction. Hence
no jump in temperature dependence of resistivity along $b$ direction
was detected \cite{Vuletic:2005}.

While the electron-phonon interaction is crucial for development of
CDWs, pure electron-electron interaction can drive the
antiferromagnetic ordering of the conduction electrons, leading to a
spin density wave (SDW). It was nicely observed as a drop in the
far-infrared reflectivity of Cr \cite{Barker68}, URu$_2$Si$_2$
\cite{Bonn88}, and (TMTSF)$_2$PF$_6$ \cite{Degiorgi96} [inset of
Fig.~\ref{fig:tmtsf}(c)]; and more recently in iron-pnictides
\cite{Dong:2008,hu:257005} [see Fig.~\ref{optFeAs}(a) in
Sec.~\ref{subsec:Iron-pnictides}].

\section{Optical Probes of Insulator-to-Metal Transitions}
\label{sec:Probing Metal Insulator Transition in Frequency and Time Domain}

\subsection{Emergence of conducting state in correlated insulators}
\label{subsec:Emergence of conducting state in correlated insulators}
\begin{figure}
\centering
\includegraphics[width=\columnwidth]{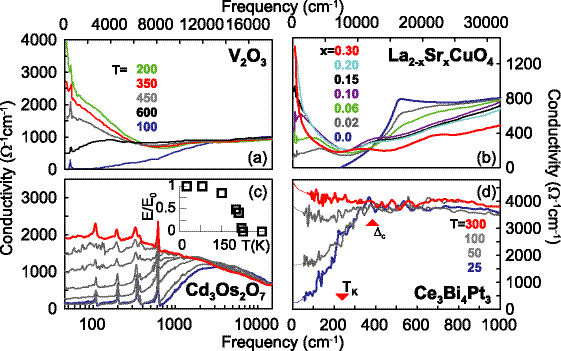}
\caption
  {(Color online) Evolution of  the optical conductivity across insulator-to-metal transitions. (a) Band-width controlled Mott transition in V$_2$O$_3$ \cite{baldassarre-v2o3}; (b) filling controlled Mott transition in La$_{2-x}$Sr$_x$CuO$_4$ \cite{UchidaPhysRevB.43.7942}. (c) The Slater transition in Cd$_2$Os$_2$O$_7$ \cite{PhysRevB.66.035120}(data plotted at the following temperatures: red - 300 K; grey - 250, 220, 210, 200 K, blue - 30 K; inset - the temperature dependence of the energy gap  determined from the kink in $\sigma(\omega)$ ). (d) Kondo system Ce$_3$Bi$_4$Pt$_3$ \cite{Bucher94}.}
\label{fig:IMTs}
\end{figure}

In Fig.~\ref{fig:IMTs} we display data revealing the evolution of
the electromagnetic response across the insulator-to-metal
transition (IMT) in four important examples of correlated electron
systems. The top row presents results for V$_2$O$_3$ and
La$_{2-x}$Sr$_x$CuO$_4$ which are two classes of Mott insulators.
Transport and thermodynamics experiments revealed that V$_2$O$_3$ is
driven towards the insulating state at $T<150\,$K by control of the
electronic bandwidth $W$ \cite{PhysRevLett.27.941,PhysRevB.48.16841}
so that the strength of electronic correlations $\propto U/W$ is
modified.  In La$_{2-x}$Sr$_x$CuO$_4$ diavalent Sr substitutes
trivalent La and the IMT is best understood in terms of band filling
\cite{Kumagi1993}. The electromagnetic
response of bandwidth controlled and filling controlled Mott systems
reveals a number of commonalities: (i) As expected, in the
insulating state one finds clear evidence for the electronic gap
where $\sigma_1(\omega)\rightarrow 0$ followed by a rapid increase
of the conductivity. (ii) There is no obvious suppression of the gap
in either bandwidth- or filling-controlled systems. Instead, the
energy region below the gap is gradually ``filled up'' with states
before any remanence of the gap disappears on the metallic side of
the transition. (iii) A salient feature of both classes is that the
redistribution of the electronic spectral weight involves a broad
range of frequencies extending to several eV. In the case of
V$_2$O$_3$ one therefore finds a giant mismatch between the $k_B T$
and $\hbar \omega$ scales near the IMT \cite{Thomas1994a}. (iv) The
response on the metallic side of the transition is highly
unconventional. The frequency dependent conductivity is different
from that of non-interacting systems. One commonly finds a
Drude-like mode at far-IR frequencies attributable to mobile charges
followed by a broad, incoherent part at higher energies.  The
oscillator strength of the Drude-like contribution to
$\sigma_1(\omega)$ is significantly reduced compared to expectations
of the band theories ignoring electronic correlations
(Fig.~\ref{fig:KE-all}). (i) - (iv) above can be regarded as
universal attributes of the Mott transition obeyed both by oxides
(Fig.~\ref{fig:IMTs}) and also organic compounds (Fig~\ref
{fig:et1}). The evolution of optical constants upon variation of the
$U/W$ parameter has been monitored in $R$TiO$_3$ ($R$ = La, Ce, Pr,
Nd, Sm, Gd) systems
\cite{PhysRevB.44.13250,PhysRevLett.75.3497,PhysRevB.58.5384,PhysRevB.51.9581,yang:195125}.
All these systems reveal transfer of the electronic spectral weight
to lower energies with increasing doping similar to cuprates and
also consistent with the theoretical studies based on the Hubbard
Hamiltonian.\cite{RevModPhys.66.763}  The rate of this spectral
weight transfer allows one to quantify the strength of electronic
correlations $U/W$\cite{PhysRevLett.75.3497,yang:195125}.

The energy gap in the optical data for Mott insulators is a direct
consequence of the high energetic cost of double occupancy $U$.
Doping of a Mott insulator partially releases the restriction for
double occupancy and thus ``unjams'' the electronic conduction.
Since doping typically impacts some but not all sites, doped Mott
insulators are inherently inhomogeneous on the atomic scale. Their
electromagnetic response combines features characteristics of an
undoped insulator and a doped conductor. A partial electronic gap
(or pseudogap) often persists in doped Mott insulators over an
extended region of their phase diagram
\cite{timusk1999a,Puchkov1996a,lee:054529}and has been linked to
local inhomogeneities \cite{Homes2003a}. An extended Drude analysis
commonly uncovers very strong dissipation on the metallic side of
the transition
\cite{Orenstein1990a,Rotter1991a,Elazrak1994a,Puchkov1996a,basov-prl96,Dodge00a,lee:054529,
vandermarel03,Zaanen2004a,qazilbash:205118}. Routinely, one finds
that $1/\tau(\omega)$ exceeds the energy in the frequency region
corresponding to the incoherent component. These results challenge
the notion of well defined quasiparticles.

A counterpart in dc transport is an exceptionally short electronic
mean free path of the order of interatomic spacing commonly
registered in doped Mott insulators \cite{Emery95}. An
interdependence between the dc transport and incoherent
(mid-infrared) response has been analyzed within an interesting
framework based on the oscillator strength sum rule
\cite{Gunnarsson03,qazilbash:205118}. As of today, there is no
consensus on a microscopic scenario for the mid-infrared band.

The electronic spectral weight in filling-controlled systems
proportional to $n/m^*$ is vanishingly small on the insulating side
of the transition and varies linearly with
doping \cite{UchidaPhysRevB.43.7942,heumen:184512}. Both bandwidth-
and filling-controlled classes of Mott systems may reveal the
divergence of the effective mass near the IMT. Thermodynamic
measurements uncover mass enhancements in a bandwidth-controlled
system:
V$_2$O$_3$ \cite{PhysRevLett.27.941,PhysRevB.48.16841,Limelette2003}
but not in La$_{2-x}$Sr$_x$CuO$_4$ \cite{Kumagi1993,Loram1989498}.
The behavior of the optical effective mass (Sec.~\ref{subsec:Vanadium oxides}) near the
IMT boundary in V$_2$O$_3$ has not been investigated. A closely
related VO$_2$ system \cite{Qazilbash2007a} does reveal strong
enhancement of $m^*$ near the IMT boundary and so do organic materials
with the bandwidth control of the transition (Fig.\ref{fig:et1})
\cite{Merino08,Dumm09}. Sr$_{1-x}$La$_x$TiO$_3$ is an example of a
filling-controlled system revealing nearly divergent behavior of
$m^*$ in proximity of the IMT \cite{PhysRevB.46.11167}. On the
contrary, the optical mass in high-$T_c$ superconductors, which also
belong to the filing-controlled class and includes
La$_{2-x}$Sr$_x$CuO$_4$ and YBa$_2$Cu$_3$O$_y$, show no anomalies in
proximity to the Mott phase \cite{padilla:205101}. This latter
inference relies on the Hall data to discriminate between $n$ and
$m*$ contributions to infrared spectral weight and is therefore not
unambiguous. Nevertheless, this finding is in accord  with specific
heat results \cite{Kumagi1993}. Weak doping dependence of the
optical mass in cuprates is in accord with the DMFT analysis of the
filling-controlled transition in these materials
(Sec.~\ref{subsec:Cuprates}).

Fig.\ref{fig:IMTs}(d) shows the evolution of the electromagnetic
response in a Kondo insulator Ce$_3$Bi$_4$Pt$_3$ \cite{Bucher94}.
The gross features of this insulating state are best understood
within the periodic Anderson model. This model predicts the
formation of the energy gap in the vicinity of the Fermi energy as
the result of hybridization between narrow $d-$ or $f$-levels and
conduction electrons \cite{Hewson93}. Provided the Fermi energy falls
inside the narrow hybridized band, the resultant behavior is
metallic characterized by a large effective mass
(Sec.~\ref{subsec:Intermetallic-Compounds}). However, if the Fermi
energy is located within the hybridization gap, Kondo insulating
behavior results. Quite remarkably, upon the transformation from the
insulating to metallic state the conductivity spectra repeat
universal characteristics of Mott insulators including gap filling,
redistribution of the spectral weight across a broad energy range
and dominant incoherent contribution in the metallic state (see
Sec.~\ref{subsec:Kondo-insulators} for more detailed discussion of
the recent data).

In contrast, Slater insulators reveal much more conventional
behavior across the IMT. The term refers to an insulating state
produced by antiferromagnetic order alone due to a doubling of the
magnetic unit cell \cite{PhysRev.82.538}.  While there are numerous
examples of Slater/spin-density-wave (SDW) insulators in the realm
of one-dimensional conductors \cite{Degiorgi96,Vescoli99}, very few
three dimensional systems fall under this classification. Elemental
Cr is a poster-child of a
three-dimensional SDW system. Elemental Cr does reveal an optical gap below
the Neel temperature but remains metallic \cite{RevModPhys.60.209}
since only a portion of the Fermi surface is impacted by the gap. A
rare example of a Slater insulator is Cd$_2$Os$_2$O$_7$ pyrochlore
\cite{PhysRevB.63.195104,PhysRevB.66.035120}. Unlike Mott/Kondo
systems the insulating gap is reduced with the increase of
temperature following the mean field behavior [inset
Fig.~\ref{fig:IMTs}(c)]. The conductivity on the metallic side of
the transition does not show a prominent incoherent component and is
consistent with the Drude model. A redistribution of the electronic
spectral weight primarily impacts the range of several energy gap
values. Optical data for Cd$_2$Os$_2$O$_7$ along with the band
structure calculations \cite{PhysRevB.65.155109} yield $K_{\rm
exp}/K_{\rm band}\simeq 1$ suggesting that correlations are mild in
this compound. This latter result is in accord with the notion that
proximity to antiferromagnetic order does not appreciably reduce
electronic kinetic energy of a conducting system
\cite{qazilbash-np-2009}.

\subsection{Quasiparticles at the verge of localization}
\label{subsec:Quasiparticles at the verge of localization}
As mentioned in the Introduction, 
strong correlations arise due to tendency of electrons to localize.
While the hallmark of a metal is the decrease in conductivity with
frequency, any sort of localization causes the overall conductivity
$\sigma_1(\omega)$ to increase at low frequencies: ${\rm
d}\sigma_1/{\rm d}\omega >0$ because photon (or thermal) energy is
required to overcome some barrier. This can be a geometrical
localization in clusters or grains, Anderson localization in an
disordered potential, confinement in a strong magnetic field
(quantum Hall effect), or Mott localization by Coulomb repulsion.

In the two-dimensional electron gas of a Si inversion layer,
Anderson localization is observed as the electron density is
lowered. At low temperatures \textcite{Gold82} measured a maximum in
$\sigma_1(\omega)$ around $10-20$~\cm\ as a precursor of the
metal-insulator transition which is approached when the carrier
concentration is reduced toward some critical value. In those
strongly disordered systems transport can be largely understood
without taking electronic correlations into account \cite{Gotze78},
although their fingerprints become observable at some
point.\footnote{Although most of the theoretical was done in the
1980s
\cite{Gotze78,Gotze79,Gotze81,Belitz81,Vollhardt80a,%
Vollhardt80b,Gold81,Belitz94,Evers08} experimental evidence was
collected only recently.
The frequency dependent hopping conductivity changes its power
law, for instance, when electron-electron interaction becomes
relevant \cite{ShklovskiiEfros84}, as nicely demonstrated in
heavily doped silicon
\cite{Lee01,Helgren02,Helgren04,Hering07,Ritz08,ritz:084902}.}

The superconducting state of cuprates is known to be very sensitive
to Zn impurities \cite{Alloul09,Logvenov09} and, further, the influence extends
well into the normal state. Underdoped cuprates are in particular
susceptible to localization since the Fermi-energy is located
slightly below the band edge. The optical spectra of Zn doped
YBa$_2$Cu$_4$O$_8$, for instance, are dominated by a peak around
120~\cm; hence \textcite{Basov98} concluded that the system tends to
localize. The Drude peak present in the undoped compound
YBa$_2$(Cu$_{1-y}$Zn$_y$)$_4$O$_8$ shifts to finite energies as $y$
increases. \textcite{lupi:206409} performed a similar far-infrared
study on the MIT transitions in other hole-doped cuprates.

\begin{figure}
\centering
\includegraphics[width=7cm]{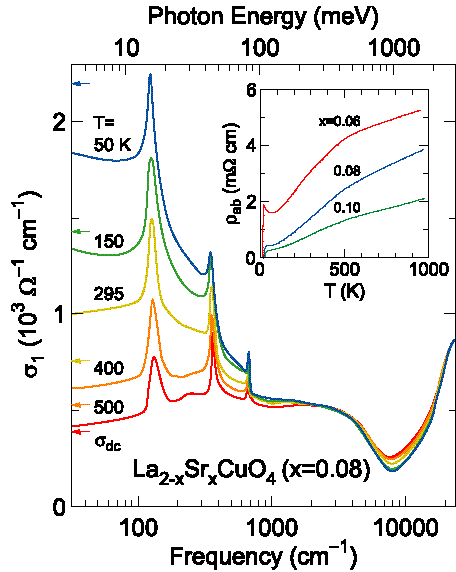}
\caption{\label{fig:LaSrCuO}(Color online) In-plane optical conductivity
of La$_{1.92}$Sr$_{0.08}$CuO$_4$ obtained from reflectivity measurements
at different temperatures. The arrows on the left axis represent the dc
conductivity. The inset shows the temperature dependence of the dc
resistivity $\rho_{ab}(T)$ of La$_{2-x}$Sr$_{x}$CuO$_4$ for different doping levels \cite{Takenaka02}.}
\end{figure}
The optical conductivity of La$_{1.92}$Sr$_{0.08}$CuO$_4$ changes
its behavior around $T^*\approx 250$~K, as seen in
Fig.~\ref{fig:LaSrCuO}. This temperature evolution is not reflected
in the dc resistivity (cf. inset). At elevated temperatures, the
far-infrared conductivity increases with frequency leading to a
broad maximum that moves to lower energies as the temperature drops.
The overall spectral weight shifts to lower energies; but only below
$T^*$  a Drude-like peak develops that evidences coherent charge
transport. At high temperatures the transport is incoherent
(diffusive) with a finite-energy peak below 400~\cm.
\textcite{Takenaka02} suggested this to be a consequence of a
so-called dynamical localization. Also, in the extended Drude
analysis, a clear distinction can be seen around $T^*$ with a
low-frequency divergence of $1/\tau(\omega)$ for $T>T^*$. As doping
with Sr proceeds, the collective mode in the far-infrared grows in
intensity and the crossover temperature $T^*$ increases. The
intraband contribution becomes larger with $x$; causing an overall
enhancement of $\sigma_1(\omega)$ for $\omega<5000$~\cm.

Phenomena very similar to the cuprates are observed in SrRuO$_3$
for which $\rho(T)$ increases linearly with $T$ beyond the
Ioffe-Regel limit without evidence of a crossover or  saturation,
suggesting a bad metallic behavior \cite{Emery95}. The optical
conductivity shows a $\sigma_1(\omega)\propto\omega^{-1/2}$
dependence (for a discussion of power laws, see Sec.~\ref{subsec:Power law behaviour of optical constants and quantum criticality}), yielding
a linear increase in the scattering rate with frequency for low
temperatures \cite{Kostic:1998}. At elevated temperatures the
scattering rate of SrRuO$_3$ finally diverges for $\omega\rightarrow
0$ corresponding to the finite energy peak (200~\cm) in the
conductivity observed above approximately 150~K.

A violation of the so-called Mott or Ioffe-Regel condition means
that $\rho(T)$ exceeds the upper limit fixed by the common
understanding of a metal for which the mean free path $\ell$ should
be larger than the atomic distance $d$ \cite{Ioffe60,Mott90}.
Numerous counterexamples, like alkali-doped fullerenes $A_3$C$_{60}$
\cite{Gunnarsson00},  weakly-doped La$_2$CuO$_4$ \cite{Calandra03g},
or some organic conductors, for instance
$\theta$-(BEDT-TTF)$_2$I$_3$ \cite{Takenaka05,Gunnarsson07}, are
discussed for different reasons. \textcite{Gunnarsson03} argue, that
in cuprates the resistivity saturates at some higher level due to a
strong reduction of the kinetic energy caused by correlation
effects, while for $A_3$C$_{60}$ (and probably in a different way
for the organics) coupling to phonons becomes important.

A general scenario can be sketched from the electrodynamics point of
view: with increasing Coulomb repulsion the low-frequency
conductivity drops before entering the insulating phase, leaving
some maximum in $\sigma_1(\omega)$ at finite frequency
\cite{Mutou06}. For two-dimensional conducting systems there is a
strong redistribution of spectral weight as the Mott transition is
approached with temperature (cf.\ Sections~\ref{subsec:Emergence of
conducting state in correlated insulators} and
\ref{sec:organics-2D-Mottinsulator}, for instance). Going from
half-filled to quarter-filled conduction band and further, the
effect of inter-site Coulomb repulsion $V$ takes the leading role.
Charge-order develops in certain ways depending on the structural
arrangement, the strength of on-site and inter-site interaction
compared to the bandwidth, but also on the electron-phonon
interaction. Although a incomplete gap develops when the system
approaches the metal-insulator transition, there always remains a
Drude-like contribution to the optical conductivity as long as
$V<V_c$ some critical value. In addition, a finite-energy peak is
found in $\sigma_1(\omega)$ that is related to charge-order
fluctuations which eventually may cause superconductivity
\cite{Merino01}. Recently collective excitations of the charge order
have been observed which are linked to lattice vibrations. Prime
examples are cuprates, doped manganites $R_{1-x}$Ca$_x$MnO$_3$ ($R$
= Pr or Ca) or organic conductors of $\alpha$-,
$\beta^{\prime\prime}$- and $\theta$-(BEDT-TTF)$_2$$X$
type.\footnote{For cuprates see
\textcite{Tajima99,Dumm02,PhysRevLett.91.077004}; perovskite-type
manganites see Sec.~\ref{subsec:Manganites}:
\textcite{Takenaka99,Takenaka00,Okimoto00,Kida02,Nucara08,Pignon08,Rusydi08};
for nickelates see \textcite{Lloyd08}; organic materials are
discussed in Sec.~\ref{sec:organics-2D-chargeorder} and by
\textcite{Calandra02,Merino06,Dressel03b,Drichko06a,Kaiser09,Dressel09c}.}

\subsection{Superconductor-insulator transition}
\label{subsec:Superconductor-insulation transition}
Disorder has a very delicate influence on the superconducting
state \cite{Goldman98,Dubi07}. While \textcite{Anderson59} showed
that weak disorder cannot destroy anisotropic $s$-wave
superconductivity, the case is more complex for anisotropic
systems; $d$- and $p$-wave superconductors are sensitive to
disorder.\footnote{Extensive investigations on high-$T_c$ cuprates
demonstrated that disorder causes localization in the
quasi-two-dimensional transport of CuO$_2$ planes and acts as an
efficient pair-breaking process; $T_c$ is suppressed and the
superconducting carrier density reduced
\cite{Basov94,Basov98,Tajima99,Dumm02}} As far as two-dimensional
systems are concerned, numerous
experiments\footnote{Investigations have been performed on thin
films of Nb$_x$Si alloys \cite{Aubin06}, amorphous InO$_x$
\cite{Sambandamurthy04,Sambandamurthy05,Sambandamurthy06},
disordered TiN \cite{Baturina07a,Baturina07b,Vinokur08,Sacepe08}
or Ta \cite{Qin06}.} have demonstrated a transition from a
superconductor to an insulating state with increasing disorder or
magnetic field. Strong disorder gives rise to spatial fluctuations
of the local complex order parameter $\Psi({\bf
r})=\Delta\exp\{i\phi\}$. Superconducting islands are surrounded
by regions with relatively small $\Psi$ and only due to tunnelling
of Cooper pairs between the islands correlations are sustained.
With increasing $B$ the superconducting island loose their phase
coherence, causing a magnetic-field-driven
superconductor-insulator transition, although the amplitude of the
order parameter vanishes only at higher fields. There seem to be
links to superconducting phase fluctuations in underdoped cuprates
and organic superconductors
\cite{Xu00,Corson99,Spivak08,Muller09,Emery95}.

\textcite{Crane07a,Crane07b} investigated superconducting InO$_x$
films with $T_c=2.28$~K (defined by a well-developed amplitude of
the order parameter) and found that below the transition temperature
the generalized superfluid stiffness acquires a distinct frequency
dependence. Superconducting amplitude fluctuations cause a peak in
the microwave dissipation [given by $\sigma_1(\omega,T)$] which
shifts to higher temperatures and decreases in amplitude as $\omega$
increases from 9 to 106~GHz. The peak occurs when the time scale of
measurement probe matches the time scale of the superconducting
fluctuations.
The complex response measured for a finite magnetic field evidences
that the superfluid stiffness is finite well into the insulating
regime. From the frequency dependent conductivity it can be inferred
that short-range correlations exist at finite $T$ while long-range
order does not. This insulating state with superconducting
correlations is called a Bose insulator
\cite{Steiner08,Fisher89,Fisher1990}.

\subsection{Conductivity scaling for metal-insulator transition}
The superconductor-insulator transition is a prime example of a
continuous quantum phase transition \cite{sachdev99,sondhi97} for
which quantum fluctuations of diverging size and duration are
important. There has been some work on the quantum critical behavior
and on the conductivity scaling around metal-insulator transitions.
Microwave and THz optical experiments on amorphous Nb$_x$Si alloys \cite{PhysRevLett.80.4261},
for instance, revealed a correspondence between the frequency and
temperature dependent conductivity on both sides of the critical
concentration $x$ of the metal-insulator transition thus
establishing a quantum critical nature of the transition.
\textcite{Lee00} could determine a scaling function and critical
exponents  that shed light on the relationships between the temporal
and spatial fluctuations; the location of the crossover and the dynamical
exponent, however, lacks a theoretical explanation.

\subsection{Photoinduced phase transitions}
\label{subsec:Photoinduced phase transitions}
\begin{figure} [ptb]
\begin{center}
\includegraphics[width=2.8in,keepaspectratio=true]%
{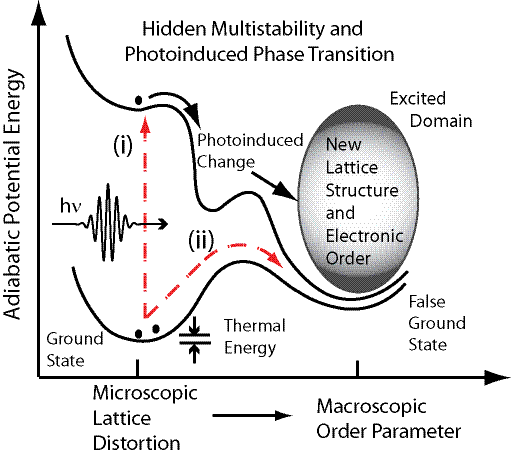}%
\caption{(Color online) Depiction of a photoinduced phase transition showing the
potential energy as a function of a generalized structural
coordinate. The basic idea is the so-called ``domino effect":
photoexcitation initiates a structural change which, in turn, drives
an electronic phase transition with an order parameter different
from that in the ground state. The energy barrier to obtain
the new long-range (though likely metastable) ordered phase is greater than k$_{b}$T. The red dashed arrows labeled
(i) and (ii) depict electronic and vibrational routes towards inducing a phase transition. After Nasu \cite{nasu2004a}. }%
\label{averitt12}
\end{center}
\end{figure}

The emerging field of photoinduced phase transitions relies on
optical techniques to initiate a cooperative response in a given
material resulting in a new macroscopically ordered phase
\cite{nasu2004a,Yonemitsu08,kuwatagonokami_koshihara2006a,JPCS2005a,JPCS2009a}.
The resultant change can be probed using x-rays, dc transport,
magnetic susceptibility, or a host of other electronic or structural
probes. Optical spectroscopy is a powerful tool to monitor induced
changes over a broad energy scale with temporal resolution from
femtoseconds to days. Ultrafast optical spectroscopy is of
particular interest as it provides the means to follow the dynamics
of photoinduced phase transitions and, additionally, interrogate a
metastable (though macroscopic) phase which may only exist for a few
ns before thermal fluctuations drive the system back towards the
true groundstate.\footnote{In recent years ultrafast optical
techniques have been extended to include time resolved x-ray
diffraction, electron diffraction, and photoemission and with
interesting results obtained on various correlated electron
materials
\cite{cavalleri2005a,gedikScience2007,baumscience2007,perfettiPRLTaS2}.}
Materials such as organics and transition metal oxides with
optically accessible on-site or inter-site charge excitations ({\em
i.e.} of a local nature) and competing degrees of freedom are ideal
candidates in which to investigate photoinduced phase transitions.

Initial work on photoinduced phase transitions was on the reversible
structural interconversion in polydiacetylenes highlighting a
nonlinear excitation intensity dependence and the necessity of a
photon energy $\sim$0.5 eV above the exciton absorption peak
\cite{koshihara1990a}. Other influential experiments include the
photoinduced phase transition on quasi 1-d organic
tetrathiafulvalene-p-chloranil (TTF-CA) where a photoinduced ionic
to neutral transition was observed and has lead to considerable
theoretical effort \cite{koshihara1990b,koshihara1999a}.
Fig.~\ref{averitt12} shows a simple picture capturing the essence of
photoinduced phase transitions put forth by
\textcite{nasu2004a,Yonemitsu08}. The basic idea is that
multistability can be investigated using photons to explore complex
energy landscapes with correlated electron materials being of
particular interest \cite{tokura2006a}. To date, the majority of
experiments have focussed on using pump photons $\gtrsim$1eV to
``photodope" a material. This is depicted as the red vertical arrow
labeled (i) in Fig.~\ref{averitt12}. While numerous excitation
pathways are possible, one of particular interest is photodoping
holes into a Mott-Hubbard band. This can be accomplished by exciting
electrons from the lower Hubbard band into a higher lying
non-Hubbard band. At sufficient excitation intensity this could lead
to a dynamic collapse of the Mott-Hubbard gap \cite{cavalleri2005a}.

\begin{figure} [ptb]
\begin{center}
\includegraphics[width=2.3in,keepaspectratio=true]%
{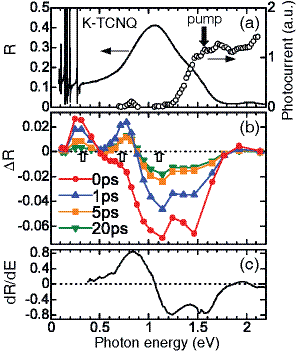}%
\caption{(Color online) (a)~Reflectivity and photocurrent measured
on K-TCNQ as a function of frequency taken at room temperature which
is well below $T_{\rm SP}=395$~K. The peak at 1 eV corresponds to an
excitonic charge transfer excitation and the onset of a photocurrent
response at approximately 0.5 eV above this is due to the excitation
of unbound electron-hole pairs. (b)~Spectral dependence of
photoinduced absolute reflectivity change $\Delta R$ at various time
delays following excitation. The lowest energy peak arises from
polaron formation while the change in the charge transfer peak
suggests ``melting" of the spin-Peierls phase. (c)~The derivative of
reflectivity in (a) with respect to photon energy shows a similar
spectral response to the time-resolved data which supports the interpretation of the dynamics. From \cite{okamoto2006a}.}%
\label{averitt13}
\end{center}
\end{figure}

A recent example highlighting the power of ultrafast optical
spectroscopy to monitor the evolution of a photoinduced phase
transition are the experimental results of the photoinduced melting
of the spin-Peierls phase in the organic charge transfer compound
K-tetracyanoquinodimethane (K-TCNQ) \cite{okamoto2006a}. K-TCNQ is a
one-dimensional organic with a half-filled $\pi$-electron band.
Strong on-site Coulomb repulsion leads to a Mott insulating state
with a transition to a dimerized spin-Peierls state below $T_{\rm SP} =
395$~K due to a strong spin-lattice interaction.

Photoexcitation creates localized carriers which destabilize the
magnetic state resulting in melting of the spin-Peierls phase in $<
400$~fs. Spectroscopic evidence for this is presented in
Fig.~\ref{averitt13}. The peak at $\sim 1$~eV(solid line) in
Fig.~\ref{averitt13}(a) is a charge-transfer transition, while the
circles are photocurrent measurements the onset of which is $\sim
0.5$~eV above the excitonic charge transfer transition and
corresponds to the creation of unbound electron hole pairs. The
photoinduced (150 fs 1.55 eV pulses) absolute reflectivity change
$\Delta R$ as a function of probe photon energy at various time
delays is shown in Fig.~\ref{averitt13}(b). There are two features.
A positive $\Delta R$ at lower energies is ascribed to small
polarons. The ultrafast decrease of $\Delta$R between 0.75 to
1.75~eV is spectrally similar to ${\rm d}R/{\rm d}E$ [see
Fig.~\ref{averitt13}(c)] determined using the static spectrum in
Fig.~\ref{averitt13}(a). This differential response is associated
with a decrease in dimerization based on analysis of the
temperature-dependent redshift of the charge transfer peak which
follows the changes in the x-ray reflection intensity associated
with the dimerization. Thus, the picture that emerges is that
photoexcitation leads to the formation of polarons which break
spin-singlet dimers that, in turn, destabilize the spin-Peierls
phase. Analysis of the oscillatory time-domain response of $\Delta
R/R$ (not shown) indicates that this ``melting'' of the spin-Peierls
phase initiates a coherent excitation of the 20 cm$^{-1}$ mode
corresponding to the lattice distortion associated with
dimerization.

\begin{figure} [ptb]
\begin{center}
\includegraphics[width=2.3in,keepaspectratio=true]%
{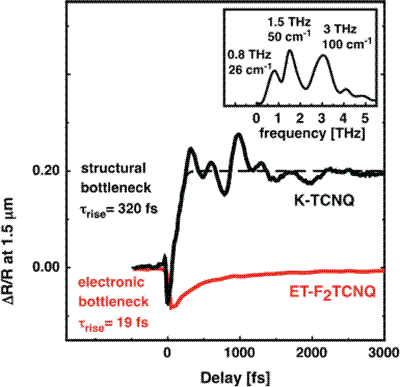}%
\caption{(Color online) Ultrafast dynamics measured on K-TCNQ and
(BEDT-TTF)(F$_{2}$TCNQ). The spin-Pierels compound K-TCNQ (black
line) shows a structural
bottleneck in that the risetime is 320~fs while the Mott-Hubbard compound K-TCNQ (red line), with a risetime of 19~fs,
does not exhibit a structural bottleneck. In addition, the oscillations in the K-TCNQ data results from the coherent
excitation of vibrational modes as shown in the inset. From \textcite{wall2009a}.}%
\label{averitt14}
\end{center}
\end{figure}

Further ultrafast studies using shorter pulses (sub-10 fs) comparing
the response of the spin-Peierls compound K-TCNQ with a pure Mott
analog (BEDT-TTF)(F$_{2}$TCNQ) (where BEDT-TTF or ET stands for
bis-(ethyl\-ene\-di\-thio)\-te\-tra\-thia\-ful\-va\-lene), show
remarkably different responses at early times
\cite{wall2009a,okamoto2007,Uemura09}, as shown in Fig.
\ref{averitt14}. These results have been interpreted as a collapse
of the Mott insulating gap in (BEDT-TTF)(F$_{2}$TCNQ) occurring on a
20-fs timescale due to photodoping. This is in contrast to a 320-fs
structural bottleneck in K-TCNQ corresponding to a quarter period
coherent structural relaxation of the dimerization. This shows that
ultrafast optical spectroscopy can track the initial steps of a
photoinduced phase transition thereby providing a powerful
discriminatory capability.

\begin{figure} [ptb]
\begin{center}
\includegraphics[width=2.8in,keepaspectratio=true]%
{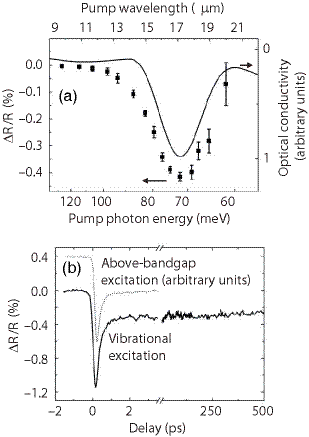}%
\caption{(a) The induced change in the near-infrared (800 nm)
reflectivity of Pr$_{0.7}$Ca$_{0.3}$MnO$_3$ as a function of pump
photon energy at 30K. The solid line is the time-integrated
reflectivity of the 17~$\mu$m mode. The magnitude of the induced
change at 800 nm tracks this mode. (b) Vibrational excitation at 17
$\mu$m initiates a dynamic response (solid black line) that is
similar to electronic excitation (i.e.
pumping directly at 800nm), shown as a dotted line. From \textcite{rini2007a}.}%
\label{averitt15}
\end{center}
\end{figure}

We mention a new area of research which has opened up with the
recent demonstration of direct vibrational control over electronic
phase transitions \cite{rini2007a}. Referring to the simple
schematic in Figure \ref{averitt12}, this would correspond to the
red arrow labeled (ii) where a phase transition is driven through
vibrational excitation in the ground state. A plausible scenario in
correlated electron materials is that of vibrationally driven
bandwidth modulation. As described in the section on manganites
(Sec. \ref{subsec:Manganites}), the narrow bandwidth manganite
Pr$_{1-x}$Ca$_{x}$MnO$_{3}$ is insulating for all values of $x$. For
intermediate values ($x \sim 0.3$ to 0.4) a ``hidden" ({\em i.e.}
thermally inaccessible) metallic phase is revealed upon application
of a magnetic field \cite{tokura2000a}. Additionally, it has been
shown that optical excitation drives Pr$_{1-x}$Ca$_{x}$MnO$_{3}$
to a metal-like state
as observed in ultrafast reflectivity studies and the observation of
stable paths for dc conduction between biased electrodes
\cite{fiebig1998a,fiebig2000a}. Recent experiments reveal that short
pulse photoexcitation of the highest-frequency optical phonon
(corresponding to a Mn-O stretching motion) at 17.5~$\mu$m leads to
an ultrafast reflectivity change as occurs with optical excitation.
This is shown in Fig.~\ref{averitt15}(b). Further, as shown in
Fig.~\ref{averitt15}(a), the magnitude of $\Delta R/R$ follows the
line shape of the optical phonon as the pump pulse is tuned across
the resonance. In addition to the reflectivity changes in
Fig.~\ref{averitt15}(a), photoexcitation of the phonons leads to an
increase in the dc electrical conductivity increases by orders of
magnitude (not shown). This demonstrates that vibrational excitation
of an insulator to metal-like transition is viable. This  motivates
future efforts to explore vibrationally induced transitions in the
electronic ground state of other correlated systems with the goal of
clarifying the influence of specific modes and thermally
inaccessible (i.e. coherent) structural distortions on the
electronic state.

To conclude this section we briefly mention the important topic of
ultrafast demagnetization and magnetization control. In
ferromagnetic metals, short pulse excitation initiates
demagnetization on a sub-ps time scale
\cite{beaurepaire1996a,bigot2009a,zhang2009NPa}. Other experiments
on antiferromagnetic orthoferrites and garnets have demonstrated
optical control of the magnetization where circularly polarized
femtosecond pulses induce coherent magnon generation through the
inverse Faraday effect \cite{kimel2005a,kimel2009a}. We mention
these results since photoinduced phase transitions through optical
manipulation of the magnetic degrees of freedom in strongly
correlated electron materials is also of significant fundamental
interest \cite{talbayev2005a}.

\subsection{Electronic phase separation}
\label{subsec:Electronic phase separation}
Electronic and magnetic phase separation is commonplace in
correlated electron systems \cite{dagotto-science05}. It is believed
to stem from the prominence of multiple simultaneously active and
competing interactions of coulomb, spin, orbital and lattice origin.
Optical studies of phase separated systems are complex. Under
special circumstances, inhomogeneities may acquire a form of
unidirectional elements extending over macroscopic dimensions (e.g.
spin and charge stripes in high-T$_c$ or organic superconductors).
In this rather exceptional situation optical experiments performed
on macroscopic specimens can be employed to probe the anisotropy
associated with this order
\cite{PhysRevLett.91.077004,PhysRev.82.538,Lee04,Kaiser09}. Provided
that the length scales associated with distinct electronic phases
present in heterogeneous specimens are smaller than the wavelength
of light one can introduce ``effective'' optical constants
$\epsilon_{\rm eff}$ for the material \cite{carr1985a}.

Effective medium theories allowed one to evaluate the effective
optical constants provided the dielectric functions and filling
fractions of constituent phases are known
\cite{ISI:000201961000004}. Interpretation of these spectra has to
be practiced with extreme caution since usual quantitative
approaches suitable for homogeneous samples may easily produce
erroneous answers.

Advances in IR/optical microscopy enabled imaging of phase separated
in correlated systems
\cite{okimotoPhysRevB.70.115104,Qazilbash2007a,qazilbashprb08:115121,nishi:014525,park-wu-ISI:000241157300026,frenzel2009}.
Nanometer-scale inhomogeneities can be registered using a host of
near-field techniques operational in different regions of the
electromagnetic spectrum with the spatial resolution reaching 8-10
nm deep in the sub-diffractional regime
\cite{keiman-ISI:000220564600007,lai-shen-rsi063702,chen-PhysRevLett.93.267401,Huber-ISI:000260888600038}.
Extended inhomogeneities occurring on the tens of microns length
scale can be detected using conventional microscopy. This latter
technique was employed to investigate the formation of stripes
induced by the electrical current in prototypical organic
Mott-Hubbard insulator (K-TCNQ) \cite{okimotoPhysRevB.70.115104}.
This material also exhibits the spin-Peierls-like structural
transition associated with the dimerization of TCNQ molecules
$T_c<395$~K. Using a combination of IR and optical  microscopies
the authors have been able to register current-induced stripes with
length scale over several $\mu$m. The striped contrast is produced
by a different degree of dimerization.

Infrared microscopy has been applied to explore the metal-insulator
transition in the correlated Mott system
$\kappa$-(BEDT-TTF)$_2$Cu[N(CN)$_2$]Br
\cite{nishi:014525,sasaki-PhysRevLett.92.227001}. Micro-reflectance
measurements in a magnetic field have identified coexisting metallic
and insulating regions revealing markedly different behavior as a
function of the applied field. Modest magnetic fields (5 T) enhanced
the areal fraction of the phase attributable to antiferromagnetic
insulator. Further increase of the field up to 10 T has triggered
the transition of the insulating antiferrromagnetic regions in the
metallic state. In Mott systems, the magnetically induced
metal-insulator transition is expected in the regime $U/W\simeq 1$
\cite{laloux-PhysRevB.50.3092}.

\begin{figure}
\centering
\includegraphics[width=0.9\columnwidth]{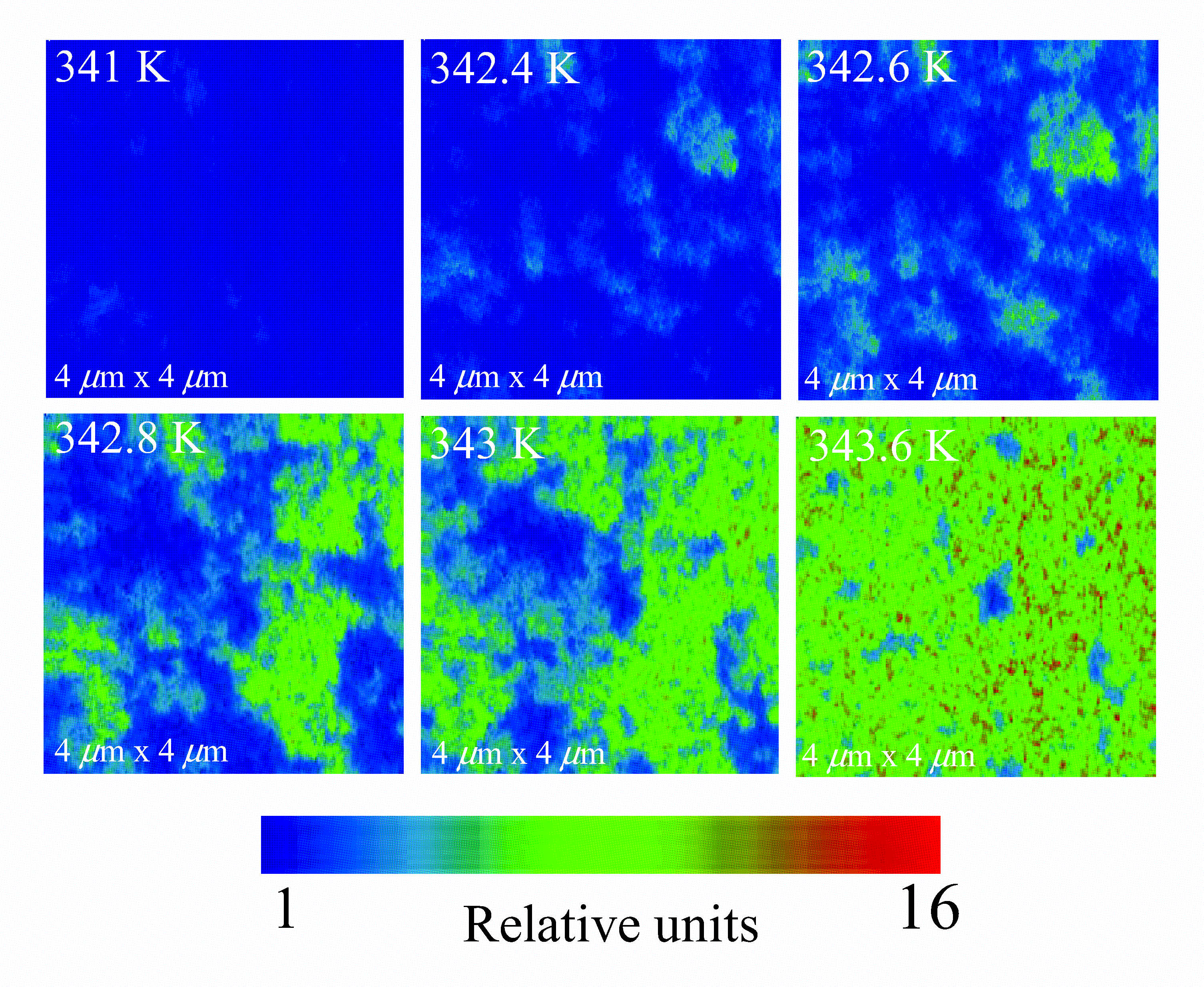}
\caption
{(Color online) Images of the nearfield scattering amplitude over the same
$4\times4$ $\mu$m$^2$ area at the infrared frequency $\omega=930$ ~cm$^{-1}$.
These images are displayed for representative temperatures in the
insulator-to-metal transition regime of VO$_2$ to show percolation
in progress. The metallic regions (light blue, green, and red colors)
give higher scattering near-field amplitude compared with the insulating
phase (dark blue color). Adapted from \textcite{Qazilbash2007a}.}
\label{fig:near-field}
\end{figure}

The highest spatial resolution of infrared experiments is a achieved
using near field instruments based on atomic force microscopes
coupled to IR lasers \cite{keilman-rev-4}. This technique enables a
contactless probe of local conductivity at the nanoscale (down to
8-10 nanometers). Near field measurements uncovered the percolative
nature of the IMT in VO$_2$
\cite{Qazilbash2007a,qazilbashprb08:115121,zhan:162110} [see
Sec.~\ref {subsec:Vanadium oxides} for background and unresolved
issues in this canonical correlated material]. Representative scans
in Fig.~\ref{fig:near-field} showed that the metallic regions
nucleate, then grow with increasing temperature, and eventually
interconnect. The observed phase separation results from an
interplay of intrinsic physics such as the first order nature of the
transition in VO$_2$ and extrinsic effects including local strain,
deviations from stoichiometry and grain boundaries in these films.
This interplay may result in enigmatic memory effects routinely
observed in correlated oxides including VO$_2$ \cite{Driscoll-2009}.

\subsection{Insights by numerical methods}
\label{subsec:DMFT prospective}
Bandwidth controlled insulator-metal transition (IMT) phenomena at
high temperatures and strong frustration is quite universal. The
paramagnetic Mott-insulating phase competes with the correlated
metallic phase of strongly renormalized quasiparticles.
At a critical pressure and critical temperature, a first order
transition occurs and is accompanied with a hysteresis behavior. The
critical exponents are of Ising type, as recently demonstrated in
Cr-doped V$_2$O$_3$ \cite{Limelette2003}.
This universality enabled one to understand the Mott transition using
a simple one band Hubbard model studied in the limit of infinite
dimension, or strong frustration. In this limit, DMFT is exact, and
can be used to compute spectroscopic quantities as a function of
temperature, pressure and chemical doping.

\begin{figure}[hbt]
\centering{
\includegraphics[width=0.9\linewidth]{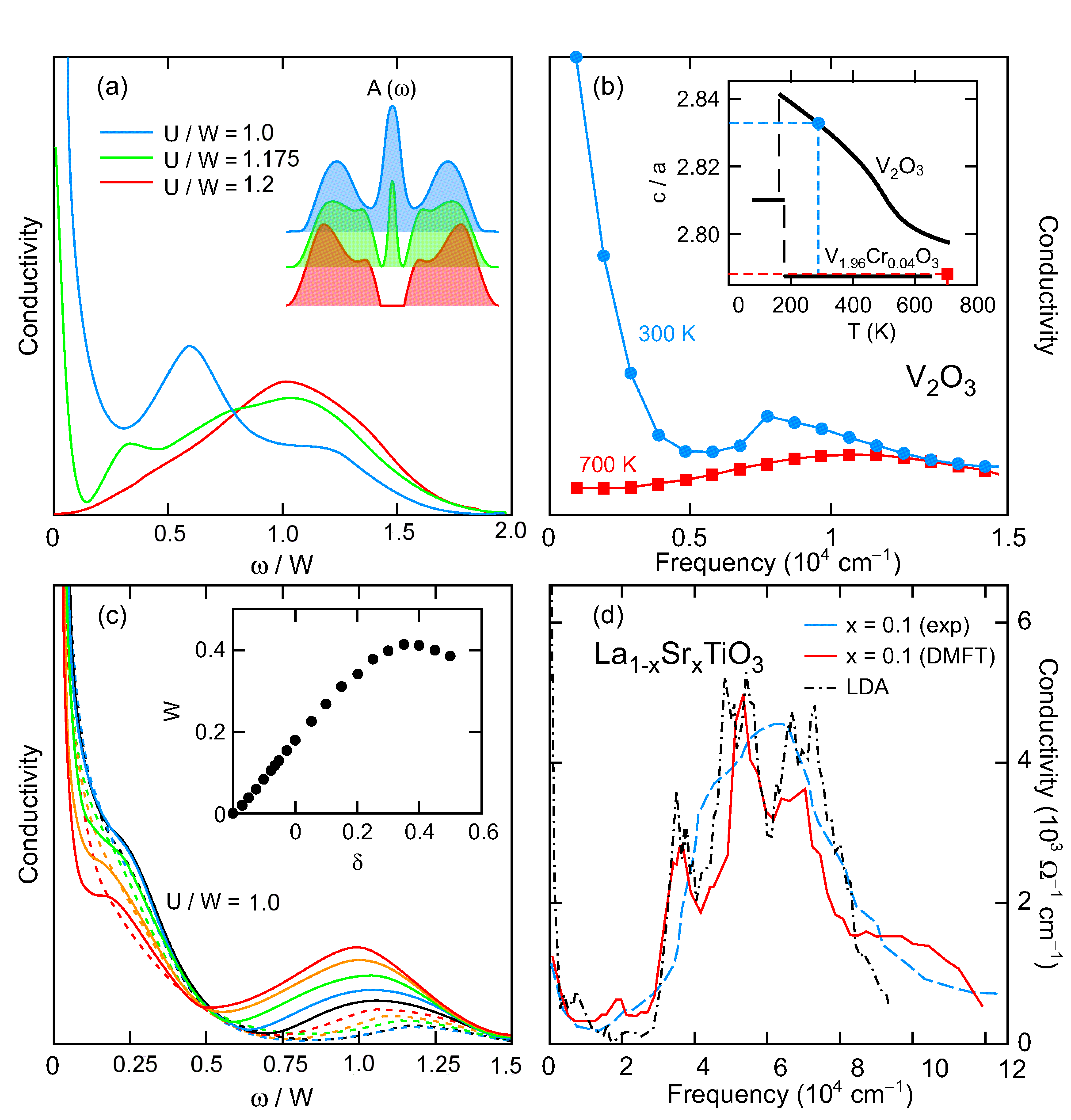}
}
\caption{(Color online)
(a) The optical conductivity of the bandwidth-controlled IMT of the
one band Hubbard model within DMFT approximation. The inset shows
the electron spectral function for the same values of $U/W$. From
\onlinecite{Blumer:2002}. (b) The realistic calculation of the optical
conductivity across the bandwidth controlled IMT of V$_2$O$_3$. The
inset shows the change of the lattice parameters with temperature,
which was important to model the transition. From
\onlinecite{baldassarre:113107}. (c) The optical conductivity of the doping
controlled IMT in the one band Hubbard model within DMFT
approximation.  The inset shows the doping dependence of the Drude
weight. From \onlinecite{Jarrel:1995}. (d) The realistic calculation of
the optical conductivity in the material exhibiting the doping
driven Mott transition. From \onlinecite{Oudovenko:2004}.  }
\label{IMTk}
\end{figure}
The inset of Fig.~\ref{IMTk}a shows the evolution of the electronic
states of the one band Hubbard model within DMFT when the ratio of
the Coulomb interaction $U$ and the bandwidth $W$ is varied. The
main panel shows the corresponding optical conductivity. Within
single site DMFT, the Mott transition is achieved by a vanishing
quasiparticle weight $Z_F$ at critical interaction $U=U_{c2}$, and
consequently the diverging effective band mass $m^*/m_b \propto
1/Z_F$. The interactions do not change the Fermi surface of a one band
model within DMFT. Shrinking of the quasiparticles leads to
decreased Drude weight, being proportional to $D\propto Z_F\propto
(U_{c2}-U)$ \cite{Georges:1996}.  However, this metallic state is
metastable at $U_{c2}$, and the first order transition to the
insulating state occurs in equilibrium before the $U_{c2}$ point is
reached. Thus, the effective mass does not truly diverge at finite
temperatures, but it is strongly enhanced at the IMT.  The optical
conductivity in  close proximity to the IMT has two additional
peaks, one which is due to excitations from the Hubbard band into
the quasiparticle peak (around $\omega \sim 0.25\,W$ for $U=1.175W$
in Fig.~\ref{IMTk}a ), and a second that is due to excitations from
the lower to the upper Hubbard band (around $\omega \sim W \sim U$).
Only the latter peak is present in the insulating state (red curve
in Fig.~\ref{IMTk}a ) where the quasiparticles cease to exist.

The qualitative features related to the IMT at finite temperature
carry over to more general models having other integer orbital
occupancies and band degeneracy, as well as including coupling to the
lattice. To illustrated that,
%
%
we show in Fig.~\ref{IMTk}b the realistic
LDA+DMFT modeling of the IMT transition in V$_2$O$_3$, which was
recently studied by \textcite{baldassarre:113107}
and \textcite{rodolakis2010a}. Similar studies in the context of
simplified models were previously carried out
by \textcite{PhysRevLett.75.105}, and by
\textcite{Kajueter:1997}.
\textcite{baldassarre:113107}  emphasized that for the theory to
quantitatively agree with experiment, it was important to include
the variation of the lattice structure with temperature. The ratio
$c/a$ was taken from experiment \cite{McWhan:1969} and is plotted in
the inset of Fig.~\ref{IMTk}b.

State of the art calculations were recently carried out for VO$_2$
by \textcite{Biermann:2008}.  The high temperature rutile phase and
the low temperature monoclinic phase were modeled and very favorable
agreement with experiments was achieved. Since the low temperature
monoclinic phase shows simultaneous Mott and Peierls correlations,
the authors had to go beyond single site DMFT and include the
tendency for dimerization by a cluster extension of DMFT
\cite{Kotliar:2001}.

The optical conductivity in the vicinity of the filling controlled
IMT was first studied by DMFT in \textcite{Jarrel:1995}. The results
are reproduced in Fig.~\ref{IMTk}c. The conductivity in the single
band Hubbard model shows the characteristic three peaks, Drude peak,
mid infrared peak, and peak at $U$. With increasing doping, the
mid-infrared peak reduces in strength, while the Drude peak
increases. The weight in the Drude peak is directly proportional to
doping $\delta$, as shown in the inset of Fig.~\ref{IMTk}c.


A well studied example of the 3D doping driven IMT is
La$_{1-x}$Sr$_x$TiO$_{3+\delta/2}$ \cite{PhysRevB.51.9581}. Its
properties can be qualitatively explained by DMFT. The resistivity
of the doped compound shows a $T^2$ Fermi liquid behavior and
the specific heat $\gamma$, which is proportional to the electron
effective mass $m^*$, is enhanced significantly near the
metal-insulator phase-transition boundary \cite{Kumagi1993}.  This
large mass enhancement suggests a divergence of the effective electron
mass due to the strong electronic correlation on approaching the
metal-insulator transition.


A realistic LDA+DMFT calculation for
La$_{1-x}$Sr$_x$TiO$_{3+\delta/2}$ was carried out by
\textcite{Oudovenko:2004} with the results reproduced in
Fig.~\ref{IMTk}d.
There are many contributions to the optical conductivity in this
material.  The Drude spectral weight is of the order of doping $\approx
\delta$, just as in the model calculations. The onset of the
Mott-Hubbard gap appears around $0.2\,$eV (see
\textcite{PhysRevB.51.9581}), and the peak around $0.5\,$eV
%
%
is due to transitions between the two Hubbard bands.
%
The rise in optical conductivity around $4\,$eV is the
charge-transfer gap between the O-$2p$ filled state and the Ti-$3d$
upper Hubbard band. The appearance of the $p-d$ transitions at the
higher energy than the excitations across the Mott-Hubbard gap in
LaTiO$_3$ confirms that LaTiO$_3$ is a Mott-Hubbard insulator rather
than a charge transfer insulator in the scheme of
\textcite{zaanen85}.

\begin{figure}[!bt]
\centering{
\includegraphics[width=0.95\linewidth]{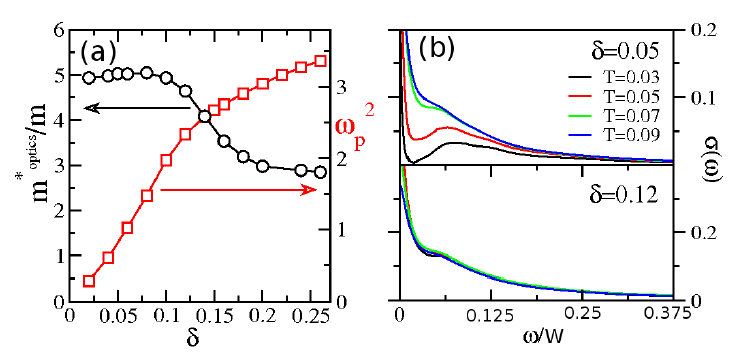}
}
\caption{(Color online) (a) The optical mass $1+\lambda=\frac{m^*}{m_{opt}}$ and the plasma frequency as a
function of doping for the t-J model within the cluster DMFT
approach for a 2D unfrustrated square
lattice. From \onlinecite{Haule:2007}. (b) The optical conductivity
of the same model at two selected doping levels and few
temperatures. From \textcite{Haule:2007} } \label{CDMFT}
\end{figure}

Owing to the success of the DMFT, we have now a firm understanding
of the appearance of conducting states in very frustrated correlated
material, or in systems with large lattice coordination, where DMFT
predictions are accurate.
The hallmark of the IMT transition in such systems is the vanishing of
the quasiparticle weight and consequently a diverging effective
mass.
%
%

Inclusion of short range non-local correlations, when these are weakly
frustrated, considerably modifies the IMT in the small doping
regime. The most prominent example is the appearance of the pseudogap
in the cuprates. In recent years, cluster extensions of the DMFT were
developed \cite{Maier:2005,Kotliar:2006}, which treat short range
correlations exactly, while the long range correlations are treated on
the mean-field level. The inclusion of commensurate short range
spin-fluctuations does not change the order of the bandwidth
controlled IMT \cite{Park:2008}, which remains first order. However,
the quasiparticle residue $Z_F$ no longer diverges at the transition.
Finite $Z_F$ at the IMT might wrongly suggest non-vanishing Drude
weight at the transition. However, the short range spin fluctuations
also strongly modify the Fermi surface in the vicinity of the
IMT \cite{Civelli:2005}, such that the plasma frequency vanishes at
the IMT. This is because the active part of the Fermi surface shrinks,
while $Z_F$ remains finite on this active part of the Fermi surface.
In Fig.~\ref{CDMFT}a we show that $\omega_p^2$, within cluster DMFT,
is an approximately linear function of doping, just as in single site
DMFT. However, the effective mass $1+\lambda$ is not diverging at the
doping controlled transition,
in accord with experimental data by
\textcite{Padila:2005} on the cuprates.
The increase in the effective mass from overdoped to underdoped regime
is of the order of $2$, in agreement with measurements
in \textcite{heumen:184512}.

In Fig.~\ref{CDMFT}b we reproduce optical conductivity for the t-J
model computed by cluster DMFT on the plaquette \cite{Haule:2007}.
Compared to single site DMFT, the mid-infrared peak is now at much
lower frequency and scales as $2J$. Hence, it comes from spin
fluctuations. This peak is quickly suppressed by doping and is
barely visible at $\delta=0.12$.


From the theoretical perspective, Mott transition can thus lead to a
diverging effective mass at the transition or not, depending on the
degree of frustration of the system.  In the absence of long range
magnetic order, and in the limit of strong frustration (or large
lattice connectivity) the Mott transition does involve divergence of the
effective mass. In this limit, the single site Dynamical Mean field
theory describes the Mott transition correctly, and the quasiparticle
renormalization amplitude $Z_F$ is only weakly momentum dependent, and
effective mass is inversely proportional to $Z_F$.

Note however that the transition is first order at finite
temperatures, hence the truly diverging effective mass occurs only if
one follows the metastable metallic state to the point where it ceases
to exist.


In the opposite limit of very weak frustration, such as in the Hubbard
model in two dimensions, the Mott transition is accompanied by a
strong differentiation in momentum space and strong momentum
dependence of quasiparticle renormalization amplitude
\cite{Parcollet04,Civelli:2005}.
In this limit, the effective mass at the Mott transition remains
finite \cite{Park:2008}, and only a part of the Fermi surface becomes
metallic at IMT \cite{PhysRevB.80.045120}. Moreover, the Fermi surface
appears gradually with increasing doping or increasing the bandwidth.


Studies of correlated electron models, such as the Hubbard and the
t-J model, have a long history. The most successful method among the
numerical methods for computing optical response was exact
diagonalization of small clusters using the Lanczos algorithm (for a
review see \textcite{RevModPhys.66.763}). A series of these
numerical studies
\footnote{\textcite{Moreo:1990,Stephan:1990,Sega:1990,Chen:1991,Inoue:1990,Dagotto:1992,Loh:1988,Tohyama:1999,
PhysRevB.16.2437,Zemljic:2005,Tohyama:2006}.} established that the
Hubbard model has a charge gap in the Mott state. The Drude weight
in the doping controlled IMT increases continuously with
doping~\cite{Dagotto:1992}, roughly linearly with doping
level~\cite{Tohyama:2006}. These studies also showed that for small
doping, a large part of the optical spectral weight is incoherent
and is located inside the original charge
gap.\footnote{\textcite{Moreo:1990,Stephan:1990,Sega:1990,Chen:1991,Inoue:1990,Zemljic:2005}.}
The mid-infrared peak appears around $2J$
\cite{Sega:1990,Moreo:1990,Stephan:1990,Chen:1991}, which was
ascribe to a propagating hole dressed by spin excitations.
Upon increasing doping, the Drude-like part of the spectra grows, and
the weight is progressively transferred from the higher energy region
above the charge gap of the insulator to the region inside the
gap.
The low frequency optical spectral weight and its doping dependence
was studied by a variety of other methods, including the variational
quantum Monte Carlo method~\cite{Millis:1990}, slave boson mean field
methods~\cite{Grilli:1990}, and memory function
approach \cite{Jackeli:1999}.

These numerical methods proved to be very useful in unraveling the
IMT in a simple one band model, such as the Hubbard and the t-J
model. Their generalization to realistic materials however are
prohibitively expensive. On the other hand, density functional
theory (DFT) methods were developed and it was empirically
established that DFT band structure of simple metals is very close
to the experimentally established excitation spectrum. It was also
realized that the high energy excitation
spectrum~\cite{Pickett:1989,Jones:1989} as well as the Fermi
surface~\cite{Andersen:1994,Schabel:1998} of many correlated
materials is well described by DFT.  This was not expected since DFT
is a ground state theory. Due to this unexpected success, the
optical conductivity of the DFT free carriers turned out to be
surprisingly similar to optical spectra of simple metals and even
transition metals \cite{Maksimov:1988,Maksimov:1989}, owing to
careful implementation of velocity matrix
elements~\cite{Uspenski:1983,Maksimov:1988,Ambrosch:2006}.  Although
the Mott insulating state cannot be described by DFT, nor the
anomalous metallic state in the vicinity of the IMT, high energy
interband transitions are usually very satisfactory given by this
method~\cite{Kircher:1993,Maksimov:1989}. The itinerant metallic
state away from the IMT in transition metal compounds is routinely
found to be well described by DFT~\cite{Maksimov:1989}, with the
exception of the reduction of the Drude weight. Given the simplicity
and speed of  DFT methods, as well as their ab-initio standing, they
are invaluable tools for understanding and predicting material
properties. Their combination with perturbative GW
method~\cite{Giovanni:2002} and Dynamical Mean Field
Theory~\cite{Kotliar:2006} makes them even more attractive for the
description of correlated materials.

\section{Transition Metal Oxides}
\label{sec:Transition Metal Oxides}

\subsection{Cuprates}
\subsubsection{Steady state spectroscopy}
\label{subsec:Cuprates}
The key unresolved issue in the physics of the high-$T_c$ cuprates
is the mechanism of superconductivity. Despite unprecedented
research effort it remains unclear if superconducting pairing is
mediated by strong coupling to bosonic modes in fashion not
dissimilar to the BCS theory or a totally new mechanism is
operational in this class of materials \cite{bonn-np-2006}.
Arguably, the most significant departure from the BCS scheme in
high-T$_c$ cuprate superconductors is revealed by optical studies
\cite{legget-np-2006,Basov-sci1999}, which indicate that electronic
processes occurring on the energy scale $10^2 - 10^3$ $k_B$T$_c$ are
often involved in the formation of the superconducting
condensate.\footnote{
\textcite{katz-PhysRevB.61.5930,PhysRevB.63.134514,Molegraaf2002,Kuzmenko2003,
mklein-PhysRevB01,Homes2004a,
Santander2004a,boris-science-ISI:000221105300037,
norman-PhysRevB.66.100506,laforge:097008}.}
These high-frequency optical effects have been observed in the
response of the CuO$_2$ planes: the key building block of all
cuprates as well as in the response along the less conducting
interplane direction. An appealing interpretation of these effects
is in terms of electronic kinetic energy savings at $T <T_c$
\cite{Hirsch1992305,chakravarty-PhysRevLett.82.2366} and is at odds
with predictions of BCS theory. The low-energy spectralweight is not
conserved in the normal state either as discussed in section
\ref{subsec:Sum rules} where we also analyze some of the caveats of
possible interpretations.

The cuprates offer the best studied example of the filling
controlled Mott transition. Superconductivity in this class of
materials occurs as undoped antiferromagnetic insulators are being
transferred into a fairly conventional Fermi liquid on the overdoped
side.  Much of currently accepted phenomenology of high-$T_c$ phases
has been established with strong involvement of optical techniques
\cite{basov:721}. This includes strong dissipation and
unconventional power law behavior of the scattering rate in the
normal state, the formation of a partial electronic gap or
pseudogap, and strong anisotropy of both normal state and
superconducting properties.

In particular, pseudogap state physics has captured unparalleled
attention. The pseudogap state is realized in the moderately doped
materials with a $T_{c}$ lower than the maximum $T_{c}$ for a
given series. Transport and spectroscopic probes reveal a
pseudogap in the normal state that resembles the superconducting
gap in magnitude and symmetry leading to the common view that the
origin of the pseudogap may be intimately related to
superconducting pairing at $T>T_{c}$
\cite{Puchkov1996a,timusk1999a,lee-rmp2006}. The pseudogap is not
unexpected by continuity with fully gapped insulating counterparts
but is in conflict with Landau Fermi liquid theory. However,
studies focusing on strongly under-doped samples uncovered
different doping trends between the superconducting gap and the
pseudogap
\cite{lee:054529,tacon-2006,shen-science06,ISI:000256544200001}.
These latter experiments point to the different microscopic
origins of superconductivity and pseudogap.  Yet another alternate
point of view asserts that the pseudogap represents a state that
competes with superconductivity \cite{lee-rmp2006,Kaminski-nat09}.
Infrared signatures of the pseudogap include a suppression of the
scattering rate at $\omega<500$ cm$^{-1}$ in the conductivity
probed along CuO$_2$ planes
\cite{Rotter1991a,Orenstein1990a,Puchkov1996a,basov-prl96,puckkov-pgap-prl}
with the simultaneous development of a gap-like structure in the
inter-plane c-axis conductivity
\cite{Homes2003a,PhysRevB.52.R13141,tajima-pgap97,Bernhard-pgap99}.
These trends are common between several different families of
hole-doped cuprates. In electron doped materials a gap-like
structure can be identified directly in the conductivity spectra
\cite{onose-el-doped,zimmers-05,zimmers-epl-05,PhysRevB.74.214515}.
The electronic kinetic energy is most strongly suppressed compared
to the $K_{\rm band}$ value in underdoped materials characterized
by the pseudogap (Fig.~\ref{fig:KE-all}).

\begin{figure}
\centering
\includegraphics[width=3.0in]{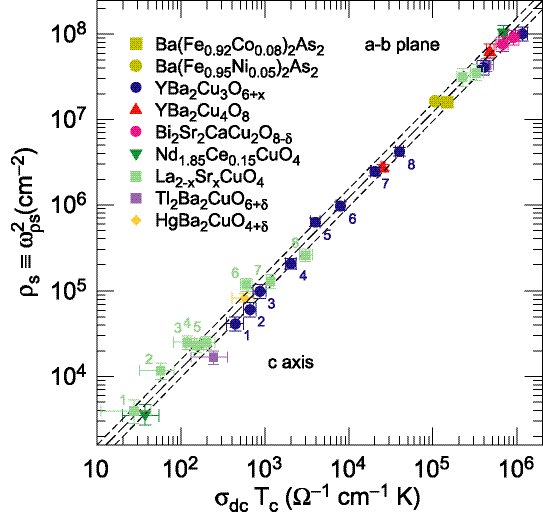}
\caption
{(Color online) The log-log plot of the spectral weight of the superfluid
density $\rho_{s,0}$ for a variety superconductors probed both
along conducting $ab$-plane direction and interplane $c$-axis direction.
After \textcite{ISI:000222946100041} supplemented by iron-pnictide data from \textcite{Wu10c}.
\label{fig:Homes-law}}
\end{figure}

Electrodynamics of the superconducting condensate has been explored at zero and finite
magnetic fields. Microwave studies of the temperature dependence of
the superfluid density within the CuO$_2$ planes have first hinted
to the unconventional $d$-wave nature of the order parameter
\cite{hardy-PhysRevLett.70.3999} later confirmed through direct
phase sensitive measurements \cite{vanharlingen-RevModPhys.67.515}.
The layered nature of the cuprates implies strong anisotropy of the
superfluid density $\rho_s$. The properties of the interlayer
components of $\rho_s$ can be understood in terms of Josephson
coupling of the CuO$_2$ planes
\cite{PhysRevB.50.3511,PhysRevLett.72.2263,PhysRevB.65.134511}. The
formation of the (stripe-like) magnetic order within the CuO$_2$
frustrates the Josephson coupling leading to two-dimensional
superconductivity within the CuO$_2$ layers in several families of
cuprates \cite{berg:127003,PhysRevLett.96.257002,PhysRevB.81.060506,PhysRevB.81.064510,PhysRevLett.104.157002},
in particular La$_2$CuO$_4$. The in-plane superfluid density reveals
universal scaling with T$_c$: the $\rho_{CuO_2}\propto T_c$
\cite{PhysRevLett.66.2665,Uemura200323}. This effect is regarded as
one of the most evident manifestations phase fluctuations in a doped Mott insulator\textcite{emery-kivelson-05}. Deviations from the Uemura plot primarily
in the overdoped crystals are captured with $\rho_{\rm CuO_2}\propto
\sigma_{DC}\times T_c$\cite{ISI:000222946100041,PhysRevB.71.184515,PhysRevB.72.134517}. Physics underlying
this latter universal behavior seen in many classes of exotic
superconductors (Fig.~\ref{fig:Homes-law}) may involve strong
dissipation in the normal state characteristic of conducting
correlated electron systems \cite{Zaanen2004a}.  Systematic studies
of $\rho_s$ in very weakly doped ultrathin films indicate that the
disappearance of superconductivity at low dopings may be due to
quantum fluctuations near a two-dimensional quantum critical point
 \cite{lemberger07}. On the contrary the behavior of the in-plane superfluid density in very weakly doped YBa$_2$Cu$_3$O$_{6+y}$ single crystals of high purity is consistent with the notion of a quantum phase transition in the (3+1)-dimensional XY universality class.\cite{Broun07} Low values of the superfluid density in
underdoped phases in combination with the short coherence length all
realized in the environment of  copper-oxygen plane give rise to
prominent fluctuations effects both below and above T$_c$
\cite{PhysRevLett.73.1845,Corson99}. One of the most remarkable
observations pertains to the survival of superconducting
fluctuations at temperature above $T_c$ by 10-30~K \cite{Corson99}
later confirmed through systematic studies of the Nernst effect
\cite{wang:024510}. These findings indicate that the energy scale
associated with the fluctuation regime is much smaller than the
pseudogap scale implying that the two phenomena may be of different
origin. Anomalies of the in-plane superfluid density in weakly doped cuprates have been interpreted by \textcite{PhysRevB.80.180509} in terms of Josephson phases: a regime of isolated superconducting regions experiencing Josephson coupling.

The electromagnetic response in the presence of the magnetic field
is dominated by the dynamics of both pancake and Josephson vortices
\cite{matsudaprl95,kojima-prl02,dordevic-vortex02,laforge-vort07}.
THz spectroscopy of the non-diagonal components of the conductivity
tensor in the magnetic field (often referred as Hall angle
experiments) \cite{Park-PhysRevB.56.115,grayson-prl02,schmad-prl}
concur that the charge dynamics near optimal doping is fairly
conventional and governed by a single relaxation rate (contrary to
earlier theoretical proposals). Magneto-optics data for moderately
underdoped samples of YBCO family are indicative of density-wave
like reconstruction of the Fermi surface \cite{schmad-prl}. This
latter result is of interest in the context of quantum oscillations
observed in high purity underdoped samples of YBCO
\cite{doiron-osc,jaudet:187005,sebast-osc,yelland:047003}. Quantum
oscillations signal the presence of coherent quasiparticles.
Virtually all attempts to explain the oscillating phenomena invoke
Fermi surface reconstruction due to some type of density-wave order
\cite{sudip-osc,Harrison-osc,millis-norman-osc}. An alternative proposal\cite{PhysRevB.81.060506,PhysRevB.81.064510} is aimed to reconcile quantum oscillations data with transport, infrared and photoemission experiments highlights the role of stripes.

\subsubsection{Pump-probe spectroscopy}
\label{subsec:High T$_{c}$ Time Domain Dynamics}
Given the ability to directly measure electron-phonon coupling in
metals and the work on nonequilibrium dynamics in BCS
superconductors, it is natural that time-resolved spectroscopy would
be utilized to investigate quasiparticle recombination in the
cuprates with a view towards obtaining insights into the pairing
mechanism. Initial studies using all-optical pump-probe spectroscopy
on YBa$_{2}$Cu$_{3}$O$_{7-\delta}$ thin films revealed a response
that changed dramatically at $T_{c}$ showing, for example, a slow
(bolometric) induced increase in $\Delta R/R$ above $T_{c}$ crossing
over to a fast induced decrease in $\Delta R/R$ below $T_{c}$
\cite{han1990a,chekalin1991a,stevens1997PRL}. Below $T_{c}$, this
data was interpreted in terms of a fast ({\em i.e.} 300 fs)
avalanche process followed by quasiparticle recombination to Cooper
pairs on a picosecond time scale limited by the $2\Delta $ phonon
relaxation time similar to BCS superconductors. Importantly, the
timescales in the cuprates are much faster than in BCS
superconductors given the larger gap
\cite{kabanov1999a,Carbotte2004a}.

The initial studies on cuprates have been extended since 1999 to
include more detailed investigations of the nonequilibrium dynamics
using all-optical pump-probe spectroscopy and, additionally, low
energy probes in the mid and far infrared.\footnote{
\textcite{gay1999a,demsar1999a,kaindl2000a,averitt2001b,schneider2002a,
demsar2001a,kaindl2005a,kaindl_averitt2007a,segre2002PRL,gedik2005a,gedik2004a,gedik2003a,chia2007a}.}
These results reveal a marked sensitivity to the onset and
subsequent evolution of the superconducting state with decreasing
temperature.
Photoinduced changes at 1.5 eV were performed on Y$_{1-x}$Ca$_{x}$Ba$_{2}$Cu$%
_{3}$O$_{7-\delta }$ single crystals for x=0 ($T_{c} = 93$~K) and
$x=0.132$ ($T_{c}=75$~K). A two exponent relaxation was observed.
The slow component (3 ps) was interpreted \cite{demsar1999a} as the
recombination time where $\tau_{\rm exp}\sim 1/\Delta $ with pair
breaking due to phonons with energies greater than $2\Delta$
limiting the recovery time. The fast component (0.5 ps) was
associated with a temperature independent gap (i.e. the pseudogap).
These results strongly show that the recovery dynamics of $\Delta
R/R$ at probe frequencies well above the gap energy are sensitive to
the superconducting gap and to pseudogap and that the reformation
time of the condensate is rapid.

Ultrafast measurements of YBa$_{2}$Cu$_{3}$O$_{7-\delta }$ thin
films in the mid-IR (60-200~meV) which probe in the vicinity of the
superconducting gap also revealed an (i) picosecond recovery of the
superconducting condensate, (ii) a subpicosecond response related
pseudogap correlations, and (iii) a temperature dependence of the
amplitude that follows the antiferromagnetic 41 meV peak observed in
neutron scattering \cite{kaindl2000a}. These dynamics are similar to
the 1.5 eV measurements except for a difference in the temperature
dependence of the amplitudes that remains unexplained.

\begin{figure} [ptb]
\begin{center}
\includegraphics[width=2.3in,keepaspectratio=true]%
{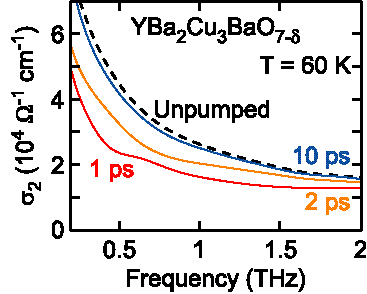}%
\caption{Imaginary part of the far-infrared conductivity in a
YBa$_{2}$Cu$_{3}$O$_{7-\delta }$ film as a function of frequency for
various times following photoexcitation. A initial reduction in the
condensate density yields a reduction
in the 1/$\omega$ response which recovers in $\sim$2 ps. After \textcite{averitt2001b}. }%
\label{averitt5}
\end{center}
\end{figure}

Optical pump Terahertz-probe studies of
YBa$_{2}$Cu$_{3}$O$_{7-\delta}$  revealed that it is possible to
simultaneously monitor the dynamics of excess quasiparticles and the
condensate recovery which manifest as a strong 1/$\omega$ component
in the imaginary conductivity. The first far-infrared optical-pump
terahertz-probe experiments were performed on near-optimally doped
and underdoped samples of YBa$_{2}$Cu$_{3}$O$_{7-\delta }$
\cite{averitt2001b,averitt2002c}. The measurements were made on
epitaxial thin films including YBa$_{2}$Cu$_{3}$O$_{7 - \delta}$
with $T_{c}  = 89$~K. Fig.~\ref{averitt5}(a) shows
$\sigma_{2}(\omega)$ as a function of frequency at various delays
following photoexcitation. There is a rapid decrease in the
condensate fraction followed by a fast, picosecond recovery. The
condensate recovery is nearly complete by 10~ps, in dramatic
contrast to conventional superconductors. In the optimally doped
films the condensate recovery is approximately 1.7 ps (in comparison
to $>$100ps in conventional superconductors) and increases near
$T_{c}$, consistent with a decrease in the superconducting gap.
Above $T_{c}$ at 95 K, the lifetime has decreased to 2 ps and is
likely a measure of electron-phonon equilibration in the normal
state. Furthermore, the lifetime is independent of fluence
indicative of the absence of bimolecular kinetics. In contrast, for
underdoped films (YBa$_{2}$Cu$_{3}$O$_{6.5}$, $T_{c}  = 50$~K) the
lifetime was constant at 3 ps even above $T_{c}$ suggestive of a
pseudogap.

All optical pump-probe experiments have been performed on high
quality YBCO Ortho II single crystals, an underdoped cuprate
superconductor with a T$_{C}$ of 60 K
\cite{segre2002PRL,gedik2004a}. The goal was to determine whether
quasiparticle relaxation is described by one or two-particle
kinetics. In one-particle (unimolecular) kinetics, the excitation
created by the pump has an intrinsic lifetime and the decay is
expected to be exponential. In two-particle (bimolecular) kinetics
the photon creates a pair of excitations, for example an electron
and hole, that inelastically scatter off each other in order to
recombine. In this case the decay rate is expected to follow a power
law in time and to become faster as the initial excitation density
increases. Direct evidence for the importance of bimolecular
kinetics in cuprates superconductors was reported by Segre et al.
\cite{segre2002PRL} and Gedik et al. \cite{gedik2004a}. Bimolecular
recombination was observed, indicative of the recombination of a
pair of opposite spin quasiparticles.

More recently, detailed optical-pump terahertz-probe experiments
were performed on Bi$_{2}$Sr$_{2}$CaCu$_{2}$O$_{8+\delta}$
\cite{kaindl2005a}. The pump-induced change $\Delta\sigma(\omega)$
of the in-plane THz conductivity was measured in 62 nm thick
optimally doped Bi$_{2}$Sr$_{2}$CaCu$_{2}$O$_{8+\delta}$ films. As
in YBa$_{2}$Cu$_{3}$O$_{7-\delta}$, the superconducting state
exhibits a $1/\omega$ component in the imaginary part of
conductivity $\sigma_{2}(\omega)$. This response, as is well known
(see Sec. \ref{subsec:BCS Superconductors}), provides a direct
measure of the condensate density and, therefore, its temporal
evolution is a direct measure of the condensate recovery dynamics.
In these studies, bimolecular recombination (i.e. a two particle
process) was also observed consistent with the pairwise interaction
of quasiparticles as they recombine into Cooper pairs.

\begin{figure} [ptb]
\begin{center}
\includegraphics[width=3.0in,keepaspectratio=true]%
{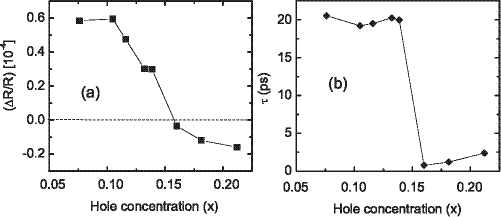}%
\caption{(a) Peak induced change in $\Delta R/R$ at 1.55 eV and (b)
recovery lifetime as a function of hole concentration in
Bi$_{2}$Sr$_{2}$Ca$_{1-y}$Dy$_{y}$Cu$_{2}$O$_{8+ \delta}$ at
$T=5$~K. An abrupt change in the dynamics occurs at optimal doping
($x \sim 0.15$). Note that the values of hole doping x are obtained
through
variation of Dy and the oxygen stoichiometry. From \textcite{gedik2005a}.}%
\label{averitt6}
\end{center}
\end{figure}

There have also been very careful all-optical pump-probe studies on
Bi$_{2}$Sr$_{2}$Ca$_{1-y}$Dy$_{y}$Cu$_{2}$O$_{8+ \delta}$ as a
function of doping. While optical-pump terahertz-probe has the
advantage of directly probing the low energy dynamics, all-optical
pump-probe spectroscopy has advantages in the sensitivity and that
small single crystals can be measured with comparable ease.
Fig.~\ref{averitt6} summarizes time-resolved measurements of the
photoinduced change in reflectivity $\Delta R/R$ at 1.5 eV in
Bi$_{2}$Sr$_{2}$Ca$_{1-y}$Dy$_{y}$Cu$_{2}$O$_{8+ \delta}$ as a
function of hole concentration \cite{gedik2005a}. The quasiparticle
decay and the sign of $\Delta R/R$ change abruptly at the maximum of
the superconducting transition temperature with respect to doping.
This reveals that a sharp transition in the quasiparticle dynamics
takes place precisely at optimal doping in
Bi$_{2}$Sr$_{2}$Ca$_{1-y}$Dy$_{y}$Cu$_{2}$O$_{8+ \delta}$. The sign
change was interpreted in terms of the change in the real part of
the dielectric function arising from spectral weight transfer from
the condensate. This is consistent with time-integrated optical
conductivity measurements experiments directly measuring the
spectral weight transfer due to thermal depletion of the condensate
as a function of doping \cite{Basov-sci1999,Molegraaf2002}.

The change in the dynamics indicates different quasiparticle
relaxation dynamics for over and underdoping [Fig.
~\ref{averitt6}(b)]. In the underdoped side, the recombination rate
depends linearly on the density, consistent with pair-wise
bimolecular kinetics. In contrast, for the overdoped side the decay
is fast and independent of excitation density. The fact that the
amplitude of $\Delta R$ depends linearly on excitation density
suggests that the photoinduced quasiparticles are antinodal, as the
spectral weight transfer due to nodal quasiparticles would lead to
$\Delta R \propto n^{0.33}$ \cite{Carbotte2004a}. Thus, the
intensity dependent dynamics might indicate that the antinodal
quasiparticles are metastable on the underdoped side of the phase
diagram while readily decaying into other excitations for
overdoping. We mention that time-integrated experiments of the
interplane response measure the antinodal conductivity
\cite{Basov-sci1999}. Thus, time-resolved studies of the c-axis
optical conductivity might provide additional insight into the
dynamics of antinodal quasiparticles.

From the optical-pump THz-probe experiments described above the
decrease in superfluid density demonstrates that Cooper pairs are
broken following photoexcitation. Nonetheless, it is not known
whether the resulting quasiparticle population is nodal, antinodal,
or both. A step towards this identification was the direct
observation  of the diffusion coefficient of the photoinduced
quasiparticles in YBCO Ortho II Gedik et al. (2003).  The diffusion
coefficient at low temperature was determined to be 24 cm$^{2}$/sec,
which is approximately 200 times less than would be expected from
nodal quasiparticle transport measurements on the same crystals.
This indicates quite strongly that antinodal quasiparticles are
present in the photoexcited state, possibly together with nodal
quasiparticles.

\subsection{Vanadium oxides}
\subsubsection{Steady state spectroscopy}
\label{subsec:Vanadium oxides}
Vanadium oxides are canonical examples of transition metal oxides
with correlated electrons. Vanadium dioxide shows a transition from
the insulating phase to metallic phase as $T$ rises across $T_{\rm
IMT}=340$~K accompanied by a structural phase transition
\cite{PhysRevLett.3.34,ISI:A1971K569100005}. Vanadium sesquioxide
V$_2$O$_3$ undergoes a first-order insulator-metal transition at
$T_{\rm IMT}=150$~K from a low temperature antiferromagnetic (AF)
insulating phase to a high temperature paramagnetic metallic phase.
The crystal structure deforms from monoclinic in the insulating
phase to rhombohedral symmetry in the metallic phase. Transport and
thermodynamic measurements present solid evidence for a band-width
controlled form of the Mott transition in V$_2$O$_3$
\cite{PhysRevLett.27.941,PhysRevB.48.16841}. The low-$T$ insulating
phase of VO$_2$ shows no AF ordering but does reveal charge ordering
of vanadium pairs along the $c$-axis. The presence of such vanadium
chains imparts a quasi one-dimensional character to what is
essentially a three-dimensional system and prompted an
interpretation of the IMT in terms of the Peierls physics.

There has been considerable controversy over the relative importance
of the Peierls scenario and electronic correlations representing
Mott
physics.\footnote{\textcite{zylb-1975,allen-94,rice-prl94,bierman-prl05,kim-prl06,kim-njp-04,
hilton2007a,haverkort:196404, Qazilbash2007a,haglund-nanolett,perucchi2009a}.} A
number of experimental studies indicates that the electronic IMT
transition in VO$_2$ precedes the structural phase transition
\cite{kim-prl06,lupi-vo2,kim-apl07}. These results motivated time
resolved optical and structural studies discussed in Sec.~\ref
{subsec:VO$_{2}$ Time-Domain Dynamics}. Ternary and quarternary
vanadium oxide reveal a number of fascinating phenomena including
heavy fermion behavior in LiV$_2$O$_4$ \cite{PhysRevLett.78.3729},
optical anisotropy induced by orbital effects in pseudocubic
La$_{1-x}$Sr$_x$VO$_3$ \cite{fujioka:196401}, 1D magnetic chains in
$\alpha '$-NaV$_2$O$_5$ \cite{PhysRevB.61.2535} and many others.

\begin{figure}
\centering
\includegraphics[width=3.0in]{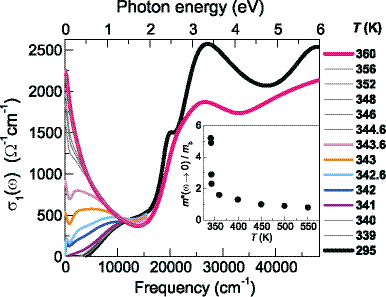}
\caption
{(Color online) The real part of the optical conductivity of VO$_2$ film extracted
from ellipsometric measurements in mid-IR -- UV in combination with
reflectance in far-IR. The transformation of these spectra across the
IMT reveal spectral weight transfer over the energy scale beyond 6 eV
common to other correlated electron systems. Spectra in the transition
region (339-352 K) are representative of electronic phase separation of
VO$_2$ in metallic and insulating regions. Applying the effective medium
theory to the analysis of these data one can extract the electromagnetic
response of metallic puddles. This response is characterized by nearly
divergent behavior of the effective mass (inset).
Adapted from \textcite{Qazilbash2007a}.}
\label{fig:VO2}
\end{figure}

Early experiments have first uncovered dramatic modification of
optical properties of both VO$_2$ and V$_2$O$_3$ across the IMT
\cite{Barker-v2o3,barker-vo2}.
\textcite{Thomas1994a,rozenberg-prl75,rozenberg-96} discovered a
redistribution of the electronic spectral weight in V$_2$O$_3$
associated with the IMT. As pointed in \ref{subsec:Emergence of
conducting state in correlated insulators} such a non-local
redistribution of spectral weight involving the conductivity range
extending up to several eV is now generally regarded as a salient
feature of a correlated electron state. Experiments for thin film
samples reaffirm these earlier findings
\cite{noh-vo2,okazaki-vo2,Qazilbash2007a,baldassarre:245108,perucchi2009a}.
Ellipsometric data for $\sigma_1(\omega)$ of VO$_2$ displayed in
Fig.~\ref{fig:VO2} visualize the issue of the spectral weight
transfer with utmost clarity. Spectra taken in the transition region
reveal the ``filling'' of a 0.6 eV gap with states.

One obstacle towards quantitative analysis of these latter spectra
is phase separation (Fig.~\ref{fig:near-field}). The effective
medium analysis of a combination of near field and broad-band
ellipsometric data \cite{qazilbash:075107} revealed strong
enhancement of the quasiparticle effective mass in newborn metallic
nano-clusters. This result supports the notion of the dominant role
of electron-electron interaction \cite{brinkman-rice}  in the IMT
physics of VO$_2$ similarly to its V$_2$O$_3$ counterpart
\cite{baldassarre:245108}. The electronic kinetic energy of the
metallic phases is reduced compared to band structure results both
in VO$_2$ and V$_2$O$_3$ attesting to correlated character of the
metallic regime (Fig.~\ref{fig:KE-all}).

Optical studies of yet another vanadium oxide: V$_3$O$_5$
\cite{baldassarre:245108} are in stark contrast with the
observations for both VO$_2$ and V$_2$O$_3$ \cite{perucchi2009a}.
Specifically, the spectral weight redistribution in V$_3$O$_5$
across $T_{\rm IMT}=420$~K is confined the region within 1 eV. The
form of the optical spectra and their evolution with temperature and
pressure are indicative of prominent polaronic effects. Thus data
for V$_3$O$_5$ show that lattice effects and structural phase
transitions can only lead to a fairly conventional picture of
electrodynamics across the insulator to metal transition. This may
imply that the role of lattice effects in other vanadium oxides is
also fairly mundane whereas exotic effects of metallic transport of
VO$_2$ and V$_2$O$_3$ likely originate from the proximity to Mott
insulating state.

The insulator-to-metal transition in both VO$_2$ and V$_2$O$_3$ can
be tuned not only by temperature but also by doping, the electric
field\cite{qazilbash-vo2-fet},
pressure\cite{PhysRevB.5.2541,PhysRevB.48.16841} and
photoexcitation\ref{subsec:VO$_{2}$ Time-Domain Dynamics}. While all
these stimuli promote metallicity it is not evident that either
nature of the transition or the end metal phase are identical in all
these cases. At least in V$_2$O$_3$ x-ray absorption data revealed
that the metallic phase reached under pressure is different from the
one obtained by changing doping or temperature
\cite{rodolakis2010a}. Since electronic phase separation is clearly
playing a prominent role in the insulator-transition
physics\ref{fig:near-field} more detailed insights can be expected
from the exploration of transport, spectroscopic and structural
aspects of the transition using experimental probes with adequate
spatial resolution at the nano-scale.

\subsubsection{Pump-probe spectroscopy}
\label{subsec:VO$_{2}$ Time-Domain Dynamics}
The insulator to metal transition in VO$_{2}$ has been extensively
investigated using ultrafast optical spectroscopy with experiments
spanning from the far-infrared through the visible including, in
addition, time resolved x-ray and electron diffraction studies.
\footnote{\textcite{cavalleri2001a,cavalleri2004a,cavalleri2005a,kubler2007a,hilton2007a,baumscience2007,rini2008a,nakajima2008a}.}
For VO$_{2}$, a primary motivation of these photoinduced phase
transition experiments (see Sec.~\ref{subsec:Photoinduced phase
transitions}) is to determine the relative influence upon the IMT
transition of structural distortions (associated with the vanadium
dimerization) and correlation effects using the temporal
discrimination ultrafast optical spectroscopy provides. The existing
body of work reveals a prompt nonthermal photoinduced transition
which occurs on a sub-picosecond timescale and a slower thermally
induced phase transition which is sensitive to a softening of the
insulating phase. A fluence threshold to drive the phase transition
is observed and is a well-known feature of photoinduced phase
transitions \cite{nasu2004a,koshihara1999a} where the cooperative
nature of the dynamics results in a nonlinear conversion efficiency
as a function of the number of absorbed photons.

The approach of these experiments is that, starting from the
insulating phase, excitation with above bandgap photons leads to a
reduction from half-filling in the lower Hubbard band
(``photodoping'') which initiates a collapse of the Hubbard gap on a
sub-picosecond timescale. For example, an all-optical pump-probe
study of the dynamics revealed that for pulses shorter than 70 fs
the transition time is constant \cite{cavalleri2004a}. This
indicates a structural bottleneck to obtaining the metallic phase in
contrast to what would be expected for a purely electronic phase
transition suggesting that lattice effects must be considered in any
complete scenario of the photoinduced IMT transition in VO$_{2}$ and
other systems. This is consistent with time-resolved x-ray
diffraction experiments where the photoinduced change in diffraction
of the rutile phase shows an initial subpicosecond increase and a
longer $\sim 15$~ps increase related to the nucleation and growth of
the metallic phase \cite{cavalleri2001a}. This is also in consistent
with the discussion at the end of Section \ref{subsec:Vanadium
oxides} discussing multiple pathways in driving a IMT
\cite{rodolakis2010a}. Along these lines, the physics of a
photoinduced IMT may be considerably different than IMTs driven by
temperature, electric or magnetic fields, or applied currents and is
a topic of considerable interest.

\begin{figure} [ptb]
\begin{center}
\includegraphics[width=2.3in,keepaspectratio=true]%
{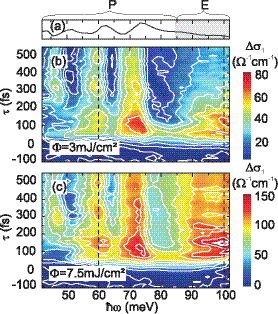}%
\caption{(Color online)
Optical-pump THz-probe on VO$_{2}$ at $T=250$~~K. (a) In the 40 - 85 meV
range, the temporal evolution of the optical phonons probes the
lattice degree of freedom, while changes above 85 meV are
representative of the electronic response. (b) and (c)
Two-dimensional evolution of the optical conductivity as a function
of time and frequency. In (b) the excitation fluence is below the
threshold to drive the transition while (c) shows the above
threshold dynamics which reveal a prompt increase in
the conductivity above 85 meV. From \textcite{kubler2007a}.}%
\label{averitt7}
\end{center}
\end{figure}

Fig.~\ref{averitt7} shows results using electric-field resolved
mid-IR spectroscopy which have been used to probe the nonthermal
photoinduced phase transition \cite{kubler2007a}. The
two-dimensional scans highlight the power of this technique showing
the temporal evolution of the optical conductivity.
Fig.~\ref{averitt7}(b) is the response at 3mJ/cm$^{2}$ which (at
300K) is below the fluence threshold for to drive VO$_{2}$ to the
metallic state while Fig.~\ref{averitt7}(c) reveals the response at
an excitation fluence of 7.5mJ/cm$^{2}$. There is a clear increase
in conductivity above 85 meV that is established within 100 fs. A
simple model was proposed whereby photoexcitation from bonding to
antibonding orbitals initiates coherent wavepacket motion that,
above threshold, overcomes correlations which stabilize the
insulating state. This is similar to the case of K-TCNQ discussed in
Sec.~\ref{subsec:Photoinduced phase transitions}.\footnote{In this
experiment on VO$_{2}$ the full conductivity of the metallic state
is not obtained and the role of intragap dynamics cannot be strictly
ruled out.}

Finally, we mention investigations of the photoinduced
metal-insulator transition using far-infrared pulses
\cite{hilton2007a}. In these experiments, a threshold fluence was
also observed with additional evidence of a softening of the
photoinduced transition with increasing temperature in the
insulating state. Interestingly, even at the highest excitation
fluence, the evolution towards metallicity required tens of
picoseconds. This is in contrast to other studies and also in
contrast with ultrafast electron-phonon thermalization which
typically occurs on a picosecond timescale. To interpret these
results, a model was developed based on Bruggeman's effective medium
theory where photoexcitation rapidly drives the temperature above
$T_{C}$ followed by dynamic growth and percolation of metallic
domains. An important aspect to consider in photoinduced phase
transitions is that photodoping and thermal effects must be
considered simultaneously. For VO$_{2}$, there are interesting
aspects which require further effort such as the potential role of
midgap states on the dynamics and the exact nature of the prompt
metallic state. For example, is it truly indicative of the metallic
phase in thermal equilibrium or is it more closely related to
enhanced effective mass state observed in mid-IR microscopy
experiments \cite{Qazilbash2007a}?

\subsection{Manganites}
\label{subsec:Manganites}
A strong resurgence in  manganite research occurred upon the
observation of negative magnetoresistance in lanthanum based
manganite thin films. This ``colossal" negative magnetoresistance
(CMR) was observed near the Curie temperature $T_{c}$ coinciding
with the transition from paramagnetic semiconductor to ferromagnetic
metal \cite{jin1994a}.\footnote{It was during the 1950's that mixed
valence manganites (Re$_{1-x}$D$_{x}$MnO$_{3}$, where Re is a rare
earth such as La or Nd and D is a divalent alkali such as Sr or Ca)
were first synthesized and extensively studied \cite{jonker1950a}. }
More recently, work has emphasized the diversity of phenomena in the
manganites including charge and orbital ordering, electronic phase
separation and quenched disorder, and investigations of other
families of manganites and related materials. This includes the
2-dimensional Ruddlesden-Poper phases and, to a lesser extent,
pyrochlores such as Tl$_{2}$Mn$_{2}$O$_{7}$ which also exhibit CMR
\cite{ramirez1997a,shimakawa1997a}. Numerous reviews and monographs
are available describing the properties and physics of manganites
\cite{salamon2001a,ramirez1997a,tokura2000a,dagotto2003a}.

\begin{figure} [ptb]
\begin{center}
\includegraphics[width=2.25in,keepaspectratio=true]%
{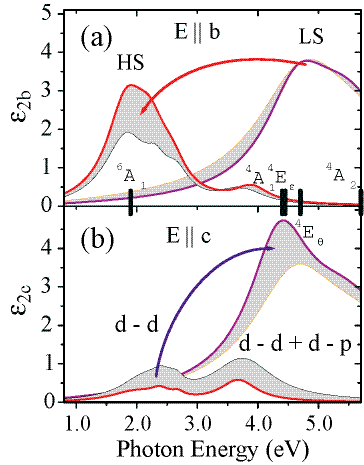}%
\caption{(Color online) Summary of the optical response of LaMnO$_{3}$ upon
crossing from the orbitally ordered state to the antiferromagnetic
state. (a) The imaginary part of the dielectric function
$\epsilon_{2b}$ along the b-axis. As the temperature decreases below
T$_{N}$, there is a transfer of spectral weight from the low-spin
(LS) to high-spin (HS) configuration which is consistent with what
is expected for d-d intersite transition. (b) The imaginary part of
the dielectric
function $\epsilon_{2c}$ along the c-axis. From \textcite{kovalevaPRL2004}.}.%
\label{averitt8}
\end{center}
\end{figure}

The parent compounds of manganites - as with the cuprates - are
antiferromagnetic Mott insulators. The canonical example is
LaMnO$_{3}$ with octahedrally coordinated Mn$^{3+}$ and a coherent
Jahn-Teller effect (i.e. orbital ordering) due to the occupation of
a single electron in the doubly degenerate e$_{g}^{1}$ level. There
is also a lower lying t$_{2g}^{3}$ level to which the e$_{g}$ levels
are slaved to through on-site ferromagnetic exchange coupling.
Divalent substitution leads to hole doping in the e$_{g}$ derived
band with transport described by double exchange between adjacent Mn
ions hybridized with O$_{2p}$ orbitals strongly influenced by
Jahn-Teller distortions \cite{MillisDE1995,MillisJT1996}. In
addition, the bandwidth of manganites can be sensitively controlled.
The Mn-O-Mn bond angle depends on the ionic sizes of the rare earth
and the dopants. As a function of decreasing ion radius the Mn-O-Mn
bond angle decreases with a gradual structural change from cubic to
rhomboheral to, eventually, orthorhombic.

The observed properties of manganites show a strong correlation to
this bond angle \cite{hwang1995PRL}. For example,
La$_{0.7}$Sr$_{0.3}$MnO$_{3}$ is classified as an intermediate
bandwidth material exhibiting a transition (T$_{c} \approx 260$~K)
from ferromagnetic metal to paramagnetic semiconductor. In
contrast, the narrow bandwidth manganites
Pr$_{0.6}$Ca$_{0.4}$MnO$_{3}$ does not exhibit metallic behavior
instead entering a charge ordered phase with decreasing
temperature though, as described below, it is very sensitive to
external perturbations. These properties manifest in dramatic
fashion in optical spectroscopy with significant redistribution of
spectral weight from the far-infrared through the visible.
Considerable insight into the electronic properties of manganites
has been obtained from optical conductivity
measurements.\footnote{\textcite{okimoto1997a,okimoto2000a,kim1998a,quijada1998a,lee2002a,jung2000a,kim2007a,cooper2001a,
Rusydi08,PhysRevB.60.R16263}.}

For example, LaMnO$_{3}$ is orbitally ordered below 780K with an
onset of antiferromagnetic ordering at T$_{N}$ = 140K where spins
ferromagnetically align in the ab plane with antiferromagnetic
ordering along the c axis. Optical conductivity measurements to
elucidate the character of the lowest lying transition
(approximately 2 eV) in LaMnO$_{3}$ must distinguish between on-site
d-d transitions (allowed through hybridization with O(2p) orbitals),
charge transfer (O(2p)-Mn(3d)), or d$_{i}$ $\rightarrow$ d$_{j}$
intersite transitions. Early studies on detwinned single crystals
revealed a strong anisotropy in the optical conductivity arising
from orbital ordering \cite{TobePRB2001}. A lack of spectral weight
transfer in the vicinity of T$_{N}$ was taken as an indication of
the charge transfer character of the transition as this would be
insensitive to spin ordering.

However, a recent experiment exhibits dramatic spectral weight
transfer upon antiferromagnetic ordering \cite{kovalevaPRL2004}. The
results are summarized in Fig. \ref{averitt8} which plots the
imaginary part of the dielectric response as a function of energy
above and below T$_{N}$. Fig. \ref{averitt8}(a) is with the electric
field polarized along the b-axis (E$\parallel$b) while Fig.
\ref{averitt8}(b) is for the electric field parallel to c
(E$\parallel$c). The strong anisotropy arises from orbital ordering,
as mentioned above. For E$\parallel$b spectral weight shifts from
higher energies to lower energies with decreasing temperature while
the converse is true for E$\parallel$c. Focusing on Fig.
\ref{averitt8}(a), the spectral weight transfer is from a low spin
(LS) state to a high spin (HS) state. The HS state should, in fact,
be favored in comparison to the LS state because of ferromagnetic
spin alignment in the a-b plane below T$_{N}$.\footnote{See
\textcite{kovalevaPRL2004} and references therein and also
\textcite{kimNJP2004} for a similar analysis.} That is, the increase
in spectral weight at 2.0 eV is consistent with an intersite
transition from a singly occupied e$_{g}$ orbital to an unoccupied
e$_{g}$ orbital on an adjacent site. The peak at 3.8 eV is assigned
to a t$_{2g}$-e$_{g}$ high spin transition while the 4.7 eV peak is
likely a charge transfer transition. These results are consistent
with LaMnO$_{3}$ as a Mott-Hubbard insulator and are a
representative example of how optical spectroscopy can discern the
character of multiorbital transitions as influenced by spin
correlations.

\begin{figure} [ptb]
\begin{center}
\includegraphics[width=2.5in,keepaspectratio=true]%
{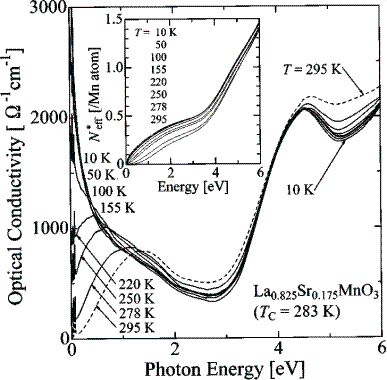}%
\caption{$\sigma_{1}$ as a function of photon energy for various
temperatures measured on a cleaved single crystal of
La$_{0.825}$Sr$_{0.175}$MnO$_{3}$. The polaron peak at approximately
1 eV gradually redshifts with decreasing temperature to a Drude
peak. The inset shows
the integrated spectral weight as a function of energy for various temperatures. From \textcite{Takenaka99}.}%
\label{averitt9}
\end{center}
\end{figure}

Hole doping of LaMnO$_{3}$ creates mobile carriers which (for
appropriate doping in intermediate bandwidth manganites) leads to
incoherent hopping of Jahn-Teller polarons in the paramagnetic phase
crossing over to coherent transport in the low temperature
ferromagnetic metallic state. This manifests in the optical
conductivity as shown in Fig. \ref{averitt9} for
La$_{0.825}$Sr$_{0.175}$MnO$_{3}$ (T$_{C}$ = 283 K). The optical
conductivity shows a redshift of an incoherent peak at approximately
1 eV at 293K to lower energies with the clear onset of a Drude
response below 155K \cite{Takenaka99}. These data were obtained from
reflection measurements on a cleaved single crystal yielding a
considerably larger Drude spectral weight extending to higher
energies than previously obtained on polished samples
\cite{okimoto2000a}. Fig. \ref{averitt9} shows that even on a
pristine cleaved crystal the incoherent response persists well into
ferromagnetic phase suggestive of residual polaronic effects which
may be strongly influenced by the orbital degrees of freedom.

Contrasting with this are narrower bandwidth manganites that are not
metallic at any temperature. For example with decreasing temperature
Pr$_{0.6}$Ca$_{0.4}$MnO$_{3}$ transitions from a paramagnetic
semiconductor to a charge ordered insulator (T$_{co}$ = 235K).
However, an applied magnetic field ``melts'' the charge order with a
Drude-like peak emerging in $\sigma_{1}(\omega)$ between 6 and 7
Tesla \cite{okimoto2000a}. This highlights the sensitivity of the
optical and electronic properties of manganites resulting from
nearly degenerate ground states with differing order parameters.

\begin{figure} [ptb]
\begin{center}
\includegraphics[width=3.0in,keepaspectratio=true]%
{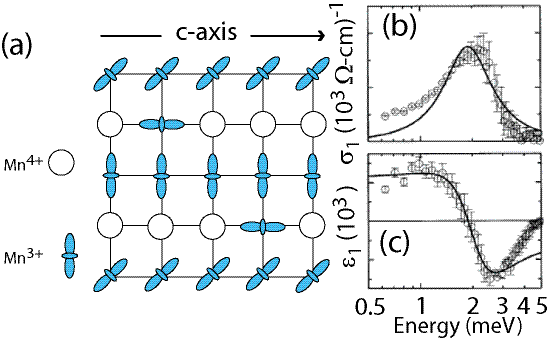}%
\caption{(Color online)
(a) Depiction of charge and orbitally ordered state. (b)
$\sigma_{1}(\omega)$ and (c) $\epsilon_{1}(\omega)$
measured on Pr$_{0.7}$C$_{0.3}$MnO$_{3}$ thin film at 4K. From \textcite{Kida02}.}%
\label{averitt10}
\end{center}
\end{figure}

In the charge ordered manganites a subgap collective mode excitation
of the charge density condensate has been reported
\textcite{Kida02}. Specifically, a pinned phason mode occurs at
terahertz frequencies. Figure \ref{averitt10} shows the first
observation of a collective mode of the charge ordered state
measured on epitaxial Pr$_{0.7}$Ca$_{0.3}$MnO$_{3}$ thin films. The
observed response persisted above the charge ordering temperature
which could arise from mixed phase behavior, though other effects
could not be ruled out. Quite recently, optical conductivity
measurements on commensurate charge-ordered manganites including
single crystal Nd$_{0.5}$Ca$_{0.5}$MnO$_{3}$ and polycrystalline
pellets of La$_{1-n/8}$Sr$_{n/8}$MnO$_{3}$ (n = 5, 6, 7) have also
shown a pinned phason response, that however, did not persist above
the charge ordering transition \cite{Nucara08}. An alternative
explanation assigns the mode to former acoustic phonons folded back
optically activated due to charge ordering \cite{Zhukova09,Zhang10}.
Future studies of collective mode excitations in the manganites as a
function of applied magnetic field or in the vicinity of a phase
transition could provide further insights into the nature of charge
localization and fluctuations as has been the case for two
dimensional metal organics (Sec.~\ref{subsec:Two-dimensional
molecular crystals}).

\textcite{PhysRevB.62.12354} performed optical conductivity
measurements on the layered manganites
La$_{2-2x}$Sr$_{1+2x}$Mn$_{2}$O$_{7}$. Similar to the cuprates, the
charge transport is highly anisotropic with metallic conductivity in
plane and activated conduction along the c-axis. The results of
optical conductivity measurements coupled with ARPES and scanning
tunneling microscopy suggest a picture of a metal with a highly
anisotropic bandstructure and very strong electron-phonon coupling
\cite{sun2006PRL,mannella2005a,ronnow2006a,Sun07}.

There have also been interesting time-resolved studies on the
manganites and related materials with a focus on probing the
quasiparticle dynamics within a given
phase.\footnote{\textcite{lobad2000a,lobad2001a,averitt2001d,prasankumar2005a,
ren2008a,talbayev2005a,ogasawara2003a,ogasawara2001a,kise2000a,
prasankumar2007a,matsubara2007a,Polli2007a,tomitomo2003a,mazurenko2008a,mcgill2004a}.}
Photoexcitation with a probe pulse results in a dynamic
redistribution of spectral weight whose subsequent temporal
evolution is monitored with a probe pulse. The timescales over
which this occurs provides information about which degrees of
freedom are involved in the dynamic spectral weight transfer
\cite{lobad2000a,lobad2001a}. In the perovskite manganites,
optical-pump terahertz-probe measurements in the ferromagnetic
metallic phase revealed a two-exponential decrease in the optical
conductivity \cite{averitt2001d}. A short $\sim$1 picosecond
response is associated with electron-phonon equilibration while
the longer ($>$10 ps) relaxation is due to spin-lattice
thermalization.

\begin{figure} [ptb]
\begin{center}
\includegraphics[width=2.8in,keepaspectratio=true]%
{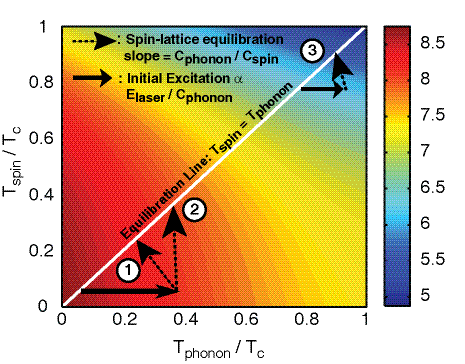}%
\caption{(Color online) Plot of the natural logarithm of the conductivity in the
phonon-spin temperature plane [the color bar denotes the magnitude
of ln$(\sigma)$]. The white diagonal line denotes the equilibrium
line ($T_{\rm spin} = T_{\rm phonon}$) of conventional
time-integrated measurements. Photoexcitation provides access to
the lower half of the plane. The dynamics can be dominated by
changes in the phonon temperature as is the case for the arrow one
or
dominated by changes in the spin temperature as indicated by arrow three. From \textcite{averitt2001d}.}%
\label{averitt11}
\end{center}
\end{figure}

It is possible to use time-resolved data to extrapolate the
conductivity in the $T_{S}-T_{L}$ plane where $T_{S}$ and $T_{L}$
are the spin and lattice temperatures, respectively. Results are
shown in Fig.~\ref{averitt11} as contours of constant
$\ln(\sigma_{1})$ in the $T_{S}-T_{L}$ plane \cite{averitt2001d}.
Conventional measurement techniques do not deviate from equilibrium
as indicated by the white diagonal line. However, optical-pump
terahertz-probe experiments, while starting from a point on the
equilibrium line, allow for access to the portion of the
$T_{S}-T_{L}$ plane below the diagonal equilibrium line since the
excited electrons couple preferentially to the phonons during the
initial 2 ps. Depending on the initial temperature the observed
conductivity decrease can depend predominantly on $T_{L}$ and/or
$T_{S}$. The ultrafast conductivity dynamics in LCMO and LSMO thin
films show that $\partial \sigma /\partial T$ is determined
primarily by phonons at low temperatures and by spin fluctuations at
higher temperatures. Other manganites including charge-order
materials have been investigated using similar techniques. These
studies are distinct from those described in
Sec.~\ref{subsec:Photoinduced phase transitions} on photoinduced
phase transitions in that the goal is to probe the quasiparticles
within a given phase as opposed to dynamically initiating a phase
change with photoexcitation.

\subsection{Ruthenates}
\label{subsec:Ruthenates}
\label{subsec:Layered Ruthenate Sr$_2$RuO$_4$}

Although Sr$_2$RuO$_4$ -- a member of the Ruddlesden-Popper series
-- possesses a crystal structure very similar to cuprates, the
electronic properties are distinctively different
\cite{Mackenzie03,Lichtenberg02}. The resistivity is strongly
anisotropic by about three orders of magnitude, but both in-plane
$\rho_{ab}$ and out-of-plane $\rho_c$ exhibit a $T^2$ dependence at
low temperatures consistent with the Fermi-liquid theory of metals
\cite{Hussey98}. This implies coherent conduction in all three
directions with an anisotropic effective mass, found in various
low-dimensional compounds
\cite{Ruzicka01,Dordevic01b,PhysRevB.65.134511}, like the example of
(TMTSF)$_2$PF$_6$ discussed in Sec.~\ref{subsec:One-dimensional
molecular crystals}. This behavior is opposite to that of the
cuprates where Fermi-liquid-like in-plane properties are accompanied
with incoherent $c$-axis transport.

\begin{figure}
\centering
\includegraphics[width=7cm]{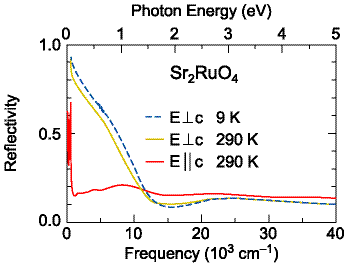}
\caption{\label{fig:Sr2RuO4a}(Color online) (Reflectivity
of Sr$_2$RuO$_4$ for $E\parallel c$ and $E\perp
c$ at different temperatures as indicated \cite{Katsufuji96}.}
\end{figure}

\begin{figure}
\centering
\includegraphics[width=6cm]{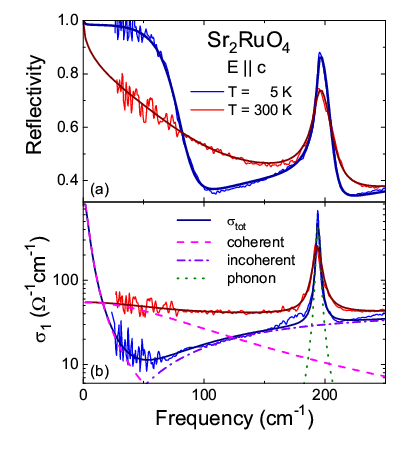}
\caption{\label{fig:Sr2RuO4b}(Color online)  The far-infrared
properties of Sr$_2$RuO$_4$ show the development of a plasma edge
around 70~\cm\ for $E\parallel c$ related to a Drude-contribution
of the low-temperature conductivity \cite{Pucher03}. The total
conductivity $\sigma_{\rm tot}(\omega)$ can be separated into a
coherent (dashed line) and incoherent (dashed-dotted line)
contribution.}
\end{figure}

Optical investigations by \textcite{Katsufuji96}  support this idea:
as seen in Fig.~\ref{fig:Sr2RuO4a}, for the in-plane polarization
there is a sharp increase in reflectivity below 15\,000~\cm\ and it
increases even more for low temperatures owing to a strong Drude
contribution. In the perpendicular direction ($E\parallel c$),
$\sigma_1(\omega)$ remains basically constant below 4000~\cm, except
for some phonons. A closer inspection of the low-frequencies
properties depicted in Fig.~\ref{fig:Sr2RuO4b}, however, reveals
that a narrow Drude-like contribution develops below 130~K
\cite{Katsufuji96,Hildebrand01,Pucher03}. The c-axis spectral weight
and charge carrier density responsible for the coherent transport
are only weakly temperature dependent, much lower than expected from
band structure calculations \cite{Singh95}. Slight doping by Ti
destroys the coherent transport \cite{Minakata01,Pucher03}.

From the extended Drude analysis a large mass enhancement of
$m^*/m_c \approx 30$ is found below 200~\cm\ leading to
$m_c/m_{ab}\approx 100$. The scattering rate $1/\tau_c(\omega)$ exhibits a
maximum around 100~\cm\ and is strongly suppressed below
\cite{Katsufuji96,Hildebrand01}.

The low transition temperature of up to 1.4~K makes the
superconducting state difficult to explore by optical experiments,
even when the gap is considerably larger than expected from mean
field theory \cite{Rao06}. Microwave experiments find a drop in
surface resistance and a small peak in the ac conductivity right
below $T_c$ \cite{Ormeno06,Gough04,Thoms04}; a finite quasiparticle
fraction is inferred with a temperature-independent relaxation rate.
Due its spin-triplet superconducting state with a $d=\Delta_0
\hat{z}(k_x\pm ik_y)$ symmetry, order-parameter collective modes
were predicted in Sr$_2$RuO$_4$ similar to the clapping mode in the
$A$ phase of superfluid $^3$He \cite{Higashitani00,Kee00}. They
should show up in the acoustic properties as well as in the
electromagnetic absorption at $\hbar\omega=\sqrt{2}\Delta$, i.e.\ in
the GHz range, but have so far defied experimental verification.

\subsection{Multiferroics}
\label{subsec:multiferroics}
Due to their fundamentally different behavior with respect to time-
and space reversal, no linear coupling can exist between static
polarization and magnetization. On the other hand, space and time
dependent polarization and magnetization {\em do} couple provided
that certain special conditions are met. Multiferroics, such as
TbMnO$_3$, Ni$_3$V$_2$O$_8$, MnWO$_4$ and CuO, are materials where
these special conditions are present, thus offering the prospect of
controlling charges by applied magnetic fields and spins by applied
voltages \cite{cheong07}. A crucial role in the phenomenological
description of magneto-electric coupling is played by the symmetry
of the crystal lattice, the symmetry of the unit cell, and of the
magnetic order \cite{kenzelmann05}. For example, a spiral spin state
\cite{sushkov08} induces, through the Dzyaloshinski-Moriya exchange
\cite{dzyaloshinski58,moriya60}, a polar lattice distortion and
accordingly a static electric polarization. In the ordered spin
state, one of the two magnons in the Hamiltonian is replaced by the
static modulation of spin density. Together, the symmetry breaking
caused by the static electric polarization and the spin-orbit
interaction, render magnons electric dipole active (see Sec.
\ref{subsec:Optical excitation of magnons}). Consequently, optical
phonons and single magnon waves of the same representation will mix.
Also, two-magnon and single magnons can be excited by the electric
field component of electromagnetic radiation \cite{katsura07}. This
is at the heart of the phenomenon of electromagnons, and it offers
interesting perspectives for the coupling of electric and magnetic
polarization in multiferroic materials \cite{cheong07}.

\begin{figure}
\centering
\includegraphics[width=\columnwidth]{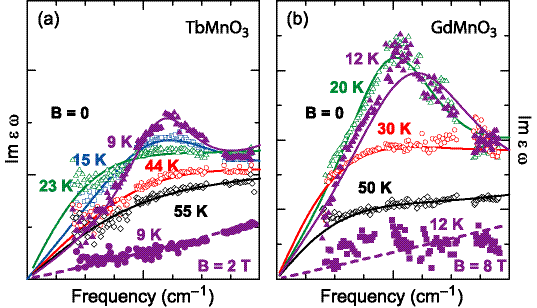}
\caption{\label{fig:electromagnons}(Color online)  Frequency dependence of the
imaginary parts of the terahertz-dielectric function in (a)
GdMnO$_3$ and  (b) TbMnO$_3$ with $e \parallel a$ and $b \parallel
c$. Open symbols represent experimental data in zero external
magnetic field and in the IC-AFM phase. Solid lines represent
model calculations adding an over-damped Lorentzian (dashed lines)
to the residual high-frequency contribution. Filled spheres
represent the data in the CA-AFM state obtained by applying
$B=2$~T (GdMnO$_3$) and $B=8$~T (TbMnO$_3$) along the c axis. The
corresponding zero-field data are shown by filled triangles
\cite{pimenov06}.}.
\end{figure}
Indeed strong ``electromagnon'' modes are observed in the infrared
transmission spectra of GdMnO$_3$, TbMnO$_3$,
Gd$_{0.7}$Tb$_{0.3}$MnO$_{3}$ and Eu$_{0.75}$Y$_{0.25}$MnO$_{3}$ at
approximately 20 cm$^{-1}$ and 60 cm$^{-1}$
\cite{pimenov06,pimenov06PRB,pimenov09,sushkov07,sushkov08,kida08,aguilar07,aguilar09}.
Recent observation of the coincidence of two antiferromagnetic resonance (AFMR) modes with
electromagnons at 18 and 26~cm$^{-1}$, illustrates the close
relationship of electromagnons to AFMR \cite{pimenov09}. Whereas a
single zone-boundary magnon seems the most plausible interpretation
of the 60 cm$^{-1}$ peak \cite{aguilar09}, the interpretation of the
25 cm$^{-1}$ peak as either a rotation mode of the spiral spin plane
\cite{katsura07} or a two-magnon process\cite{kida08} is still the
subject of discussions. \textcite{talbayev08} observed AFR at 43
cm$^{-1}$ in multiferroic hexagonal HoMnO$_3$, and demonstrated the
ferromagnetic nature of the rare-earth/Mn exchange. The
magneto-electric response in a multi-ferroic material enables
monitoring the oscillation of coherent magnons in the time domain
following femtosecond excitation: The magnetic precession modulates
the material's dielectric tensor, and this is seen as a modulation
of the intensity of a lightbeam reflected at the surface of the
sample \cite{talbayev08PRL}.

\subsection{Iridates}
\label{subsec:Iridates}
\begin{figure}[!bt]
\centering{
\includegraphics[width=0.75\linewidth]{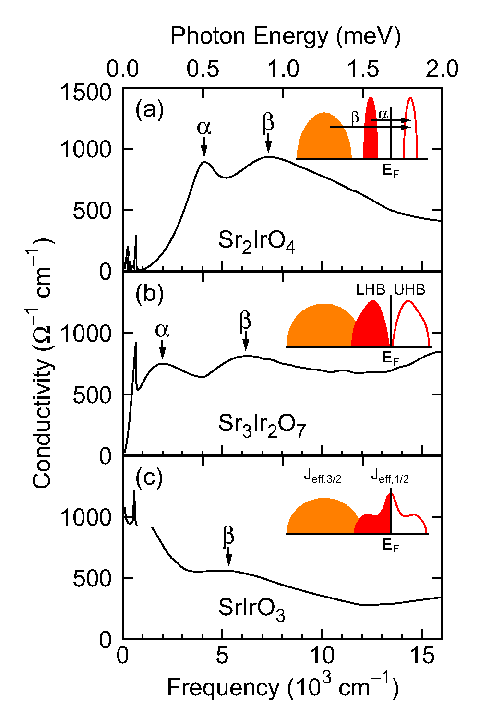}
}
\caption{(Color online)  Optical conductivity of the Ruddlesden-Popper series
Sr$_{n+1}$Ir$_n$O$_{3n+1}$ where $n=1$, $n=2$, and $n=\infty$.  The
insets sketch the $t2g$ density of states in the three
materials. From \onlinecite{Moon:2008}.  }
\label{IrO}
\end{figure}

\onlinecite{Moon:2008} studied a Ruddlesden-Popper series of Ir
oxides with chemical formula Sr$_{n+1}$Ir$_n$O$_{3n+1}$ where $n=1$,
$n=2$, and $n=\infty$. The optical conductivity of Sr$_2$IrO$_4$ and
Sr$_3$Ir$_2$O$_7$ single crystals, as well as epitaxially thin films
of SrIrO$_3$ grown on cubic MgO substrate, is reproduced in
Fig.~\ref{IrO}. The role of the substrate was to ensure the perovskite
phase of SrIrO$_3$, which is otherwise stable only at higher
pressure and temperature \cite{Longo:1971}.  To measure the
SrIrO$_3$ optical response, far infrared ellipsometry was combined
with transmittance and reflectance measurements to obtain accurate
results over an extended energy range.

Fig.~\ref{IrO}a shows that Sr$_2$IrO$_4$ ($n=1$) has an optical gap
of $\sim$0.1$\,$eV, Sr$_3$Ir$_2$O$_7$ ($n=2$) has a much smaller
gap, and SrIrO$_3$ ($n=\infty$) is a metal.  Hence there is a
metal-insulator transition in the Ruddlesden-Popper series for $n$
between $2<n<\infty$.

The optical conductivity in Fig.~\ref{IrO}a and b display a
pronounced two peak structure in both insulators, with peaks
$\alpha$ and $\beta$, which slightly decreasing with $n$. In the
metal, only the higher energy peak $\beta$ is identified.
\onlinecite{Moon:2008} interpreted these peaks as excitations across
the Hubbard bands, sketched in Fig.~\ref{IrO}b.

The band-structure LDA+U calculations for ferromagnetic
Sr$_2$IrO$_4$ \cite{Kim:2008} suggest that the $t_{2g}$ bands split
due to large spin-orbit coupling ($\xi\approx 0.4\,$eV) into two
sets of states, a set of bands with an effective angular momentum
$J_{\rm eff}=3/2$, and a band with an effective $J_{\rm eff}=1/2$.
The former states are lower in energy and thus completely filled,
while the $J_{\rm eff}=1/2$ band is half-filled (see
Fig.~\ref{IrO}b). Although the Hubbard $U$ in LDA+U calculation was
only $U=2\,$eV, it opened the gap in the half-filled $J_{\rm
eff}=1/2$ band, and split the Hubbard bands of $J_{\rm eff}=1/2$ for
roughly 0.5$\,$eV. \onlinecite{Kim:2008} suggested that the
excitations across these $J_{\rm eff}=1/2$ bands gives rise to the
peak $\alpha$ in optical conductivity, while the excitations from
$J_{\rm eff}=3/2$ into the unoccupied $J_{\rm eff}=1/2$ gives rise
to the peak marked $\beta$ in Fig.~\ref{IrO}a.

\onlinecite{Moon:2008} suggested that the bandwidth $W$ of the
Ruddlesden-Popper series increases with $n$. They argued that this
is a natural consequence of the connectivity $z$ of the Ir atom. The
conectivity is only $z=4$ in Sr$_2$IrO$_4$, but becomes $z=5$ and
$z=6$ in Sr$_3$Ir$_2$O$_7$ and SrIrO$_3$, respectively. Hence, using
the same small $U\sim 2\,$eV within LDA+U, \onlinecite{Moon:2008}
showed that the gap in Sr$_2$IrO$_4$ is indeed very tiny and that it
dissapears in SrIrO$_3$, in qualitative agreement with experiment.
Moreover, the optical conductivity data of \onlinecite{Moon:2008}
suggests that the metallic state of SrIrO$_3$ is very correlated
with heavy effective mass of the order of $m^*/m_b\sim 7$.  This
enhancement of the mass cannot be captured by LDA+U method, but it
is expected for a correlated metal.
%


\subsection{Oxide heterostructures}
\label{subsec:Oxide heterostructures}
\begin{figure}[htb]
\centering{
\includegraphics[width=0.7\linewidth]{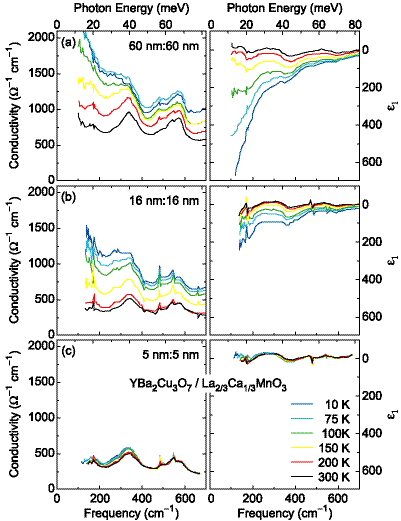}
}
\caption{(Color online)  Optical conductivity and dielectric constant of
superlattices
consisting of equal thickness of YBCO and LCMO layer. The thickness
of each layer in (a), (b) and (c) is 60$\,$nm, 16$\,$nm and 5$\,$nm,
respectively.  From \textcite{Holden:2004}.  }
\label{heteros}
\end{figure}

Artificial multilayers of complex transition metal oxides attract
much attention  as possible building blocks of novel and
useful functional materials.  In particular, heterostructures of
superconducting YBa$_2$Cu$_3$O$_7$ (YBCO) and ferromagnetic
La$_{3/2}$Ca$_{1/3}$MnO$_3$ (LCMO) offer a unique opportunity to
study the interplay between two antagonistic orders, ferromagnetism
and high temperature superconductivity.

The similar lattice constants of perovskite materials YBCO and LCMO allow
one to grow high quality superlattices of any layer
thickness, and with perfect atomically sharp interface. The pure
cuprate high-T$_C$ YBCO is superconducting below $T_c=90\,$K and is
metallic above $T_C$. Bulk LCMO exhibits colossal magnetoresistance and
is a ferromagnetic metal below $T_{\rm Curie}=245\,$K and a paramagnetic
insulator above $T_{\rm Curie}$.

Using spectral ellipsometry, \textcite{Holden:2004} measured the
effective dielectric function $\epsilon_1$ and effective optical
conductivity $\sigma_1$ of superlattices, which are equal to the
volume average of the superlattice components \cite{Aspnes:1982}.
Fig.~\ref{heteros} shows $\sigma_1$ and $\epsilon_1$ for several
YBCO/LCMO superlattices with thickness ratio
$d_{\textrm{YBCO}}:d_{\textrm{LCMO}}$ of 60nm:60nm, 16nm:16nm, and
5nm:5nm.
Given the metallic properties of the pure YBCO and LCMO in the
extended range of temperatures, one would expect that the
superlattice also should exhibit a strong metallic response. Instead
Fig.~\ref{heteros}a-c highlights that the YBCO/LCMO superlattice
exhibits a drastic decrease of carrier concentration and their
mobility.
For the fairly thick superlattice of 60nm:60nm the decrease of
metallicity is not yet very pronounced, and one can even observe
signatures of superconducting gapping in 10$\,$K data of
Fig.~\ref{heteros}a.

The metallic response is strongly suppressed in other superlattices
with thinner layers. The absolute value of $\sigma_1$ and
$\epsilon_1$ is drastically reduced in 5nm:5nm superlattice at all
temperatures (see Fig.~\ref{heteros}c). The Curie temperature and
superconducting temperature of this superlattice are approximately
120$\,$K and 60$\,$K, respectively. Hence insulating-like behavior
below 120$\,$K is completely unexpected and very surprising.

To shed some light on the origin of such strong suppression of the
free carrier response, the optical
conductivity was calculated
for a model of a superlattice consisting of 16$\,$nm of
LCMO and 16$\,$nm of a material with a small Drude-like response
with $\omega_p^2=2\times 10^6$~cm$^{-2}$. The idea
for such a ``fit layer" comes from the results on 5nm:5nm
superlattice, which gives roughly the Drude weight of comparable
magnitude $\omega_p^2=2\times 10^6$~cm$^{-2}$.
The agreement between the measurements and such a model calculation
was found to be very good \cite{Holden:2004}, which might suggest that
the LCMO response does not change much with the layer thickness, while
superconducting YBCO becomes almost insulating in thin layers.
Note that the superconducting $T_C$ remains relatively high around
60$\,$K.
A model for a
superlattice with 16$\,$nm of YBCO and 16$\,$nm of a Drude-like
layer with arbitrary $\omega_p$ does not give satisfactory agreement
with measurements.

\textcite{Holden:2004} also measured the conductivity of
superlattices consisting of YBCO and material $X$, where $X$ was
insulating PrBa$_2$Cu$_3$O$_7$, paramagnetic metal LaNiO$_3$ and
ferromagnetic metal SrRuO$_3$. Ferromagnetic SrRuO$_3$ shows a
similar suppression of charge carriers as LCMO, while this effect is
absent for paramagnetic metal LaNiO$_3$ or insulator
PrBa$_2$Cu$_3$O$_7$. Hence the competition of ferromagnetism and
superconductivity is likely responsible for the measured suppression
of the conductivity. It should be noted that superlattices of
YBCO/LCMO with YBCO layer much thicker than LCMO, such as
$60\textrm{nm}:15\textrm{nm}$, $30\textrm{nm}:15\textrm{nm}$, and
$8\textrm{nm}:3\textrm{nm}$, do not show a strong suppression of
conductivity, hence only comparable thickness of the two layers
gives the intriguing effect of suppressed conductivity
\cite{Holden:2004}. Similar suppression of the superconducting
condensate density in the superlattices of comparable thickness was
observed in terahertz measurements by \textcite{Chen:2004}.

Understanding of the electronic states at the interface between
a high-Tc material and a ferromagnet is far from complete, hence the
origin of the conductivity suppression is at present unknown.
However, there are various proposals for the origin of these
phenomena. In the most straightforward interpretation, there is a
massive transfer of holes from YBCO to LCMO, such that YBCO becomes
underdoped and LCMO is driven into a charge ordered state, similar
to the one observed for Ca content of $x>0.45$
\cite{Kancharla:2007,Holden:2004}. However, metallic LaNiO$_3$ does
not give rise to this effect, while ferromagnetic SrRuO$_3$ does.
Thus, the magnetic proximity effect might play a major role in
affecting the YBCO superconductor \cite{Buzdin:2005}.

Indeed there are some indications of very interesting interfacial
phenomena in YBCO/LCMO superlattices. X-ray spectroscopy and neutron
measurements by \textcite{Chakhalian-Nature:2006} showed that Cu
atoms in the first Cu-O layer of YBCO acquire a ferromagnetic
polarization, likely due to canting of Cu magnetic moments. This is
due to coupling between the Cu-O layer and Mn-O layer at the
interface. The LCMO ferromagnetic layer at the interface has a
somewhat suppressed ferromagnetic moment, and is coupled
antiferromagnetically to the net Cu-polarized moment
\cite{Stahn:2005}.  Resonant x-ray spectroscopy by
\textcite{Chakhalian-Science:2007} furthermore suggests a major
change in orbital occupation of the electronic states on the Cu-atom
at the interface. The $3d_{3z^2-r^2}$ Cu-orbital is almost fully
occupied and inactive in bulk YBCO, but becomes partially occupied
at the interface.

There are indications of a strong modification of the ferromagnetic
LCMO layers, obtained by neutron spectroscopy experiments
\cite{Hoppler:2009}. It was suggested that every second LCMO layer
might loose as much as 90\% of the magnetic moment, while the
remaining half of the LCMO layers might have strongly enhanced
magnetic moments, such that the average magnetization remains
unchanged.
While these unusual interfacial effects do not directly explain the
origin of the strong suppression of the conductivity, they show that
the interface physics and proximity effects might be far more
complicated than previously thought, and might be relevant for
correct interpretation of the measured conductivity suppression.

Numerous other oxide heterostructures were synthesized recently, such
as LaTiO$_3$/SrTiO$_3$ \cite{Ohtomo:2002}, CaRuO$_3$/CaMnO$_3$
\cite{Takahashi:2001}, and La$_2$CuO$_4$/La$_{1.55}$Sr$_{0.45}$CuO$_4$
\cite{Gozar:2008}. For a recent review see \textcite{Ahn:2006}. Very
recently, far infrared spectral ellipsometry was applied to
superlattice of correlated paramagnetic metal CaRuO$_3$ and the
antiferromagnetic insulator CaMnO$_3$ \cite{Freeland:2009}. It was
found that the ferromagnetic polarization is due to canted Mn spins in
CaMnO$_3$ penetrated unexpectedly deep into the CaMnO$_3$ layer (3-4
unit cells).

\section{Intermetallic Compounds and Diluted Magnetic Semiconductors}
\label{sec:Intermetallic Compounds and Diluted Magnetic Semiconductors}

\label{subsec:Intermetallic Compounds}
\label{subsec:Intermetallic-Compounds}
The properties of intermetallic compounds containing elements with
$f$ electrons (like U, Ce or Yb) are governed by the competition
between Kondo and Ruderman-Kittel-Kasuya-Yosida (RKKY)
interactions \cite{Hewson93,Doniach77,Fulde06}. While magnetic
ground states occur when  RKKY ordering dominates, Kondo
interaction links localized $f$- elec\-trons and conduction
electrons. Their hybridization severely influences the density of
states. The Anderson model contains the essential physics of $d$-
and $f$- states in heavy fermions, but is also crucial for the
understanding of magnetic impurities in simpler systems like
(Ga,Mn)As.
After early reviews by \textcite{Millis91} and
\textcite{Wachter94}, the electrodynamic properties have been
extensively discussed by \textcite{Degiorgi99}; thus we confine
ourselves to some recent developments.

\subsection{Heavy-fermion metals}
The heavy-fermion phenomenon exists in a number of lanthanide and
actinide compounds\footnote{\textcite{PhysRevLett.78.3729} also
discovered heavy fermion characteristics in the $d$-electron system
LiV$_2$O$_4$. Optical experiments reveal that it behaves more like a
``bad metal'' close to a correlation-driven insulating state and
that the spectral weight is transferred over an extremely wide
energy range \cite{jonsson:167402}. Hydrostatic pressure suppresses
the Drude response and a charge-order insulator develops
\cite{Irizawa09}.} and manifests in the apparent existence of
quasiparticles with very large effective mass $m^*$ below some
characteristic temperature $T^*$.
These materials have partially filled $f$ orbitals, which hybridize
with lighter and more spatially extended $s$, $p$, and $d$ orbitals.
At low temperature, the electrons can either form a \textit{heavy}
Fermi-liquid state, a composite of $f$-electron spins and
conduction electron charges. Alternatively, the $f$-electrons can
magnetically order, or the heavy quasiparticles can superconduct
[for reviews see
\cite{RevModPhys.73.797,RevModPhys.78.743}].

In a simple picture these correlations reduce the scattering rate
$1/\tau$ and spectral weight
\cite{Varma85a,Varma85b,Millis87a,Millis87b,Coleman87}. In other
words, the Drude response becomes extremely narrow. As pointed out
by \textcite{Scheffler05,Scheffler06,Scheffler09,Scheffler10}, the Fermi
velocity $v_F$ is  small (cf. Sec.~\ref{sec:Drude},
Fig.~\ref{fig:Drude}).

In addition to the narrow Drude-like response, a mid-infrared
absorption peak around $\omega_{\rm mir}$ is commonly observed that
scales as $\omega_{\rm mir}\propto\sqrt{T^*}$
\cite{Garner00,Degiorgi01,Hancock04,Hancock06,Okamura07a}. From
optical and magneto-optical experiments on numerous non-magnetic
systems, \textcite{Dordevic01,Dordevic06b} could confirm the scaling
relation $m^*/m_b = \left(\Delta/k_BT^*\right)^2$ between the
magnitude of $m^*$ and the hybridization gap $\Delta$.

\begin{figure}
\centering
\includegraphics[width=5.8cm]{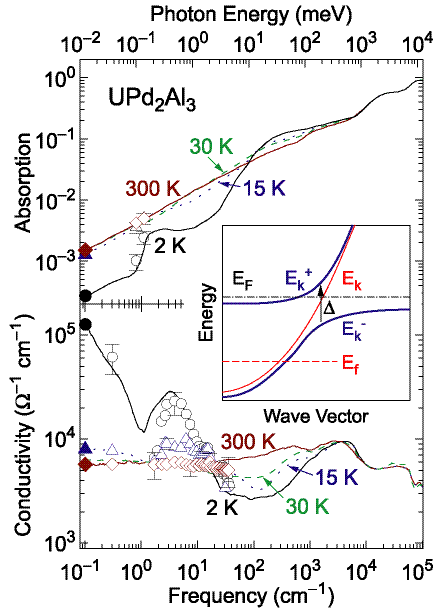}
\caption{\label{fig:UPdAl}(Color online) Frequency dependent
absorptivity $A(\omega)=1-R(\omega)$ and optical conductivity of
UPd$_2$Al$_3$ at different temperatures  shown over a wide frequency range.
The solid symbols on the left axis are from dc measurements; the
open symbols in the microwave range are obtained by cavity
perturbation technique. In the THz range the optical conductivity
is directly determined using the transmission and the phase shift
measured by the Mach-Zehnder interferometer. The lines are
from various optical investigations
[transmission through films and reflection of bulk samples
\cite{Degiorgi94b}] and simultaneously match the directly
measured conductivity and dielectric constant
\cite{Dressel02a,Dressel02b}. Inset: Renormalized band structure
calculated from the Anderson lattice Hamiltonian. $E_k$ and $E_f$
denote bands of free carriers and localized $f$-electrons. At low
temperatures, a direct gap $\Delta$ opens. The Fermi level $E_F$
is near the bottom of the upper band $E^+_k$, resulting in
enhanced effective mass of the quasiparticles.}
\end{figure}

Extending previous experiments \cite{Bonn88c,Bommeli97} to lower
frequencies and temperatures, \textcite{Holden03} observed
coherent transport in UBe$_{13}$ with an abrupt decrease in
scattering rate and a strong increase of the effective mass. The
strongest enhancement of the effective mass is observed in
heavy-fermion systems with a magnetically ordered ground state.
UPd$_2$Al$_3$ is a prime example for which
\textcite{Dressel00,Dressel02a,Dressel02b} measured the
low-tem\-per\-ature optical properties in a wide frequency range.
As seen in Fig.~\ref{fig:UPdAl}, below $T^*\approx 50$~K, the
hybridization gap opens around 10~meV. As the temperature
decreases further ($T\leq 20$~K), a well pronounced pseudogap of
approximately $0.2$~meV develops in the optical response that may
be related to the antiferromagnetic ordering, $T_N\approx 14$~K.
Similar observations are reported for UPt$_3$
\cite{Donovan97,Tran02}, UNi$_2$Al$_3$ \cite{Scheffler10} and
URu$_2$Si$_2$ \cite{Morales09}.

The heavy-fermion compound CeCoIn$_5$ with the highest superconducting transition, $T_c=2.3$~K, is subject to intense optical investigations \cite{Singley:2002,Mena:2005,Burch:2007,Sudhakar09} without giving insight into the superconducting state yet.

\begin{figure}
\centering
\includegraphics[width=5cm]{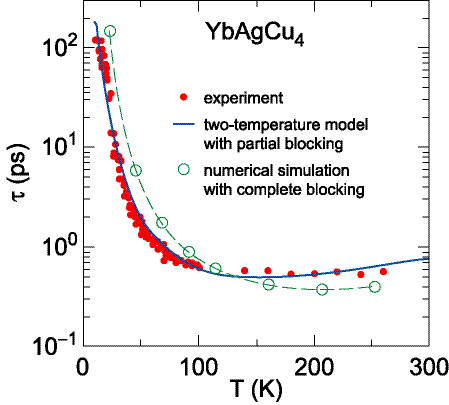}
\caption{\label{fig:YbAgCu}(Color online)
Temperature dependence of the relaxation time on
YbAgCu4. Numerical simulations have been performed
assuming suppressed scattering of heavy electrons
by phonons; the solid line corresponds to simulations
by a two-temperature model with a partial reduction of
the scattering. An equally good description is obtained
by coupled Boltzmann equations when the electron-phonon
coupling is suppressed for electronic states within the
Abrikosov-Suhl peak in the density of states \cite{Demsar03,Ahn04}. }
\end{figure}

Time-resolved optical investigations on YbAgCu$_4$ reveal that the
electron-phonon thermalization increases below $T^*$ by more than
two orders of magnitude (Fig.~\ref{fig:YbAgCu}) because the heavy
quasiparticles acquire a large specific heat and their scattering on
phonons is suppressed \cite{Demsar03}. In conventional metals
\cite{Groeneveld95,Hase05,Hase05e} the relaxation time can be well
described by coupling the electrons to the lattice bath
(two-temperature model), but it fails at low temperatures when
electron-electron thermalization becomes the limiting factor. In the
case of heavy fermions, the latter process is much faster; however
the scattering of the heavy-electrons on phonons is suppressed due
to the extremely slow Fermi velocity $v_F$ compared to the sound
velocity. An alternative scenario is based on the bottleneck idea of
\textcite{Rothwarf67} (cf.\ Sec.~\ref{subsec:BCS Superconductors}):
the recombination of photoexcited quasiparticle across the
hybridization gap is the limiting factor in heavy fermion metals as
well as in Kondo insulators or spin-density wave systems
\cite{Demsar06a,Demsar06b,chia2006a}.

\subsection{Kondo insulators}
\label{subsec:Kondo insulators}
\label{subsec:Kondo-insulators} In a few cases the hybridization of
conduction electrons and $f$-electrons leads to semiconducting
characteristics  with a small energy gap (of the order of 10 meV)
and a van Vleck-like susceptibility at low temperatures
\cite{Aeppli92,Wachter94,Degiorgi99,Riseborough00}. A canonical
example of a Kondo insulator is Ce$_3$Bi$_4$Pt$_4$, for which the
gap opening below 100~K is shown in Fig.~\ref{fig:IMTs}(d). The
depleted spectral weight grows linearly with temperature and is
displaced to energies much larger than the gap.

\begin{figure}
\centering
\includegraphics[width=6cm]{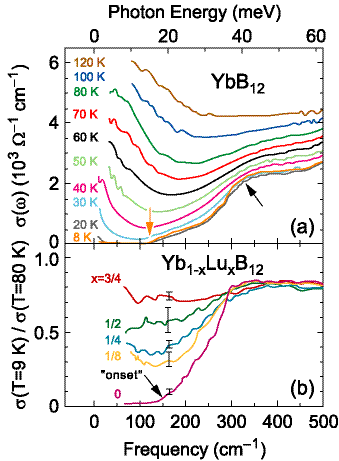}
\caption{\label{fig:Okamura}(Color online) (a) Optical
conductivity of YbB$_{12}$ at different temperatures. The black and orange
arrows indicate the sholder and the conductivity onset.
(b) Substitution of Yb by Lu decreases correlations;
the corresponding energy gap does not shift but disappears  \cite{Okamura00,Okamura05}. }
\end{figure}

Numerous efforts have been undertaken to elucidate the nature of
the gap and possible states inside \cite{Riseborough03}.
Experiments by \textcite{Okamura00,Okamura04,Okamura05,Okamura98}
shed light on the optical properties of the ytterbium compounds.
The infrared conductivity of YbB$_{12}$ and YbAl$_{3}$ is governed
by a broad peak centered at 2000~\cm\ related to direct
transitions between the Yb $4f$-derived narrow band and the broad
conduction band. The behavior can be reproduced by calculations
based on the electronic band structure
\cite{Antonov02a,Antonov02b,Saso04}. As depicted in
Fig.~\ref{fig:Okamura}(a) for the case of YbB$_{12}$, below
$T=80$~K an energy gap opens around 320~\cm. In YbAl$_{3}$ a
similar behavior is found at 500~\cm, although the compound
remains metallic. It is argued that these excitations are indirect
transitions within the hybridization state. The correlated nature
of the low-temperature state is reflected in the mass enhancement
$m^*/m_b\approx 12$ and $1/\tau \propto\omega^2$ dependence
of YbB$_{12}$ obtained from a generalized Drude analysis
\cite{Gorshunov06,Gorshunov06a}. For $T\leq 10$~K the conductivity
onset is 15~meV, which agrees with the renormalized-band model
\cite{Hewson93}. Resistivity, Hall effect, photoemission and
specific heat measurements yield comparable values of the gap
\cite{Iga99,Takeda04}. Substituting Yb$^{3+}$ by
nonmagnetic Lu$^{3+}$ lowers the electronic correlations: similar
to the temperature increase, the gap gradually fills in
[Fig.~\ref{fig:Okamura}(b)] without shifting the shoulder at
300~\cm. The mid-infrared peak, however, moves to lower energies
in Yb$_{1-x}$Lu$_x$B$_{12}$ with $x$ increasing until it is lost
for $x\geq 0.75$ because the coherence among the Kondo singlets
vanishes \cite{Okamura00}; the same effect is reached by rising
temperature.

\textcite{Gorshunov99} and \textcite{Dressel99} extended previous
optical experiments on the intermediate-valence compound SmB$_6$
by \textcite{Nanba93} and \textcite{Travaglini84} to extremely low
frequencies ($\nu>1$~\cm). Within the 19 meV energy gap in the
density of states they found an additional narrow donor-type band
only 3 meV below the bottom of the upper conduction band, seen as
an absorption peak at 24~\cm. At $T < 5$~K, only the charge
carriers within the narrow band contribute to the ac conductivity.
Correlation effects cause a sizeable effective-mass enhancement
$m^*\approx 30 m_b$ which is discussed in the frame of a specific
exciton-polaron formation at low temperatures when moving carriers
get self-trapped \cite{Kikoin90,Curnoe00,Sluchanko00}. However,
field and pressure-dependent NMR and transport experiments
\cite{Caldwell07,Derr08} evidence the intrinsic nature and
magnetic origin of the in-gap states. The observations are
explained by antiferromagnetic correlations based on a localized
Kondo model \cite{Kasuya96,Riseborough00,Riseborough03}.

\textcite{Matsunami09} succeeded  to tune the Kondo insulator to a
heavy-fermion metal by apply high pressure. In the well studied
insulator \cite{Dordevic01,Degiorgi01} the gap between the $f$
band and conduction electrons closes above 8~GPa and metallic
behavior with heavy carriers is observed.

While Yb$_4$As$_3$, TmSe, or FeSi are often subsumed under Kondo
insulators, the viewpoint has shifted in recent years.
\textcite{Fulde95} pointed out that the unusual properties of
Yb$_4$As$_3$ can be explained by charge ordering of the Yb ions
which at low temperatures are self-doped leading to strong
electronic correlations of the $4f$-holes in charge-ordered Yb
chains. In accord with broadband reflection measurements by
\textcite{Kimura96,Kimura97,Kimura02}, magnetooptical investigations
on Sb-substituted crystals yield  an energy difference of 0.42 eV
between the occupied and empty $4f^{14}$ state \cite{Pittini98}. In
TmSe, $\sigma(\omega)$ reveals a gap-like feature below
$100$~cm$^{-1}$  for $5~{\rm K}<T<50$~K, which is accounted for as a
mobility gap due to localization of $d$ electrons on local Kondo
singlets rather than a hybridization gap in the density of states
\cite{Dumm05,Gorshunov05}.
As discussed in Sec.~\ref{subsec:Transition metal silicides},
extensive investigations of the narrow-gap semiconductor FeSi and
related compounds, like Fe$_{1-x}$Co$_x$Si , MnSi and FeGe, consider
FeSi as an itinerant semiconductor whose properties can be explained
without a local Kondo-like interaction.\footnote{Previous
investigations \cite{schlesinger93,Degiorgi94,Paschen97} were
hampered by sample quality;
 but also the point of view has changed over the years \cite{Damascelli97,Marel98,Mena03,Mena06,Guritanu07,Zur07,Klein08}.}

\subsection{Beyond the Anderson model}
\label{subsec:DMFT prospective on Heavy Fermion behavior}
\begin{figure}[hbt]
\centering{
\includegraphics[width=1.00\linewidth]{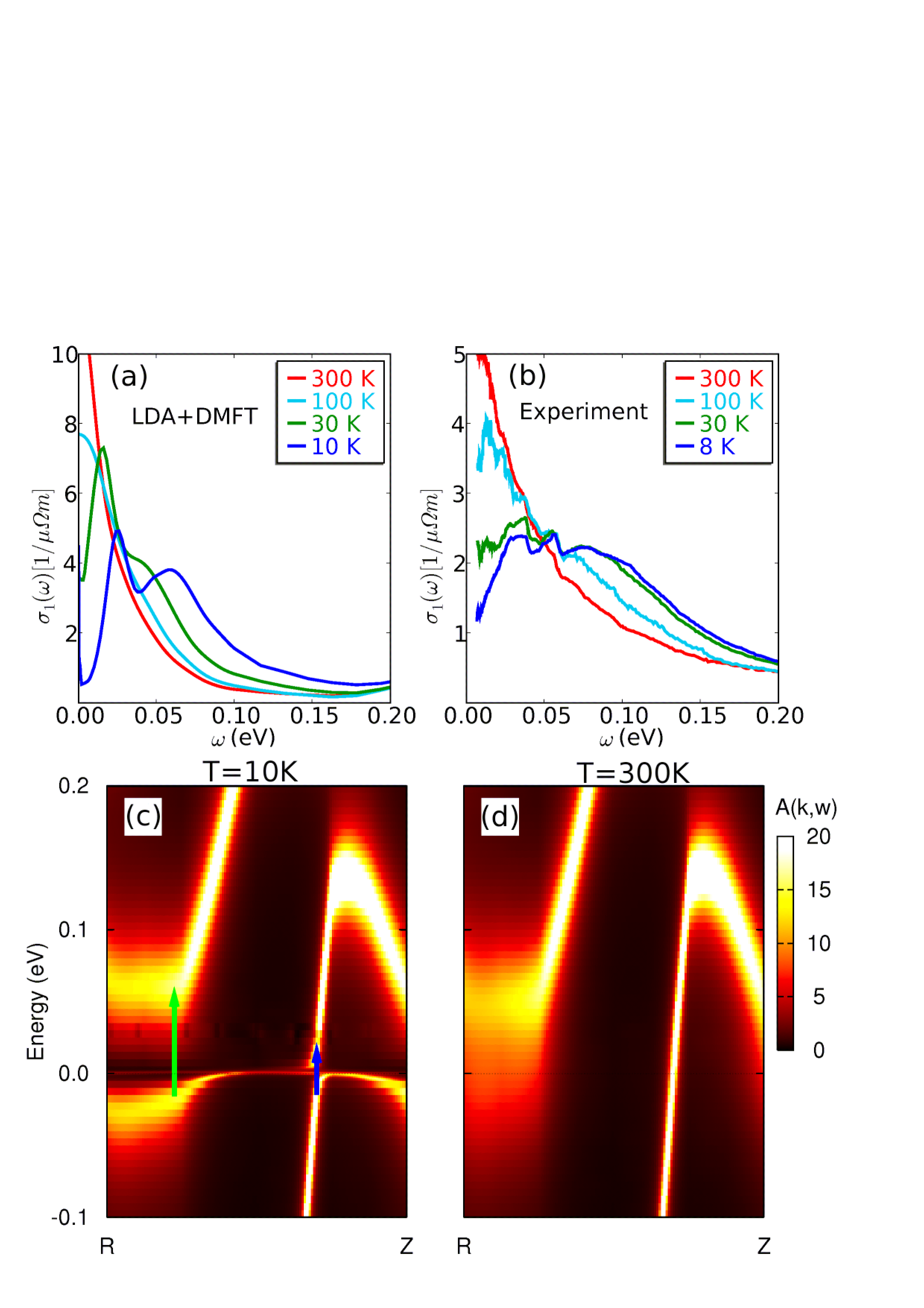}
}
\caption{(Color online)
The development of the hybridization gap in the heavy fermion
compound CeIrIn$_5$.
(a) The optical conductivity calculated by LDA+DMFT method (from
\onlinecite{Shim:2007}). (b) Experimental optical conductivity for
the same compound measured by \onlinecite{Mena:2005}. (c) Theoretical
momentum resolved non-$f$ spectral function at low temperature. The
two types of hybridization gaps are marked by arrows in panel (c).
(d) same as (c) but for higher temperature.  }
\label{heavy115}
\end{figure}

The Anderson lattice model (ALM) adequately describes the gross
features of the heavy electron physics (\ref{subsec:Intermetallic
Compounds}) and Kondo insulating behavior (\ref{subsec:Kondo
insulators}). Specifically, the model accounts for the fingerprints
of electrodynamics of a heavy fermion metal including massive
quasiparticles, the hybridization gap \cite{Degiorgi99} as well as
the scaling of these observables with the coherence temperature
$T^*$. DMFT results \cite{Grenzebach:2006} reproduce all of the
qualitative features of the complex conductivity characteristic to a
typical heavy fermion system. Nevertheless, optical data for a class
of Ce-based heavy fermions with chemical formula Ce$X$In$_5$
($X$=Co,Rh,Ir)\cite{Singley:2002,Mena:2005,Burch:2007} reveal
signifiicant departures from these established trends. Spectra
displayed in Fig.~\ref{heavy115}b do not reveal a clear
hybridization gap. Instead a continuum of states extends down to
lowest energies; this later behavior is most clearly seen in data by
\onlinecite{Singley:2002}. \onlinecite{Burch:2007} interpreted this
unusual response in terms of the distribution of the energy gaps and
conjectured that the strength of the hybridization may be
momentum-dependent.

Microscopically, a momentum dependence in the hybridization is
hardly surprising given the fact that the local-moment orbitals are
of $f$-type and may hybridize with several conduction-electron
orbitals. \onlinecite{weber:125118} considered heavy fermion metals
with hybridization nodes and concluded that the low-temperature
specific heat of these type of systems is dominated by heavy
quasiparticles whereas the electrical conductivity at intermediate
temperatures is carried by unhybridized light electrons.
Calculations of the optical conductivity carried out by
\cite{weber:125118} confirm smearing of the hybridization gap
feature.  The LDA+DMFT calculations by \onlinecite{Shim:2007},
reproduced in Fig.~\ref{heavy115}a, revealed that the ``in-gap"
states are related to excitations across a second hybridization gap.
Namely, at low temperature, the Ce local moment is strongly coupled
to electrons on neighboring In atoms. The coupling is strong with
the out-of-plane In, and weaker with the in-plane In, which results
in variation of the hybridization gap in momentum space (see
Fig.~\ref{heavy115}c). The larger (smaller) hybridization gap gives
rise to a peak at higher (lower) frequency of $\sim 0.075\,$eV
($\sim 0.03\,$eV). At higher temperature, the electronic states
which are strongly coupled to Ce moments (and result in large
hybridization gap at low temperature) become highly scattered and
acquire large broadening of the bands (see Fig.~\ref{heavy115}d), a
signature of the local moment regime.

\subsection{Magnetic semiconductors}
\label{subsec:Magnetic semiconductors}
In this section we will overview optical properties of several
different classes of ferromagnetic semiconductors including
III-Mn-As, EuB$_6$ and transition metal silicides. A common
denominator between these systems is low carrier density $n$ and
plasma frequency $\omega_p$. Furthermore, in all of these systems
the formation of magnetic order is associated with significant
changes of the plasma frequency and low-energy conductivity. In both
borides and III-Mn-As the plasma frequency increases below the Curie
temperature $T_C$ revealing scaling with the magnetization. In
silicides the dominant contribution to the  transformation of
optical properties below $T_C$ is due to magnetic disorder leading
to suppressed metalicity in the ferromagnetic state.

\subsubsection{III-Mn-As}
\label{subsec:III-Mn-As}
The discovery of ferromagnetism in III-V hosts heavily doped with Mn
has propelled research in this class of materials \cite{ohno-1996}.
From the applications point of view, ferromagnetic (FM)
semiconductors are appealing because magnetic, electronic and
optical effects in these systems are intimately entangled. These
properties, combined with Curie temperatures as high as $T_{\rm
Curie} =170~K$, may enable new device functionalities
\cite{zutic-rmp,awsch-flatt-np}. The fundamental physics of FM
semiconductors is equally exciting. The detailed understanding of
complex behavior of III-Mn-V ferromagnetic semiconductors relies on
resolving the roles played by electron-electron interaction and
disorder in the previously unexplored regime of exceptionally high
concentration of magnetic dopants. Magnetic impurities radically
modify the IMT in this class of materials
\cite{jungwirth-rmp,burch2008a}. The most studied system,
Ga$_{1-x}$Mn$_{x}$As, undergoes the IMT in the impurity band that
survives well on the metallic side of the transition
\cite{burch2008a}. On general grounds, one expects strong electronic
correlations to be an essential element of physics of a system where
transport phenomena are dominated by states in the impurity band
\cite{jungwirth-rmp}.

Ga$_{1-x}$Mn$_{x}$As is commonly referred to as a ``prototypical''
ferromagnetic semiconductor. An Mn ion in a GaAs host has a
half-filled $d$-shell and acts as a $S=5/2$ local moment. The spin
degeneracy of Mn$_{Ga}$ acceptor is lifted due to the large on-site
Coulomb repulsion $U=$3.5 eV. Itinerant carriers produced by Mn
substituting Ga are locally magnetically coupled to the Mn spins via
an exchange coupling. The exchange between the Mn local moments and
the carriers they produce, plays a key role in the physics of
III-Mn-V diluted magnetic semiconductors and is responsible for
mediating ferromagnetism. Another important aspect of the exchange
coupling is its tendency to localize the holes around the Mn. The
effect of disorder tends to be stronger in magnetic semiconductors
compared to nommagnetic counterparts
\cite{Ohno-dietl2008,timm-review}. Mn doping introduces the
insulator-to-metal transition near $x=1-2$\%. The IMT concentration
depends on the presence of compensating donors (As antisites and
interstitially doped Mn) and disorder.

\begin{figure}
\centering
\includegraphics[width=3.0in]{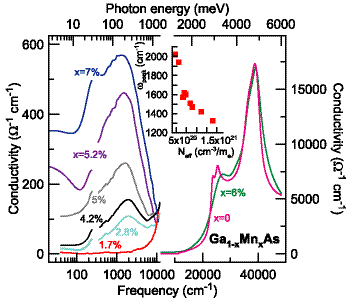}
\caption {(Color online) The optical conductivity of Ga$_{1-x}$Mn$_{x}$As
in the energy region below the band gap of the GaAs host (left
panel) and above the gap (right panels). Data reveals systematic
enhancement of the intra-gap conductivity with Mn doping and only
weak modification of the van Hove singularities due to interband
transitions. Data from \textcite{burch-annealed,singley-prl02}.
Inset: the peak position of the MIR resonance plotted versus the
spectral weight below 6450 cm$^{-1}$, which is proportional to the
number of holes added through the doping process as described in
details in \cite{burch-annealed}. A prominent redshift of the
resonance is apparent. All infrared data (left panel and the inset)
at $T=8$~K. Ellipsometry data in the right panel are at room temperature.}
\label{fig:GaMnAs}
\end{figure}

Infrared and optical properties of Ga$_{1-x}$Mn$_{x}$As were a
subject of detailed experimental
\cite{nagai-JJAP,singley-prl02,singley-prb03,burch-digital,burch-ellips,seo-noh,burch-annealed}
and theoretical \cite{sinova-02,Moca09} investigations.
Fig.~\ref{fig:GaMnAs} displays the evolution of the electromagnetic
response of Ga$_{1-x}$Mn$_{x}$As in the process of doping. A brief
inspection of these data shows a dramatic change of the optical
conductivity upon doping in  the frequency range within the band gap
of the GaAs host. Ferromagnetic films of Ga$_{1-x}$Mn$_{x}$As reveal
two new features in the intra-gap conductivity. The first is a broad
resonance initially centered at approximately 2000 cm$^{-1}$, whose
center energy redshifts with doping (see inset of Fig.\ref{fig:GaMnAs}). The second key feature is the
presence of finite conductivity in the limit of $\omega\rightarrow
0$: a signature of metallic behavior. The oscillator strength of both
features increases with additional Mn doping. The redshift of the
mid-infrared resonance \cite{burch-annealed} and further theoretical
analysis of this behavior\cite{Moca09} established
that the formation of the metallic state occurs within the impurity
band most likely overlapping the valence band of the GaAs host. This
viewpoint on the electronic structure of ferromagnetic
Ga$_{1-x}$Mn$_{x}$As ($x<7$\%) is supported by magneto-optics
measurements, time-resolved optical techniques and also
photoemission studies \cite{burch2008a}. One can anticipate that
with increasing doping concentration the Fermi energy will
eventually move into the valence band
\cite{jungwirth-rmp}.\footnote{In non-magnetic semiconductors (Si:P,
for example) the latter effect occurs at carrier densities exceeding
the critical IMT concentration by the factor of 8-10 \cite{SiP-rmp}.
Then similar ``valence band transition''  in
Ga$_{1-x}$Mn$_{x}$As can be anticipated at Mn concetrations near
20\%.}

The analysis of the optical effective masses associated
with the free carrier absorption is  indicative of
correlation effects in Ga$_{1-x}$Mn$_{x}$As. IR measurements reveal
optical masses of the order  $10 m_{e}$ \cite{singley-prb03,burch-annealed}, a result recently
confirmed by studies of the mobility and IMT in
Ga$_{1-x-y}$Mn$_{x}$AsBe$_{y}$ and
Ga$_{1-x}$Mn$_{x}$As$_{1-y}$P$_{y}$ \cite{alberi:075201}. It
is yet to be determined if these heavy masses solely originate from
impurity band physics or if many body effects play a role in mass
enhancement as well. Furthermore, in the ferromagnetic state the
optical mass is reduced and scales with the magnetization \cite{singley-prb03}. This latter finding is in accord with the
results reported for colossal magnetoresistance manganites: another
class of correlated carrier mediated ferromagnets (Sec.~\ref {subsec:Manganites}).

\subsubsection{EuB$_6$}
\label{subsec:EuB$_6$}
%
Magnetic semiconductors have attracted interest not only for their
potential use in spintronics, but also because of the fundamental
question of how a magnetic metal can be derived from a
paramagnetic insulator. The most common method, as described
above, involves insertion of transition metal atoms into common
semiconductors such as GaAs. Another important magnetic
semiconductor is EuB$_6$ and its alloys. The Eu$^{2+}$ ions have
$S=7/2$ magnetic moments. The material is a ferromagnetic
semimetal at low temperature containing 10$^{-2}$ carriers per
formula unit. The transition to the paramagnetic state takes place
in two steps, at $T_{\rm Curie} = 12.5$~K and $T_M$=15.3 K. If no
magnetic field is applied the unscreened plasma frequency shrinks
spectacularly from 5200 cm$^{-1}$ at low temperature to 2200
cm$^{-1}$ at $T_{\rm Curie}$ where it stabilizes
\cite{PhysRevLett.79.5134}. At all temperatures an externally
applied magnetic field of a few Tesla increases the plasma
frequency. Moreover, $\omega_p$ and the magnetization $M$ satisfy
the simple scaling relation \cite{broderick02} $\omega_p^2=cM$, an
effect which \textcite{pereira04} explained using a double
exchange model: The itinerant carriers move in a spin-background
potential which is disordered by thermal fluctuations. The
disorder is suppressed when the temperature is lowered and/or and
the magnetic field is increased, and localized charge carriers are
released into itinerant states. Consequently the Drude weight
grows upon magnetizing the system. Upon substituting Ca on the Eu
site the plasma frequency of Eu$_{1-x}$Ca$_x$B$_6$ decreases to a
small value at $x_c=0.35$, and a finite free carrier density
remains for $x>x_c$ \cite{kim05}. The scattering rate decreases
strongly as a function of magnetization
\cite{PhysRevLett.96.016403,PhysRevLett.92.067401}, and the
spectral weight (described by the parameter $\omega_p^2$) follows
the exponential relation $\omega_p^2(M)=\omega_{p0}^2 e^{c M}$.
This behavior was ascribed to the inhibiting effect on the charge
transport of magnetic domain walls, and as such represents a
spin-filter effect. The preponderance of the domain walls
decreases when $M$ increases, so that the scattering diminishes as
is observed experimentally.

\begin{figure}[t]
\centering
\includegraphics[width=\columnwidth]{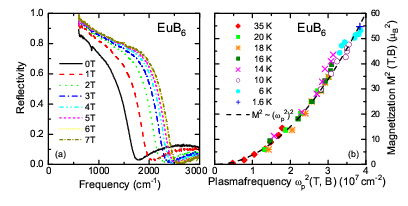}
\caption{(Color online) (a) Magneto-optical reflectivity of EuB$_6$ at 16 K; (b)
Scaling of magnetization and plasma frequency. After
\textcite{broderick02}.}
\label{fig:eub6}
\end{figure}

\subsubsection{Transition metal silicides}
\label{subsec:Transition metal silicides}

Other routes to create magnetic semiconductors are based on tuning
the free charge carrier density of stoichiometric semiconducting
compounds of transition metals or rare earth ions and other elements
such as Si or boron. The itinerant ferromagnets FeGe and MnSi belong
to the same class of transition metal silicides and germanates.
Substituting Ge on the Si site of FeSi results in a the metal
insulator transition at the composition FeSi$_{0.75}$Ge$_{0.25}$. It
has been predicted \cite{anisimov02} that application of a high
magnetic field to the semiconductor induces a highly spin-polarized
ferromagnetic state. Pure FeGe is a good metal, which undergoes a
transition to a heli-magnetic ordered state when cooled below at
$T_{\rm Curie} = 280$~K. The stoichimetric ferromagnets MnSi
($T_{\rm Curie}=29.5$~K) \cite{Mena03} and FeGe present an evolution
of the scattering rate, which resembles more that of EuB$_6$ than
that of the disordered ferromagnets Fe$_{1-x}$Co$_x$Si: At the
temperature where magnetic order occurs, a distinct and narrow
free-carrier response develops, with a strong decrease of the
frequency-dependent scattering rate in the zero-frequency limit.
$m^*(\omega,T)$ for $\omega\rightarrow 0$ is enhanced at low
temperatures and falls gradually as a function of increasing
frequency. Similar trends of $m^*(\omega,T)$ and $1/\tau(\omega,T)$
have been observed in the heavy fermion uniaxial ferromagnet
UGe$_2$($T_{\rm Curie}=53$~K) \cite{guritanu08}, the nearly ferromagnetic metal
SrFe$_4$Sb$_{12}$ (T$_{\rm Curie}=53$~K) \cite{kimura06}, and the
itinerant ferromagnet ZrZn$_2$ (T$_{\rm Curie}=28$~K) \cite{kimura07}.

\begin{figure}[t]
\centering
\includegraphics[width=\columnwidth]{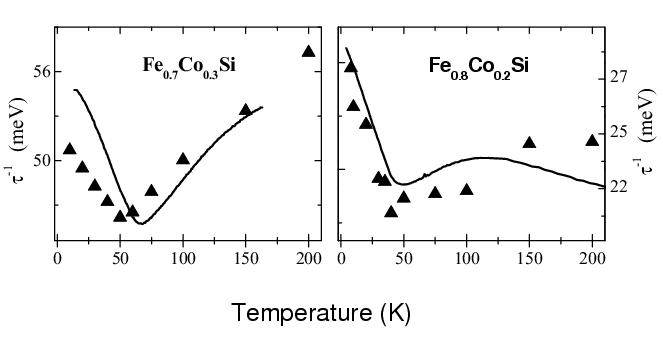}
\caption{Temperature dependence of the optical scattering rates
$\hbar/\tau$ (triangles) of Fe$_{1-x}$Co$_{x}$Si for two different
doping concentrations, and DC resistivities scaled as to overlay the
scattering rates (solid curves). From \textcite{Mena06}.}
\label{fig:fecosi}
\end{figure}

The far infrared reflectivity of Co-doped FeSi samples is suppressed
and the scattering increases when magnetic order sets in
\cite{Mena06} (see Fig.~\ref{fig:fecosi}). The physics differs in an
essential way from that of Ca-doped EuB$_6$ because FeSi is
characterized by a small gap in the density of states and a {\em
non-magnetic} ground state. It has a large 300 K response to
magnetic fields that vanishes as $T$ approaches zero, due to the
opening of a correlation gap at low temperature
\cite{schlesinger93}. The substitution of Co on the Fe site dopes
one hole per Co atom, and, contrary to Ca-doped EuB$_6$ where the
pristine material is already ferromagnetic, here the spin-polarized
state is created by Co-doping \cite{manyala08}. Through the exchange
interaction the spin polarization deepens the potential wells
presented by the randomly distributed Co atoms to the majority spin
carriers \cite{Mena06}. Consequently, the scattering increases (Fig.
\ref{fig:fecosi}), causing the gradual suppression of the metallic
conductivity.
%


\subsection{Iron-pnictides}
\label{subsec:Iron-pnictides}
The iron based superconductor LaFeAsO$_{1-x}$F$_x$ discovered by
\textcite{Kamihara:2008} opened the way to a new class of
materials with interesting magnetic and superconducting
properties. The common building block of iron-pnictides are layers
of edge-shared tetrahedra, where the central Fe atoms are
surrounded by four As, P, or Se atoms, respectively. Four major
groups can be distinguished: (i)~1111 materials with chemical
formula $Re$FeAsO (with $Re$ = La, Ce, Nd, Pr, Sm) under electron
doping; (ii)~122 materials with chemical formula $A$Fe$_2$As$_2$
(with $A$ = Ca, Sr, Ba, Eu) under hole doping, or, substitution of
Fe by Co or Ni (electron doping); and (iii)~11 materials of type
FeSe$_{1-x}$ and Fe(Se$_{1-x}$Te$_x$)$_{0.82}$; and (iv)~111
material LiFeAs. Many other chemically similar compounds with
lower superconducting transition temperature were synthesized,
including nickelates and phosphorus-based Fe-oxypnictide LaFePO
\cite{Ishida09}.

\begin{figure}[h]
\centering{
\includegraphics[width=0.8\linewidth]{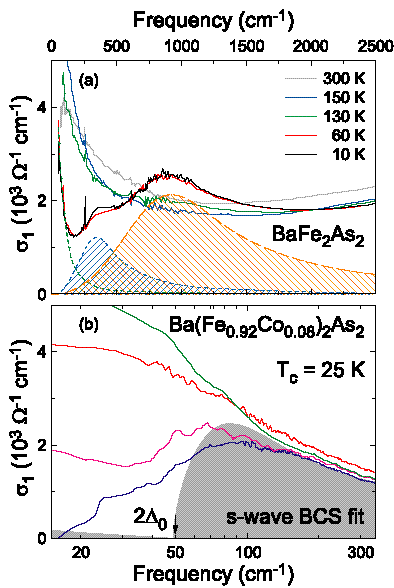}
}
\caption{(Color online)  (a) Optical conductivity of
BaFe$_2$As$_2$ from room
temperature, and across the antiferromagnetic transition around
140~K, down to 10~K.  The partial gap due to SDW is clearly  visible.
\textcite{hu:257005,Hu2009545} associates the two peaks depicted with the
excitations across the SDW gap.
(b) Low-frequency conductivity of
Ba(Fe$_{0.98}$Co$_{0.08}$)$_2$\-As$_2$ above and below the
superconducting transition $T_c=25$~K. The curve given by the
shaded area is calculated using the theory of
\textcite{mattis1958a} for the lowest temperature with a gap of
$2\Delta_0^{(1)}=50$~\cm. After \textcite{Wu10b}.} \label{optFeAs}
\end{figure}

The parent compounds of the 1111 and 122 iron-pnictides are
semimetals with an antiferromagnetic transition in the temperature
range between 130 to 200~K that is accompanied by a structural
transition. Substitutional doping --~but also pressure~--
gradually suppresses the magnetic order and finally the materials
become superconducting with $T_c$'s up to 56~K
\cite{Rotter09,chu:014506,0295-5075-85-1-17006,0256-307X-25-6-080}.
Optical experiments by \textcite{Dong:2008} first showed the
development of a spin-density wave. However, the nesting affects
only part of the Fermi surface, and the systems remain metallic.
In the parent compounds, below $T_{\rm SDW}$ the SDW gap opens
around 1000~\cm\ with the spectral weight piling up right above;
Fig.~\ref{optFeAs}(a) displays the example of BaFe$_2$As$_2$
\cite{hu:257005,Hu2009545}. A prominent in-plane infrared-active
phonon mode, likely connected with orbital ordering, was also
observed in parent compound at
253~cm$^{-1}$~\cite{wu:155103,PhysRevB.80.180502}.
Besides a large background,
a sizeable Drude contribution is present at all temperatures and
narrows upon cooling. This general feature was confirmed by
measurements on SrFe$_2$As$_2$, EuFe$_2$As$_2$, and BaNi$_2$As$_2$
\cite{wu:155103,Hu2009545,Chen09,Wu10b,ChenYuan:2010}. The extended Drude
analysis yields a linear behavior of the frequency-dependent
scattering rate below $T_{\rm SDW}$, indicating an interaction
between the charge carriers and spin fluctuations in the
spin-density-wave state \cite{wu:155103,yang:187003}.
For the superconducting compounds, like
Ba(Fe$_{1-x}M_{x}$)$_2$\-As$_2$, its resistivity follows a $T^2$
behavior, implying that superconductivity develops out of a Fermi
liquid.
%
Optical studies of both LaFePO and BaFe$_2$As$_2$
(Fig.~\ref{fig:KE-all}) reveal suppression of the electronic kinetic
energy comparable to that of other strongly correlated superconductors
including high-$T_c$ cuprates \cite{qazilbash-np-2009,ChenYuan:2010},
emphasizing the importance of correlation effects in iron
pnictides~\cite{Haule:2008,QSi:2009}.

In accord with theory, most experimental methods evidence a fully
gapped superconductor with no nodes of the order parameter. Its
symmetry might be $s_{\pm}$ wave, i.e.\ reverses sign for electron
and hole pockets of the Fermi surface. Reflection experiments
\cite{li:107004,Hu2009545,Wu10b} and transmission through films
\cite{Gorshunov10}  yield a reduction of the optical conductivity
for $T<T_c$. In accordance with ARPES measurements
\cite{evtushinsky:054517} two gaps --~different in energy by a
factor of 2~-- in certain parts of the Fermi surface evolve
simultaneously below $T_c$. In Fig.~\ref{optFeAs}(b) the optical
conductivity of Ba(Fe$_{0.98}$Co$_{0.08}$)$_2$\-As$_2$ is plotted
for various temperatures. It can be sufficiently well described by
the BCS theory (shaded area). The missing spectral weight extends
up to $6\Delta\approx 150$~\cm\ and according to the
Ferrell-Glover-Tinkham sum-rule [Eq.~(\ref{eq:FGT})] it corresponds
to a superconducting density in accord \cite{Wu10c} with Homes' scaling plotted
in Fig.~\ref{fig:Homes-law}.

\section{Organic and Molecular Conductors}
\label{sec:Organic and Molecular Conductors}
\label{sec:organics}
Organic ligands are utilized to arrange metal ions in chains
similar to K$_2$Pt(CN)$_4$Br$_{0.3}\cdot$3H$_2$O (KCP), the
canonical example of a one-dimensional metal
\cite{Kagoshima88,Keller75,Schuster75}. In organic solids extended
molecules with delocalized $\pi$ electrons form stacks or
layers with orbital overlap in certain directions causing a large
anisotropy of the electronic properties
\cite{SchwoererWolf07,Farges94,Ishiguro98}.
They serve as model
systems for investigating the physics in reduced dimension \cite{Dressel03a,Dressel07,Toyota07,Lebed08}.

The discovery of the first organic metal TTF-TNCQ in the early
1970s kicked-off a broad endeavor to understand the
one-dimensional metallic properties as well as the charge-density
wave ground state below 54~K. Optical experiments by
\textcite{PhysRevLett.32.1301,PhysRevB.13.3381,PhysRevLett.47.597}
and others \cite{
PhysRevB.28.6972,PhysRevB.31.5465,Gorshunov1986681,PhysRevB.42.4088}
turned out to be challenging and even today no agreement has been
reached about collective modes below 100~\cm.\footnote{The
electrodynamic properties of conducting polymers also continue to
attract attention concerning coherent transport
\cite{PhysRevLett.74.773,PhysRevLett.90.176602}, CDW instabilities
\cite{PhysRevB.61.1635}, and solitons and polarons
\cite{Kaiser01,Tanner04}. We skip these topics as well as fullerenes
and carbon nanotubes \cite{Wu04,Kamaras06,Kamaras08}, and limit
ourselves to the discussion of graphene in
Sec.~\ref{subsec:Graphene}.}

\subsection{One-dimensional molecular crystals}
\label{subsec:One-dimensional molecular crystals}
\label{subsec:One-dimensional-molecular-crystals} 
\begin{figure}
\centering
\includegraphics[width=6.5cm]{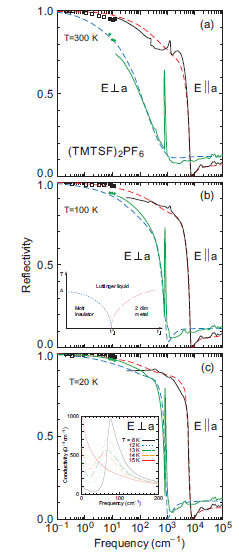}
\caption{\label{fig:tmtsf}(Color online)
Reflectivity spectra of (TMTSF)$_2$PF$_6$ measured at different
temperatures along the stacking axis $a$ (solid black line) and
perpendicular to it (solid green line). The filled symbols are
obtained by a coherent source THz spectrometer, the open symbols
are calculated from microwave experiments
\cite{Dressel96,Donovan94}. The dashed lines represent a Drude
fit, respectively. The inset of panel (b) shows the schematic
phase diagram of the deconfinement transition for a system of
weakly coupled conducting chains as suggested by
\textcite{Biermann01} and \textcite{Giamarchi04a}. The transition
from a Mott insulator to a two- or three-dimensional metallic
state occurs at $T=0$ when $t_\perp$ reaches a critical value
$t_\perp^*$. At high enough temperature, the increase in $t_\perp$
leads to a transition from a Mott insulating to a one-dimensional
Luttinger liquid and further to a dimensional crossover into a
metallic state. The development of the SDW gap at 70~\cm\ is seen
from the low-temperature  conductivity $E\perp a$ plotted in the
inset of panel (c) \cite{Degiorgi96}.}
\end{figure}

In the Bechgaard
salts (TMTSF)$_2$$X$ coherent electronic transport  develops along
the stacks of tetra\-methyl-tetra\-selena-fulvalene molecules; the
conduction band is split due to dimerization, yielding a
half-filled system. The interaction between the stacks can be
varied when selenium is replaced by sulphur or anions $X$ of
different size are selected. In the extremely one-dimensional case
of (TMTTF)$_2$AsF$_6$, for instance, Coulomb repulsion drives the
system Mott insulating. With increasing interchain coupling, a
deconfinement transition occurs
to a Luttinger liquid and two-dimensional metal
\cite{Vescoli98,Biermann01,Giamarchi04a,Giamarchi04b}; it
corresponds to a horizontal movement in the phase diagram depicted
in the inset of Fig.~\ref{fig:tmtsf}(b). The application of
pressure is a way to continuously tune the interaction between
chains \cite{pashkin06,Pashkin09}. Eventually at $p=12$~kbar (TMTSF)$_2$PF$_6$ becomes superconducting around $T_c=1$~K.

The quasi-one-dimensional metal (TMTSF)$_2$PF$_6$ exhibits a large
anisotropy  in the plasma frequency of a factor of 10
(Fig.~\ref{fig:tmtsf}). As first shown by
\textcite{Jacobsen81a,Jacobsen83}, with decreasing temperature  the
system becomes metallic even in the $b$ direction [development of a
plasma edge in Fig.~\ref{fig:tmtsf}(b,c) compared to (a)],
indicating a crossover from a one- to a two-dimensional metal
(vertical movement in the phase diagram). At high frequencies
($\hbar\omega
> t_\perp$, the transfer integral perpendicular to the chains),
the low-temperature optical conductivity follows a power-law
$\sigma_1(\omega)\propto \omega^{-1.3}$ \cite{schwartz98} in
agreement with transport measurements \cite{Dressel05,Moser98} and
with calculations based on an interacting Luttinger liquid
\cite{Giamarchi91,Giamarchi97}.\footnote{It is interesting to note
that in one-dimensional Cu-O chains of YBa$_2$Cu$_3$O$_{7-\delta}$ a
similar power law was observed \cite{Lee04,lee05}. See also
Sec.~\ref{subsec:Power law behaviour of optical constants and
quantum criticality}.}

At $T_{\rm SDW}=12$~K (TMTSF)$_2$PF$_6$ enters a spin-density-wave
(SDW) state with a sharp increase in resistivity due to the
opening of an energy gap over the entire Fermi surface
\cite{Jacobsen81b,Dressel05}.\footnote{Similar investigations have
been  performed on various sister compounds (TMTSF)$_2X$
\cite{Jacobsen83,Ng84,Ng85,Eldridge86,Kornelsen87}.}
\textcite{Degiorgi96} discovered a single-particle gap around
$2\Delta=70$~\cm\ in the optical properties
\cite{Vescoli99,Henderson99} as shown in the inset of
Fig.~\ref{fig:tmtsf}(c). In addition collective excitations of the
SDW are observed along the nesting vector leading to a pinned mode
resonance in the microwave range \cite{Donovan94,Petukhov05}. It
does not make up for the spectral weight lost upon entering the
insulating state at $T_{\rm SDW}$.

\subsection{$MX$ chains}
\label{subsec:$MX$ chains}
Halogene-bridged metal complexes forming -$M$-$X$-$M$-$X$- linear
chains constitute one-dimensional Peierls-Hubbard systems, where
the electron-phonon interaction, the electron transfer, and the
on-site and inter-site Coulomb repulsion energies compete and
cooperate with one another. In [Ni(chxn)$_2$Br]Br$_2$ (chxn =
cylohexanediamine), for instance, four nitrogen atoms of two ligand units
coordinating a Ni ion [cf.\ Fig.~\ref{fig:MX1}(a)] produce such a
strong ligand field that the Ni$^{3+}$ ion is in a low-spin state
with an unpaired electron in the $d_{z^2}$ orbital.
The strong one-site Coulomb repulsion among the Ni $3d$ electrons
causes a Mott-Hubbard gap (approximately 5~eV) with the occupied Br
$4p$ band located inside. Thus the lowest-energy electronic
excitation goes from the Br $4p$ band to the Ni $3d$ upper Hubbard
band indicated by the sharp absorption band around 1.3~eV as shown
by \textcite{Takaishi08}.
For $M$ = Pd and Pt the ionic radius is larger and thus the
electron-electron interaction is weak compared to the
electron-lattice interaction. The bridging halogen ions are
distorted from midpoint between the neighboring two metal ions,
giving rise to the CDW states or $M^{II}$–$M^{IV}$ mixed-valence
states (-$M^{II}$-$X$–$M^{IV}$–$X$-$M^{II}$-). Accordingly, the
half-filled metallic bands split  by a finite Peierls gap into the
occupied valence bands and the unoccupied conduction bands. The
compounds exhibit unique optical and dynamical properties, such as
dichroic and intense intervalence charge-transfer bands
\cite{Tanaka84}, resonance Raman spectra \cite{Clark83,Clark90},
luminescences with large Stokes shift \cite{Tanino83}, midgap
absorptions attributable to the solitons and polarons
\cite{Okamoto92,Okamoto98}.

\subsubsection{Mott insulators}
\begin{figure}
\centering
\includegraphics[width=8cm]{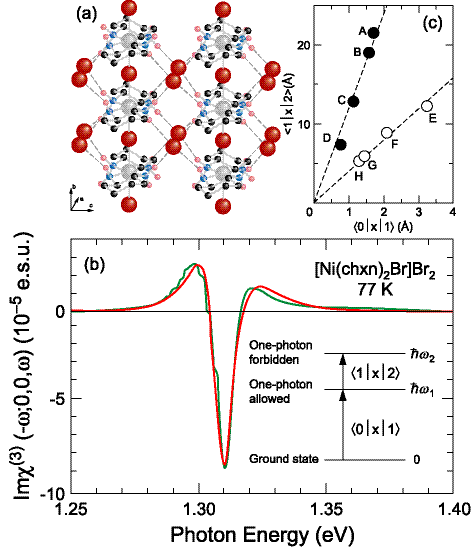}
\caption{\label{fig:MX1}(Color online) (a)~Arrangement of the
Ni (gray) and Br (red) ions along chains in [Ni(chxn)$_2$Br]Br$_2$ leads to an overlap of the Ni $3d_{x^2}$ and
Br $4p_z$ orbitals \cite{Takaishi08}. (b)~Im$\chi^{(3)}(-\omega;0,0,\omega)$
spectra of [Ni(chxn)$_2$Br]Br$_2$ at $T=77$~K. Green and red lines are the experimental and
calculated results, respectively. The energy levels are sketched in the lower panel.
(c)~Relation between the transition dipole moments of $\langle
0|x|1\rangle$ and $\langle 1|x|2\rangle$ for one-dimensional Mott
insulators (solid circles: A, [Ni(chxn)$_2$Br]Br$_2$; B,
[Ni(chxn)$_2$Cl]Cl$_2$; C, [Ni(chxn)$_2$Cl](NO$_3$)$_2$; D,
Sr$_2$CuO$_3$) compared to other one-dimensional materials (empty
circles: E, [Pt(en)$_2$]\-[Pt(en)$_2$I$_2$]\-(ClO$_4$)$_4$; F,
[Pt(en)$_2$]\-[Pt(en)$_2$Br$_2$]\-(ClO$_4$)$_4$; G,
[Pt(en)$_2$]\-[Pt(en)$_2$Cl$_2$]\-(ClO$_4$)$_4$; H,
polydihexylsilane (PDHS); I, polyacetylene (PA); J,
polydiacetylene (PDA); K, polythienylvinylene (PTV); L,
poly(p-phenylenevinylene) (PPV)) \cite{Kishida00}.}
\end{figure}
\textcite{Iwano02} explained the small Raman Stokes shift observed in the
nickel-chain compound by a suppression of the
electron-lattice interaction, in agreement with
dynamical density-matrix renormalization-group
calculations incorporating lattice fluctuations \cite{Iwano06}. The
dominance of strong electronic correlations
enhances the nonlinear optical properties, including the
third-order susceptibility $\chi^{(3)}$ \cite{Kishida00,Ono04}
shown in Fig.~\ref{fig:MX1}(b). The main reasons are the small
energy splitting of about 10~meV between the two excited states
$\omega_1$ and $\omega_2$ and the large transition dipole moments
$\langle 0|x|1\rangle$ and $\langle1|x|2\rangle$ between the
ground state $|0\rangle$, the one-photon allowed state $|1\rangle$
and the one-photon forbidden state $|2\rangle$; here $\langle
1|x|2\rangle$ basically describes the spatial extension of the
electron-hole wavefunction in the excited state, as summarized in
Fig.~\ref{fig:MX1}(c). Thus one-dimensional Mott insulators have a
larger potential for nonlinear optical devices than
one-dimensional band insulators, such as silicon polymers and
Peierls insulators of $\pi$-conjungated polymers
\cite{Takaishi08}.

\begin{figure}
\centering
\includegraphics[width=5cm]{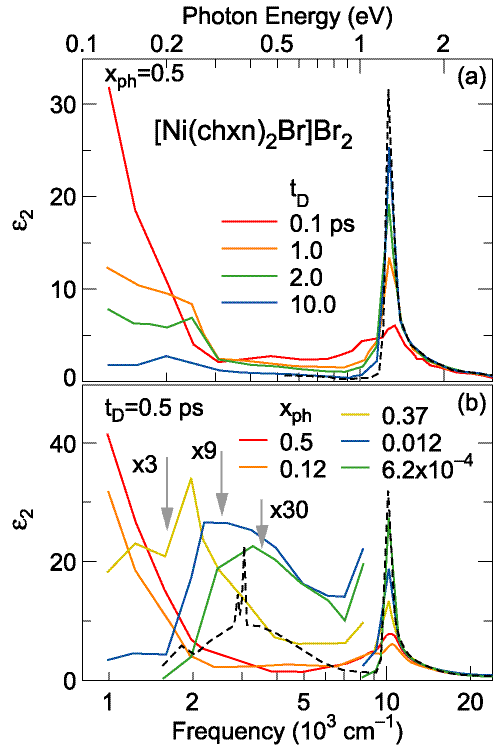}
\caption{\label{fig:MX2}(Color online)
Room-temperature spectra of the imaginary part of the dielectric
constant $\epsilon_2$ of [Ni\-(chxn)$_2$\-Br]\-Br$_2$. (a)~Data
are taken prior to the photoexcitation (dashed line) and at delay
times $t_d$ after the photoexcitation (solid lines) with an
excitation density $x_{ph}$ is 0.5 photon per Ni site.
Polarizations of the pump and probe lights are both parallel to
the chain axis. (b) Dependence of $\epsilon_2(\omega)$ on the
excitation density as indicated ($t_d=01$~ps). The dotted line
shows the spectrum before the photoexcitation obtained by direct
measurements of the polarized absorption \cite{Iwai03}.}
\end{figure}
When [Ni(chxn)$_2$Br]Br$_2$ is irradiated by light, electrons are
excited, leading to an enhancement of the Drude-like low-energy
component in the optical conductivity immediately after
photoirradiation
as demonstrated by  \textcite{Iwai03} and shown in
Fig.~\ref{fig:MX2}. This suggests a Mott transition by photodoping
with an ultrashort lifetime $t_d=0.5$~ps of the metallic state. For
very low excitation density $x_{ph}$ of about $10^{-3}$ photons per
Ni site a midgap absorption is observed around 0.4 - 0.5~eV.
Following the analysis of chemical doped Mott insulators, the
effective number of carriers $N_{\rm eff}(\omega)$ is obtained by
integrating the optical conductivity to the measurement frequency
$\omega$. This yields the total spectral-weight transfer from the
charge-transfer band to the innergap region and indicates that the
photoinduced midgap absorption is due solely to electron-type charge
carriers since the hole-type carriers are localized by
electron-lattice interaction. Interestingly, the palladium-chain
compounds remain insulating with a finite optical gap even after
photoexcitation \cite{Yonemitsu08}.

Theoretical studies using a one- and two-band extended
Peierls-Hubbard models deal with ground state, excitation
spectrum, the non-linear optical properties of $MX$ chains
\cite{Gammel92,Weber92,Saxena97}. Other platium-halide ladder compounds
are treated by a multiband extended Peierls-Hubbard Hamiltonian to
reproduce the optical spectra \cite{Yamamoto07}.

\subsubsection{Peierls systems}
\begin{figure}
\centering
\includegraphics[width=6cm]{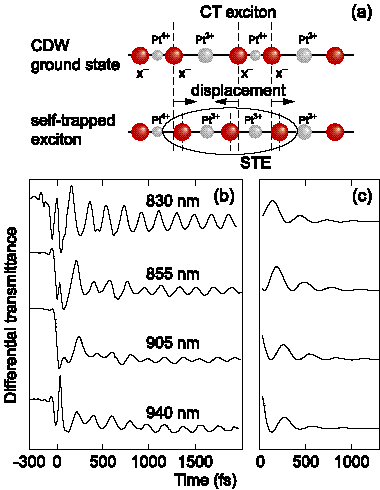}
\caption{\label{fig:MX3} (Color online)
(a)~Schematic drawing of the linear chain in a halogen bridged
Pt-complex. In the ground state a charge density is commensurately
modulated; a self-trapped exciton is formed in the excited state
\cite{Suemoto02}. (b) Time-resolved differential transmittance of
[Pt(en)$_2$][Pt(en)$_2$Br$_2$]$\cdot$(PF$_6$)$_4$ following
excitation of the intervalence charge transfer band with 35~fs
pulses centered at 800~nm. Measurements were taken at a series of
detection wavelengths selected from a broadband femtosecond
continuum. (c) The sum of the low-frequency component and the
zero-frequency component extracted by linear prediction/singular
value decomposition, showing the excited state contributions to the
response and the systematic phase shift of the wave packet
oscillation with detection wavelength \cite{Dexheimer00a}.}
\end{figure}

[Pt(en)$_2$][Pt(en)$_2$Br$_2$]$\cdot$(PF$_6$)$_4$ (en =
ethylene-diamine) shows both the periodic charge disproportionation
(mixed valence, commensurate CDW) and the periodic bond-length
distortion (Peierls distortion) as depicted in
Fig.~\ref{fig:MX3}(a). Femtosecond impulsive excitation causes a
coherent oscillation at the ground-state vibrational frequency of
180~\cm\ [known from Raman experiments \cite{Love93}] and its
harmonics.  In addition a self-trapped excition state is observed at
110~\cm\ \cite{Dexheimer00a,Dexheimer00b}. The phase of this rapidly
dampled component shifts systematically with detection wavelength
changing by approximately 180$^\circ$ between 830 and 940~nm. Using
even shorter pulses (5~fs) transmission measurements reveal
low-frequency modes around 60-70~\cm\ assigned to asymmetric
vibrational modes of the self-trapped excition state
\cite{Araoka07}. Similar findings have been made by time-resolved
reflectivity \cite{Sugita01} and luminescence spectroscopy
\cite{Tomimoto02} on various Pt$X$ chains with $X$ = Cl, Br, or I
with an decrease in lifetime from 30 to 0.65~ps.

Recent investigations by ultrafast laser pulses could establish
low-dimensional organic compounds as model compounds for
photoinduced phase transitions, as discussed in
Sec.~\ref{subsec:Photoinduced phase transitions}.

\subsection{Two-dimensional molecular crystals}
\label{subsec:Two-dimensional molecular crystals}
\label{sec:organics-2D} Among the layered organic crystals,
bis-(ethyl\-ene\-di\-thio)\-te\-tra\-thia\-ful\-va\-lene
(BEDT-TTF) molecules are of paramount interest due to their
versatility of forming (super-)conducting salts with different
anions in a large variety of patterns
\cite{Mori98,Mori99a,Mori99b,Mori00}. The fine tuning of molecular
interactions provides the possibility to study bandwidth
controlled Mott transitions, the interplay of charge order and
superconductivity, etc. \cite{Dressel04a}.\footnote{The search for
indications of a superconducting gap in the optical properties
\cite{Kornelsen91,Ugawa00} led to success only recently
\cite{Drichko02,Kaiser09}. There is some debate on whether the
organic superconductor follow the universal scaling
presented in Fig.~\ref{fig:Homes-law}
\cite{PhysRevLett.94.097006}.}

\subsubsection{Mott insulator {\em versus} Fermi liquid}
\label{sec:organics-2D-Mottinsulator}
The $\kappa$-phase BEDT-TTF
compounds, for instance, form an anisotropic triangular lattice;
the upper band is half-filled leading to a Mott transition when
$U/W$ exceeds some critical value. The antiferromagnetic
insulating ground state is next to superconductivity with the
maximum $T_c$ of 14~K which triggered numerous theoretical studies
\cite{Watanabe06a,Powell05,Powell07,Clay08,Mazumdar08,Peters09}.
The optical spectra are dominated by a strong charge-transfer band
in the mid-infrared and electron-molecular vibrational (emv)
coupled modes \cite{Eldridge91,Kornelsen92a,Kornelsen92b}. These
excitations can be separated from contributions of the itinerant
electrons \cite{Faltermeier07,Dressel09} giving insight into the
dynamics of charge carriers at the verge of localization.
Replacing Cl by Br in the anion layer of
$\kappa$-(BEDT-TTF)$_2$Cu[N(CN)$_2$]Br$_{x}$Cl$_{1-x}$ serves as
chemical pressure that increases the bandwidth.
\begin{figure}
\centering
\includegraphics[width=7.5cm]{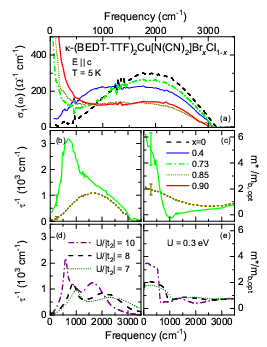}
\caption{\label{fig:et1}(Color online)
(a)~Low-temperature optical conductivity of
$\kappa$-(BEDT-TTF)$_2$Cu[N(CN)$_2$]Br$_{x}$Cl$_{1-x}$ for
different Br content $x$, which serves as chemical pressure and
decreases the effective Coulomb interaction $U/W$. The
contributions from intradimer transitions and vibrational modes
are subtracted; the $\sigma_1(\omega)$ spectra plotted here
represent the correlated charge carriers. Panels (b) and (c) show
the frequency dependence of the scattering rate and effective mass
extracted from an extended Drude model analysis of the
conductivity (a). In (d) and (e) the corresponding results of DMFT
calculations are plotted for different $U/W$ and $T=50$~K. Clearly
as the Mott insulating phase is approached, the effective mass and
the scattering rate increase significantly
\cite{Merino08,Dumm09}.}
\end{figure}
As demonstrated in Fig.~\ref{fig:et1}(a), spectral weight is
redistributed from high to low frequencies as the Mott transition
is approached at $x\approx 0.7$; a similar shift is caused by
temperature. The effective mass of the quasiparticles also
increases considerably; at low frequencies the scattering rate follows a
$1/\tau(\omega)\propto\omega^2$
 behavior \cite{Dumm09} in accord with the quadratic temperature dependence of the resistivity \cite{Dressel97,Yasin08}.

A dynamical mean-field-theory treatment of the relevant Hubbard
model gives a good quantitative description of the experimental
data as demonstrated in Fig.~\ref{fig:et1}. The calculations are
performed on a frustrated square lattice at half filling taking
the nearest-neighbor hopping amplitudes to be $t_2=-0.03$~eV and
$t_1=0.8t_2$ (as known from bandstructure calculations) that leads to a non-interacting bandwidth of $W
\approx 0.3$~eV, comparable to values from density functional
theory calculations \cite{Merino00,Merino08,Dumm09}.

\begin{figure}
\centering
\includegraphics[width=7.5cm]{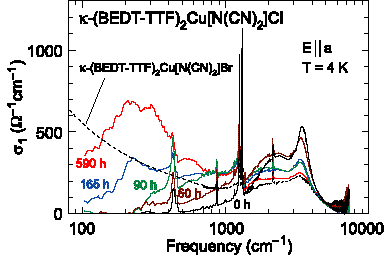}
\caption{\label{fig:et2}(Color online) Low-temperature optical conductivity of
$\kappa$-(BEDT-TTF)$_2$\-Cu[N(CN)$_2$]Cl before and after x-ray irradiation.
The dashed curve represents $\sigma_1(\omega)$ of the non-irradiated
$\kappa$-(BEDT-TTF)$_2$\-Cu[N(CN)$_2$]Br \cite{Sasaki08}.}
\end{figure}
\textcite{Sasaki07,Sasaki08} suggest that the Mott insulator
$\kappa$-(BEDT-TTF)$_2$\-Cu[N(CN)$_2$]Cl can be effectively doped
by charge carriers when irradiated by x-rays. Considerable spectral weight is
transferred from the mid-infrared region to low frequencies as the Mott gap collapses
with increasing irradiation time as displayed in Fig.~\ref{fig:et2}.
Nevertheless, no Drude-like peak is present even after 590~h at a dose of 0.5~MGy/h,
suggesting that the crystals transforms to a weakly disordered metal.

\subsubsection{Charge order and superconductivity}
\label{sec:organics-2D-chargeorder}
In the $A_2B$ stoichiometry the conduction band is quarter-filled when the BEDT-TTF
molecules are not arranged in dimers. Due to strong intersite Coulomb repulsion $V$,
the materials are subject to an electronically-driven charge order \cite{Calandra02,Seo04,Seo06},
which --~besides NMR and x-ray scattering~-- can be nicely seen
from the splitting of the charge-sensitive intra-molecular  BEDT-TTF
vibrations.
Raman and infrared investigations can quantitatively estimate
the charge disproportionation and yield information on the charge-order
pattern \cite{Yamamoto02,Wojciechowski03,PhysRevB.72.014516,Drichko09}.

\begin{figure}
\centering
\includegraphics[width=8.5cm]{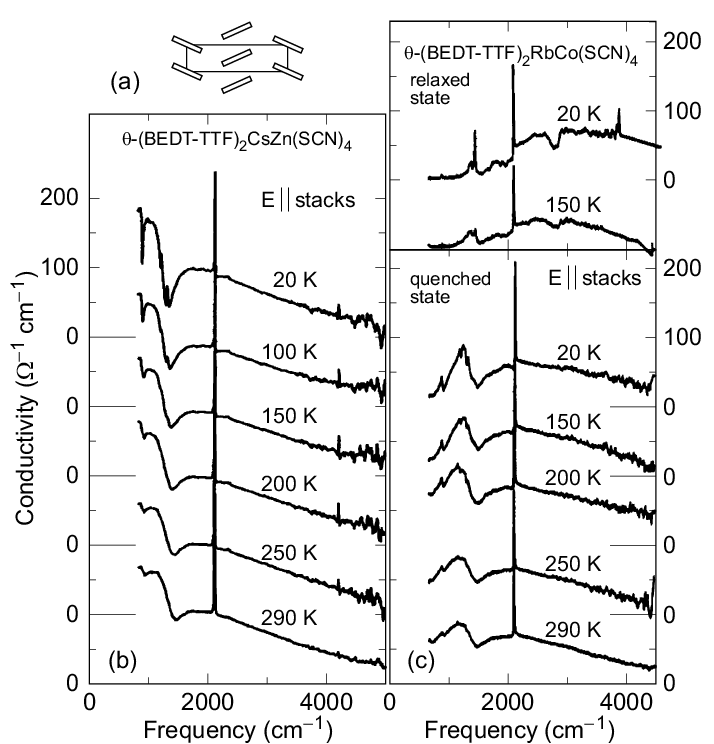}
\caption{\label{fig:et3}(a) The structural arrangement of the
$\theta$-phase reveals two stacks of BEDT-TTF molecules in a herring-bone fashion. Optical conductivity
of (b) $\theta$-(BEDT-TTF)$_2$CsZn(SCN)$_4$ and (c) $\theta$-(BEDT-TTF)$_2$RbCo(SCN)$_4$
measured along the stacks ($c$-axis) for different temperatures.
The charge-ordered state in the latter is only reached when slowly cooled (0.1~K/min),
for rapid cooling (1~K/min) the crystal remains metallic
\cite{Tajima00}. The curves for different temperatures are displaced.
}
\end{figure}
\textcite{Mori95,Mori98a,Mori98b} introduced the
$\theta$-(BEDT-TTF)$_2$$MM^{\prime}$(SCN)$_4$ family for which by
variation of the metal ions $M$ and $M^{\prime}$ the transfer
integrals $t_1$ and $t_2$ can be chosen in such a way that
horizontal and vertical charge order patterns form [cf.\
Fig.~\ref{fig:et3}(a)]. While $\theta$-(BEDT-TTF)$_2$CsZn(SCN)$_4$
remains metallic down to 20~K, the Drude contribution to the optical
response of $\theta$-(BEDT-TTF)$_2$RbCo(SCN)$_4$ vanishes upon
passing through the charge-order transition at 190~K, and the
spectrum becomes semiconductor-like, as demonstrated in
Fig.~\ref{fig:et3} \cite{Tajima00}; similar results are obtained for
$\theta$-(BEDT-TTF)$_2$RbZn(SCN)$_4$.

Within the two-dimensional conducting layer, the organic BEDT-TTF
molecules are arranged in stacks which are more or less coupled;
hence the charge disproportionation in the stacks can form certain
patterns of stripes. The actual arrangement of horizontal,
vertical or diagonal stripes very much depends upon the interplay
of the different interactions. A detailed understanding needs to
go beyond a pure electronic model because coupling of the charge
order to the underlying lattice has to be taken into account
\cite{Tanaka07,Tanaka08,Tanaka09,Miyashita08}. These effects are
weak in the $\alpha$-compounds where the molecular rearrangement
is small compared with those in the $\theta$-phase. Their
photoinduced dynamics studied by femtosecond reflection
spectroscopy \cite{Iwai07,Iwai08,Tajima05} are qualitatively
different: $\theta$-(BEDT-TTF)$_2$RbZn(SCN)$_4$ exhibits local
melting of the charge order and ultrafast recovery, while
$\alpha$-(BEDT-TTF)$_2$I$_3$ exhibits critical slowing down.

Comprehensive optical studies on metallic compounds in the vicinity
of the charge-ordered phase -- by varying $V/W$ -- reveal the
development of a pseudogap, charge-order fluctuations and collective
charge-order excitations which are coupled to lattice vibrations
\cite{Dressel03b,Dressel04b,Drichko06a,Drichko06b,Kaiser09}.
Evidence has accumulated that charge-order fluctuations induce
superconductivity in these organic charge-ordered systems similar to
Na$_{0.35}$CoO$_2$$\cdot$1.3H$_2$O
\cite{Merino01,Greco05,Watanabe05,Watanabe06b}. Changing the
band-filling beyond one quarter results in a strong increase in the
spectral weight of the Drude term \cite{Drichko05,Drichko07}.

\subsection{Graphene}
\label{subsec:Graphene}
Apart from organic conductors reviewed in
Sec.~\ref{subsec:One-dimensional molecular crystals} through
\ref{subsec:Two-dimensional molecular crystals}, hallmarks of
electronic correlations are found in a variety of carbon-based
systems. Examples include Luttinger liquid behavior of carbon
nanotubes \cite{cnt-LL} and polymers\cite{ISI:000267204600016}as well as a Mott-Hubbard state in
A$_4$C$_{60}$ (A=Na, K, Rb, Cs) \cite{fink-mh-c60}.  Graphene - a
one-atom-thick sheet of carbons - is emerging as an extremely
interesting electronic system to investigate the role of
correlations and many body physics in optical and transport
properties \cite{peres-06,neto-rmp09}. The ``relativistic''
nature of the quasiparticles in graphene, albeit with a speed of
propagation 300 times smaller than the speed of light, is expected
to give rise to unusual spectroscopic, transport, and thermodynamic
properties that are at odds with the standard Fermi liquid theory of
metals \cite{Gonzalez-99,sarma:121406,Polini200758}.

Electronic phenomena in single- and multi-layered graphene can be
readily altered by applied voltage. Importantly, the interaction of
electrons/holes with each other and with the honeycomb lattice also
can be controlled by the gate voltage
\cite{yan:166802,neto-el-phonon,goerbig:087402,charged-phonon}.
Interactions among ``mass-less'' Dirac quasiparticles in graphene
are of fundamental interest and are also of relevance for the
understanding of superconductivity with relatively high transition
temperature in various other forms of carbon including nanotubes,
doped C$_{60}$ crystals, doped diamond, as well as graphite intercalation compounds.
Spectroscopic investigations of graphene physics, to large extent,
rely on measurements of gated structures. Infrared experiments of
the gated structures are relatively scarce in view of technical
complexity of monitoring subtle changes of properties in ultrathin
accumulation/depletion layers. Following the pioneering work for Si
MOSFETs \cite{Tsui-mosfet}, infrared studies of gated devices were
extended to field effect transistors with active elements made of
polymers \cite{brown-p3ht-fet,li-p3ht-fet}, molecular crystals
\cite{fischer:182103,li:rubrene}, oxides
\cite{qazilbash-vo2-fet,choi-ZnO-fet} and most recently graphene.

\begin{figure*}
\centering
\includegraphics[width=6.6in]{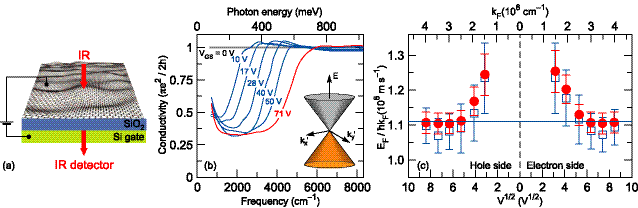}
\caption
{(Color online) Panel A: schematics of
graphene-based gated structure. Panel B: the optical conductivity of
graphene extracted from $R(\omega)$ and $T(\omega)$
synchrotron-based microscopy for various gate voltages. The
threshold feature is due to interband transitions at $2E_F$
schematically displayed in the inset. Panel C:  magnitude of  $v_F$
extracted from the conductivity data in panel B. An enhancement of
$v_F$ at small biases is indicative of many body effects in graphene
as discussed in the text. From \textcite{ISI:000257984600011}.}
\label{fig:graphene}
\end{figure*}

In charge-neutral monolayer graphene the Fermi energy is located
exactly between the two linearly dispersing cones characteristic of
Dirac quasiparticles. This dispersion leads to the frequency
independent conductivity leveling at the universal value
\cite{ando-JPSJ-02,peres-06,gusynin:245411,nair-science,li-graphene,mak:196405,kuzmenko:117401}.
Infrared data taken under applied gate voltage $V_g$ revealed
significant modification of optical properties consistent with the
expectations based on the electronic structure
\cite{ISI:000254836700036,li-graphene}. The dominant feature of the
conductivity data (Fig.\ref{fig:graphene}) is a formation of a
threshold feature in $\sigma_1(\omega, V_g)$ that systematically
hardens with the increase of $V_g$. This form of the conductivity is
consistent with the notion of Pauli blocking: direct interband
transitions between the bottom and top cones are prohibited by
momentum conservation for $\hbar\omega<2E_F$.

The spectral weight lost from the region below $2E_F$ is transferred
to the Drude conductivity due to mobile Dirac quasiparticles in
partially filled bands. The Pauli blocking in graphene is not
complete and substantial absorption can be recognized in the data
down to the lowest frequencies. At least in part the residual
response can be attributed to impurities and interaction with
phonons \cite{stauber:085418}. A different proposal accounts for
Pauli-prohibited absorption within the marginal Fermi liquid theory
of Dirac quasiparticles \cite{maria-09} originally proposed to
explain anomalous scattering processes in high-T$_c$
superconductors. Residual absorption observed in single-layer
graphene is also found bilayer graphene samples
\cite{li-graphene,fogler-zhang, kuzmenko-bilayer}. A new property of
the latter systems is gate-induced energy gap between valence and
conduction band revealed by a number of recent experiments
\cite{zhang-gap,mak-gap,kuzmenko-gap}.

The evolution of the $2E_F$ feature with $V_g$ allows one to probe
the Fermi velocity of Dirac quasiparticles. Experiments are in
accord with nearly linear variation of $2E_F(V_g)$ implied by the
linear dispersion with $v_F\simeq$ 1.12-1.2 $\times 10^6$ ms$^{-1}$.
However, at small biases one witnesses a systematic enhancement of
$v_F$. Thus, in graphene, Coulomb interaction favors electron/hole
delocalization offering an intriguing counterexample to properties
of most other systems discussed in this review where strong
interactions typically impede electronic transport. The Fermi
velocity in graphene can be independently obtained from infrared
studies of cyclotron resonance
\cite{jiang:197403,henriksen:087403,deacon:081406}. These latter
experiments carried out for both single- and bi-layer graphene yield
$v_F$ which is enhanced by 20-30\% compared to that of the bulk
graphite.

The renormalization of $v_F$ is interesting in the context of
electronic correlations in graphene since it can be attributed to
Coulomb interaction of Dirac quasiparticles \cite{Gonzalez-99}. A
salient feature of magneto-optics data for both single- and bi-layer
graphene is the violation to the Kohn theorem \cite{kohn-theorem}.
The theorem predicts only negligible role of the electron-electron
interaction in the properties of conventional 2D electron gas with
quadratic dispersion but appears to be violated for the linearly
dispersing Dirac quasiparticles. The theoretical analysis of CR
absorption of graphene supports the notion of strong
electron-electron interaction \cite{iyengar:125430,bychkov:125417}.

Several groups reported studies of both zero-field and high magnetic
field response of epitaxial graphene
\cite{plochocka:087401,sadowski:266405,dawlaty:131905,choi:172102}. One potential problem with the quantitative
interpretation of these latter data is that epitaxial graphene
obtained through high-temperature thermal decomposition of SiC
substrates are not continuous and in addition reveal substantial
variation of thickness along wafers.

\section{Outlook}
\label{sec:Outlook}
The hypothesis of \textcite{mott1937a} on the paramount role of the
Coulomb interaction in insulating behavior of NiO and other
unconventional insulators has proven to be accurate. Arguably,
Mott-Hubbard insulators present the best understood example of a
strongly correlated system. Optical experiments have made preeminent
contributions towards establishing experimental picture of these
materials through direct measurements of the the energy gap and
detailed studies of  the redistribution of the spectral weight with
doping (Section~\ref{subsec:Emergence of conducting state in
correlated insulators}). The in-depth understanding of the parent
insulating systems is a precondition for the description of some of
the most enigmatic effects in {\it doped} Mott insulators including
unconventional superconductivity and ferromagnetism accompanied with
colossal magneto-resistance. A similar level of understanding is yet
to be achieved for conducting doped Mott insulators.

The term ``optical'' as applied in this review to the
electrodynamic response of complex correlated materials should be
understood colloquially since investigations of the
frequency-dependent response readily extend from microwaves
through THz and IR to UV. Largely due to technical innovations in
1990-s and 2000-s femtosecond studies of correlated matter have
now become commonplace. The parameter space of ``optical''
investigations of correlated materials often limited to
temperature during the not so distance past is now beginning to
include high pressure, static and pulsed magnetic fields and
nano-scale spatial resolution. The breadth of applicability of
``optical methods'' to investigate correlations is further
highlighted by important insights rapidly obtained for newly
discovered materials such as graphene and the iron-pnictides.\\

\noindent Several highlights stand out:

\begin{enumerate}
\item The formation of a conducting state in a correlated
insulator is associated with the development of low-energy
spectral weight at the expense of suppression of excitations in
the charge-transfer and/or $U$ region. Optical tools enable a
comprehensive inquiry into this behavior revealing common physics
between oxide and organic Mott systems.
\item Sum rule analysis of the optical conductivity provides a potent
experimental method to classify complex materials based on the
strength of their correlations (Fig.~\ref{fig:KE-all}). The
development of low-energy spectral weight with doping that can be
determined with the help of sum rules yields a reliable estimate of
$W/U$ for a correlated material (Sections~\ref{sec:Transition Metal
Oxides} and \ref{subsec:Two-dimensional molecular crystals}).
\item The analysis of optical constants offers detailed information on
renormalization of effective masses and Fermi velocities for
electrons or holes in a correlated host. The energy dependence of
these renormalized quantities is most valuable for uncovering the
fundamental interactions ultimately responsible for renormalizations
in oxides, heavy fermion and organic systems.
\item Several classes of doped correlated materials become
superconducting. Infrared optics allows one to measure both the
energy gap and the superfluid density tensor. Relatively small
values of the superfluid density are believed to be an essential
aspect of superconductivity of synthetic conductors also pointing
the prominence of phase fluctuations in these systems. Advances in
the experimental precision and reproducibility of optical
spectroscopy have made it possible to routinely obtain detailed
spectra of the glue to which the electrons near the Fermi energy
are coupled (Sec. \ref{subsec:Electron-boson interaction}).
Correlations between the glue-spectra and $T_c$ are being
established.
\item Optical methods have shown that the gross features of both
Kondo insulating and Heavy-Fermion behavior are understood with
the Anderson lattice model. New measurements extending data to the
very far-IR and microwave regions uncovered systematic deviations
from this model some of which are captured by DMFT analysis
(Sec.~\ref{subsec:Intermetallic Compounds}).
\item Pump-probe spectroscopy of correlated electron materials has made
considerable headway during the past decade. This has primarily been
through the judicious application of femtosecond studies to all of
the material classes discussed in this review coupled with
developments in generating short pulses through the far to
mid-infrared portions of the spectrum. Initial experiments suggest
that the sensitivity of correlated materials to external
perturbations make them promising candidates to investigate the
physics of photoinduced phase transitions (Sec.
\ref{subsec:Photoinduced phase transitions}).
\item Frequency-domain spectroscopy rests upon a well-developed
theoretical foundation (\ref{subsec:general optic theory}) with DMFT
emerging as a powerful tool to calculate the optical response of
correlated electron materials. The qualitative agreement of the main
experimental spectral features with DMFT calculations for
Hubbard-like and Kondo-like systems is an important step forward
(Fig. 5). Of perhaps even greater importance is its combination with
(ab-initio) structure, like LDA calculations and the potential for
quantitative material-specific predictions with first encouraging successes
V$_{2}$O$_{3}$
(Fig. 24(b))  and CeIrIn$_{5}$ (Fig. 44).
\end{enumerate}

One challenge for future work, of course, is to advance  our basic
understanding of the role of correlations with the longer term
view towards developing, in conjunction with theory, predictive
capabilities of the electronic properties of specific materials.
As the understanding of correlated electron materials advances
hand-in-hand with the ability to synthesize new materials with
specific properties, it will be crucial to investigate possible
technological applications. The exquisite sensitivity of numerous
correlated electron materials to external perturbations would seem
to be of some promise for applications ranging from novel switches
to chemical sensors. Recent examples in this direction include
oxide heterostructures (Fig.~\ref{heteros}) and graphene
(Fig.~\ref{fig:graphene}). Time-integrated and time-resolved
optical spectroscopy will undoubtedly play an important role
towards investigating potential applications as has been the case
in characterizing semiconductor heterostructures and devices.

Apart from predictive capabilities it is imperative to develop ways
to tune and control properties of correlated materials. Again, oxide
heterostructures offer some interesting approaches. A 2D electron
gas at interfaces can have high mobility and can be gated. Recent
advances have allowed local control of this 2D gas with scanning
probes' structures with writable properties in real space. Memory
effects are equally appealing for applications and control. For
example, it has only recently been demonstrated that ferromagnetics
can be demagnetized on a sub-picosecond timescale using femtosceond
pulses
\cite{beaurepaire1996a,koopmansPRL2005,bigot2009a,zhang2009NPa,kimel2005a}.
The ramification of such a possibility are under active
investigation and include the possibility of ultrafast memory
storage. It can be envisioned that, in a similar vein, interesting
possibilities exist in correlated materials including multiferroics
and organics.

There are nascent experimental techniques that build on the
intellectual and technological developments associated with
``conventional" optical spectroscopy that are of considerable
promise for future studies of electronic correlations. For
example, as discussed in section V. F, electronic phase separation
and intrinsic (or extrinsic) inhomogeneities are of considerable
importance. Spatially resolved optical probes provide an exciting
approach in the study of electronic phase separation complementing
well-developed techniques such as scanning-probe tunnelling
spectroscopy. In the future, it will be important to broaden the
spectral range over which such studies can be carried out and also
to extend nano-scopy to cryogenic temperatures.

To date, the majority of time-resolved optical experiments have
been limited to pumping in the 1.5 to 3.0~eV range, leading to a
cascade of scattering processes as highly energetic quasiparticles
relax to low energy states which can (though not always) result in
a fairly indirect and uncontrolled way to perturb a material.
However, the development of intense pulses at lower photon
energies (mid-IR to THz pulses) will help to alleviate this issue.
For example, interesting experiments on semiconductors have probed
the non-equilibrium physics of polarons and, in the future,
resonant excitation of superconductors with intense THz pulses at
the gap energy will likely provide new insights into their
non-equilibrium properties.  Further, as discussed in section IV.
E on photoinduced phase transitions, high-intensity coupled with
spectral agility provides the ability to pump phonon modes with a
view towards vibrational excitation and control of the electronic
properties of complex materials. These pump-probe studies may also
provide means to manipulate components of the pairing "glue" in
novel superconductors. There is also a need to increase the
sensitivity of time-resolved experiments as this directly
translates to the ability to photexcite at lower fluences. This is
important in the investigation of materials with low transition
temperatures and in delicately probing the dynamics within a given
phase of particular material.  We also stress that the theoretical
underpinnings of time-domain spectroscopy of correlated electron
materials is far less developed and, while presenting a
considerable challenge, offers numerous opportunities. This
includes, as examples, basic questions ranging from the
information content of time-resolved in comparison to steady-state
spectroscopy to the physics of photodoping induced phase
transitions.

Increased spatial and temporal resolution of optical probes enables
experiments away from steady state and homogeneous samples. These
previously unattainable regimes may require a revision of basic
ideas behind an optical probe of phenomena introduced within
conventional description of electrons in the momentum space.
Progress with the studies of inhomogeneous systems critically relies
on advances enabling to deal with the spatial variation of
properties literally at the nanoscale. Ultrashort pulses of extreme
intensity changes our common understanding of electronic excitations
in the frequency domain. For optics this means to reconsider many
fundamental issues, including but not limited to locality,
equilibrium, linearity, and Kramers-Kronig consistency.

\section*{Acknowledgements}
Over the last years we had the privilege to work with a large
number of collaborators, postdocs and students and their
contributions found access to many parts of this review. We had
many valuable discussions with colleagues at different occasions
and we want to thank all of them. D.B. enjoyed support by the
Alexander von Humboldt Foundation during his stay in Stuttgart. We
also want to acknowledge support by the National Science
Foundation (NSF), the Department of Energy (DOE), the Airforce
Office of Scientific Research (AFOSR), the Office of Naval
Research (ONR), the Electronics and Telecommunications Research
Institute (ETRI), the Deutsche Forschungsgemeinschaft (DFG), the
Swiss National Science Foundation and the National Center of
Competence in Research (MaNEP).

\end{document}